\documentclass[10pt]{iopart}
\usepackage{epsfig,graphicx,color}
\usepackage[mathscr]{euscript}

\newcommand{\vsp}{\vspace*{3mm}}
\newcommand{\hsp}{\hspace*{3mm}}
\newcommand{\be}{\begin{equation}}
\newcommand{\ee}{\end{equation}}
\newcommand{\bea}{\begin{eqnarray}}
\newcommand{\eea}{\end{eqnarray}}
\newcommand{\bean}{\begin{eqnarray*}}
\newcommand{\eean}{\end{eqnarray*}}
\newcommand{\one}{1\!\!{\rm I}}

\newcommand{\bra}{\langle}
\newcommand{\ket}{\rangle}

\newcommand{\order}{{\mathcal O}}

\newcommand{\bnull}{\mbox{\boldmath $0$}}

\newcommand{\R}{{\rm I\!R}}
\newcommand{\C}{{\rm |\!\!C}}

\newcommand{\bu}{\mbox{\protect\boldmath $u$}}
\newcommand{\bv}{\mbox{\protect\boldmath $v$}}

\newcommand{\bx}{\mbox{\protect\boldmath $x$}}
\newcommand{\by}{\mbox{\protect\boldmath $y$}}
\newcommand{\bz}{\mbox{\protect\boldmath $z$}}
\newcommand{\bA}{\mbox{\protect\boldmath $A$}}

\newcommand{\bC}{\mbox{\protect\boldmath $C$}}
\newcommand{\bD}{\mbox{\protect\boldmath $D$}}

\newcommand{\bF}{\mbox{\protect\boldmath $F$}}
\newcommand{\bG}{\mbox{\protect\boldmath $G$}}

\newcommand{\bM}{\mbox{\protect\boldmath $M$}}

\newcommand{\bP}{\mbox{\protect\boldmath $P$}}
\newcommand{\bQ}{\mbox{\protect\boldmath $Q$}}

\newcommand{\bZ}{\mbox{\protect\boldmath $Z$}}
\newcommand{\bbeta}{\mbox{\protect\boldmath $\beta$}}

\newcommand{\btheta}{\mbox{\protect\boldmath $\theta$}}

\newcommand{\bxi}{\mbox{\protect\boldmath $\xi$}}

\newcommand{\bmu}{\mbox{\protect\boldmath $\mu$}}

\newcommand{\bSigma}{\mbox{\protect\boldmath $\Sigma$}}

\newcommand{\bXi}{\mbox{\protect\boldmath $\Xi$}}
\newcommand{\Prob}{\mathscr{P}}
\newcommand{\Data}{\mathscr{D}}

\newcommand{\rmD}{{\rm D}}

\newcommand{\y}{\by}
\newcommand{\z}{\bz}
\renewcommand{\C}{\bC}
\newcommand{\D}{\bD}

\newcommand{\bb}{\bbeta}

\newcommand{\A}{\bA}
\newcommand{\zz}{\zeta}
\newcommand{\half}{{\frac{1}{2}}}
\newcommand{\Chat}{\hat{C}}
\newcommand{\tb}{\tilde{\beta}}
\newcommand{\tbb}{\tilde{\bbeta}}
\newcommand{\PP}{\bP}
\newcommand{\QQ}{\bQ}

\newcommand{\uu}{\bu}
\newcommand{\vv}{\bv}
\newcommand{\tS}{\tilde{S}}

\eqnobysec %equation numbering by section

\begin{document}

\title[Replica analysis of  overfitting in generalized linear regression models ]{Replica analysis of  overfitting in\\ generalized linear regression models }

\author{ACC Coolen$^{\dag\S\P}$, M Sheikh$^{\ddag}$, A Mozeika$^{\P}$,\\[0.5mm] F Aguirre-Lopez$^{\ddag}$ and F Antenucci$^\S$\\[3mm]}

\address{
$\dag$ Dept of Biophysics, Radboud University,  
6525AJ Nijmegen, The Netherlands
\\
$\S$ Saddle Point Science Ltd, 35A South St, London W1K2XF, UK
 \\
$\P$ London Inst for Mathematical Sciences, 35A South St, London W1K2XF, UK\\
$\ddag$ Dept of Mathematics,  King's College London, London WC2R 2LS, UK}
\ead{a.coolen@science.ru.nl}

\begin{abstract}
Nearly all statistical inference methods were developed for the regime where the number $N$ of data samples is much larger than the data dimension  $p$. 
Inference protocols such as maximum likelihood (ML) or maximum a posteriori probability (MAP) are  unreliable if $p=\order(N)$, due to overfitting. This limitation has  for many disciplines with increasingly high-dimensional data become a serious bottleneck. We recently showed that in Cox regression for time-to-event data the overfitting errors are not just noise but take mostly the form of a bias, and how with the replica method from statistical physics one can model and predict this bias and the noise statistics.  Here we extend our approach to arbitrary generalized linear regression models (GLM), with possibly correlated covariates. We analyse overfitting in ML/MAP inference without having to specify data types or regression models, relying only on the GLM form, and derive generic order parameter equations for the case of $L2$ priors. Second, we  derive the probabilistic relationship between true and inferred regression coefficients in GLMs, and show that, for the relevant hyperparameter scaling and correlated covariates, the $L2$ regularization causes a predictable direction change of the coefficient vector. 
Our results, illustrated by application to linear, logistic, and Cox regression, enable one  to correct ML and MAP inferences in GLMs systematically for overfitting bias, and thus extend their applicability into the hitherto forbidden regime $p\!=\!\order(N)$.

\color{black}
\end{abstract}

\vsp


%\clearpage\tableofcontents\clearpage

\section{Introduction}

Extensive quantities of data are now available in many commercial, scientific and medical settings, due to the decreasing cost of high throughput measurement devices and data storage, and rapidly increased computing power. Here we will be concerned with data where each sample is a pair $(\bz,s)$, with $\bz\in\R^p$ (the input, or covariate vector) and with an output variable $s$. The latter can be real-valued, discrete, or even composite. The aim is to  determine  from a given set $\Data=\{(\bz_1,s_1),\ldots, (\bz_N,s_N)\}$ of randomly drawn historic samples whether there is information in $\bz$ about $s$, and to predict the value of $s$ associated with any vector $\bz$.  
In parametric statistical inference one approaches this question by postulating a parametrized probabilistic model $p(s|\bz,\btheta)$ for the dependence of $s$ on  $\bz$, followed by defining a function  $\Omega(\btheta|\Data)$ whose minimization gives a sensible estimate for the parameters $\btheta$.  
In this work we will focus on generalized linear models (GLMs, \cite{MccullaghNelder}), 
which are  regression models $p(s|\bz,\btheta)$ in which the covariates $\bz$ enter strictly via a  
linear combination  $\bbeta\!\cdot\!\bz\!=\!\sum_{\mu=1}^p \beta_\mu z_\mu $, with coefficients $\bbeta\in\R^p$. 
For the $\Omega(\btheta|\Data)$ function, common choices are $\Omega(\btheta|\Data)=-\log p(\Data|\btheta)$ (ML regression) and  $\Omega(\btheta|\Data)=-\log p(\btheta|\Data)$ (MAP regression). 
Here $p(\btheta|\Data)\propto p(\Data|\btheta)p(\btheta)$, and $p(\Data|\btheta)=\prod_i p(s_i|\bz_i,\btheta)$. MAP requires the specification of a prior parameter distribution $p(\btheta)$. ML and MAP can be seen as approximations of a computationally often intractable Bayesian  approach, where  one works with the full posterior distribution $p(\btheta|\Data)$. MAP replaces this posterior by a delta peak at the most probable point, and ML follows from MAP by choosing a flat prior $p(\btheta)$.
ML performs well when $p\!\ll \!N$, but its estimators become increasingly incorrect in the high-dimensional regime,  
which for GLMs involves both the number $N$ of samples  and the number $p$ of covariates diverging,  
with finite ratio $\zeta=p/N$. 
Remedial attempts can be categorized into corrective \cite{bartlett1953approximateI, bartlett1953approximateII, haldane1956sampling, anderson1979logistic,CoxSnell,shenton1963higher,shenton1969maximum,gauss1991} or preventative ones \cite{firth1993bias}. The former route seeks to construct better estimators via a power series in $N^{-1}$ (for fixed $p$), with the ML estimator as zeroth term, but becomes unwieldy beyond the linear term \cite{BowmanShenton}. Also various computational resampling recipes were proposed \cite{Efron,Quenouille}, and a wider family of estimators beyond ML/MAP \cite{Kosmidis}. All these remedial and corrective approaches tend to work only when the data dimension $p$ is small and fixed. The most popular remedial approaches to overfitting in the statistics and machine learning communities are regularization, i.e. MAP inference with optimized priors, and variable selection, i.e. regression with low-dimensional representations of the vectors $\bz$.

We tackle the high-dimensional statistics regime adopting a statistical physics perspective. 
Statistical physics provided many contributions to our understanding of information processing 
in this regime, 
that is not captured in traditional statistics.
Statistical physics tends to deal with `typical-case' scenarios, unlike the `worst-case' analysis 
more prevalent in statistics and computer science (see e.g. \cite{FML18}). 
The two approaches are complementary, with `typical-case' behaviour becoming relevant if the `worst-case' scenario 
is very rare as $N\! \to\! \infty$. 
Within statistical physics, the techniques from the field of spin glasses have been particularly effective,  
especially replica theory and the cavity method (and related message passing algorithms).

The replica method \cite{MPV87} gives relatively simple prescriptions  
for computing asymptotic joint distributions of model parameters, and allows one to predict  asymptotic values 
of statistical estimators. It led to valuable progress in various areas of computer science \cite{mezard2009information},  
in particular in machine learning \cite{seung1992statistical,watkin1993statistical,opper1996statistical,dietrich1999statistical,gardner1988space}. 
Although it is adaptable to many machine learning problems, the replica method  is not always provably exact.  Hence considerable effort has
been dedicated to proving rigorously the replica predictions in specific settings, for
problems originating from statistical physics \cite{talagrand2003spin},  and in machine learning
(e.g. low-rank matrix factorisation \cite{dia2016mutual}). Alternative methods were also proposed to derive the replica results, based
on the elegant interpolation technique \cite{guerra2003broken},  and later extended to Bayesian inference \cite{barbier2019adaptive}.

Inference problems are intrinsically algorithmic. One ideally wants computationally efficient  methods for finding  the answer to any problem  instance. In this aspect, the insights from statistical physics originate from the
iterative procedure for computing marginals in Ising spin models with pair interactions,
known as the Thouless-Anderson-Palmer (TAP) equations \cite{ThoulessAnderson77}.  When implemented correctly  \cite{bolthausen2014iterative}, it is equivalent to the belief propagation (BP) approach in computer science \cite{gallager1968information,pearl1982reverend}, as was realised in  \cite{kabashima1998belief}. 
For continuous variables and multibody interactions, the TAP approach to inference is now commonly referred to as Approximate Message Passing (AMP)  \cite{donoho2009message}. 
AMP is computational simple, usually competitive with
the fastest algorithms, rigorously characterized
in the large system size limit by the cavity method (or state evolution), and used to predict accurately 
performance metrics such as the mean-squared error (MSE) or the detection accuracy
\cite{bayati2015universality,bayati2011dynamics,deshpande2014information,matsushita2013low,lesieur2017statistical}.
A generalization of AMP to arbitrary priors 
and component-wise output functions is found in \cite{rangan2011generalized}, 
which coined the name generalized approximate message passing (GAMP) for the generalized linear model. 
Message passing tools were also used to study logistic regression in the high-dimensional regime, for  ML and MAP \cite{sur2019modern,salehi2019impact}.  
Implications of overfitting for likelihood ratio tests in the $p=\order(N)$ regime were explored in \cite{sur2019likelihood}.

Despite its successes, the application of AMP to real-world problems has been limited by its dependence on a Bayes-optimal setting and specific structural features of the data matrix. When there is model mismatch, replica symmetry may be broken and AMP may stop converging \cite{antenucci2019approximate,takahashi2020macroscopic}. 
When the distributions are unknown, one could try to find a minimax estimator over a class of
distributions \cite{donoho2011compressed}, or combine GAMP with expectation-maximization (EM) estimation \cite{krzakala2012statistical,krzakala2012probabilistic,kamilov2012approximate,vila2011expectation,vila2013expectation}. 
Alternatively, AMP can be modified to a replica symmetric broken (RSB) structure, but this algorithm 
becomes computationally more demanding with each RSB step, and requires introducing additional parameters 
whose values are not easily determined \cite{antenucci2019approximate}.
Nevertheless, for linear models the main limitation of AMP is often the structure of the data matrix.
AMP's original version holds only for i.i.d. sub-Gaussian random data matrices \cite{bayati2015universality,bayati2011dynamics,rangan2011generalized}, 
and  AMP is fragile with regard to alternative choices. For
example, it diverges for even mildly ill-conditioned or
non-zero-mean data matrices \cite{rangan2019convergence,caltagirone2014convergence,vila2015adaptive}. 
Several heuristic strategies have been proposed for inducing AMP to converge 
\cite{rangan2019convergence,vila2015adaptive,vila2015adaptive,manoel2015swept,rangan2016inference}
%Strategies for inducing AMP to converge include damping \cite{rangan2019convergence,vila2015adaptive}, mean-removal \cite{vila2015adaptive}, sequential updating \cite{manoel2015swept}, and direct free-energy
%minimization \cite{rangan2016inference}, 
but their effectiveness is limited. 
Other algorithms for linear regression have been designed using approximations of belief propagation
(BP) and/or free-energy minimization, such as 
Adaptive TAP \cite{opper2001adaptive}, Expectation Consistent
Approximation (EC) \cite{opper2005expectation,kabashima2014signal,fletcher2016expectation}, (S-transform AMP) S-AMP \cite{cakmak2014s}, 
and (Orthogonal AMP) OAMP \cite{ma2017orthogonal}.
Numerical experiments suggest that some are indeed more robust than AMP,  but their convergence has not been rigorously determined. 
Currently,  the AMP-like algorithm with a rigorous underpinning that is able to handle the broadest class of data matrices is Vector Approximate Message Passing (VAMP) \cite{rangan2019vector}, which converges correctly 
for all large random data matrices that are right-orthogonally invariant. While this class of matrices relax the need for fully independent matrix elements, it is still an excessively specific requirement for most practical applications.

In this work we consider GLMs and focus on generalising the structure of the data matrix,
employing the replica method -- the most versatile of our theoretical tools.
We build on recent studies \cite{coolen2017replica, SheikhCoolen2019} which gave an accurate quantitative analysis of  overfitting in (regularized) Cox models \cite{Cox,Cox_book} for time-to-event data.
We show how the calculations of  \cite{coolen2017replica, SheikhCoolen2019}  can be generalized to ML or MAP regression with arbitrary GLMs.  Here we consider only models with a single  linear combination of covariates (which includes logistic and ordinal class regression, perceptrons \cite{Coolenbook}, and other survival analysis models such as frailty and random effects models \cite{survival_analysis}). Generalization to models with multiple linear combinations (e.g. neural networks with hidden layers, or survival analysis with competing risks \cite{Hougaard}) is straightforward. We analyse overfitting in ML or MAP inference with GLMs without requiring the entries of the data matrix to be uncorrelated. 
We only assume that there is no model mismatch, and that  the $N$ covariate vectors $\{\bz^i\}$ are drawn independently from some distribution $p(\bz)$.  This distribution may describe correlated entries, provided some weak conditions on the spectrum of the correlation matrix are met.  We refer to this setting as row-independent data matrix.
Using only the generalized linear form of the models, we derive generic replica order parameter equations within the replica-symmetric ansatz (RS), for the case of Gaussian priors\footnote{The present limitation to $L2$ (i.e. Gaussian) priors is not critical, alternative choices such as $L1$ (or lasso) priors simply lead to more complicated integrals.}. Only at the stage of solving order parameter equations will one have  to specify model details. We also  calculate the probabilistic relationship between true and inferred association coefficients, and show that, when covariates are correlated and $L2$ regularizers are sufficiently strong to have an effect,   the latter induce a predictable direction change of the coefficient vector.  For linear regression problems, similar studies of MAP estimators are present in literature. Again these 
assume i.i.d. elements or some form of rotational invariance for the data matrix, either within  
an RS  \cite{rangan2009asymptotic,vehkapera2016analysis}  or an  RSB ansatz \cite{bereyhi2019statistical}.
Within the setting of rotationally  invariant data matrices, the authors of  \cite{gerbelot2020asymptotic} define an oracle version of VAMP and show rigorously that the corresponding state evolution converges to the MSE state evaluated  via replica theory   in \cite{kabashima2014signal,rangan2009asymptotic,kabashima2012typical}.
Similar proofs of replica results can be found also for  MMSE estimators in linear regression, see e.g.  
 \cite{barbier2016mutuallinear,reeves2016replica,barbier2019optimal} for Gaussian data matrices and 
\cite{barbier2018mutualbeyond} for rotationally invariant data matrices. They cannot immediately be extended to the row-independent data matrices we consider in this work. In \cite{krzakala2012statistical} the replica method is used to analyse properties of the MAP estimator in compressed sensing. Our present setting differs in two key aspects: we consider generalized linear models and we remove the need for i.i.d. entries of the data matrix. 
There is presently no AMP-like algorithm that provably works for independently drawn covariate vectors with correlated entries.    
The present RS replica calculation, however,  is able to deal with such more realistic data,  and  the mathematical physics literature provides evidence that any failures of the replica solution in practical applications are most likely to reflect model mismatch, i.e. a breaking of the replica symmetry, as opposed to fundamental features of the analytical continuation. 
Our limitation to ML and MAP estimators is also not crucial, and the results could be easily extended to e.g. the minimum mean square error (MMSE) estimator. We concentrate on ML and MAP because they are less computationally demanding, in the absence of an AMP-like algorithm for row-independent data matrices, and their practical evaluation is usually feasible using standard numerical methods.

This paper is organized as follows. We first generalize in section 2 the replica analysis of \cite{coolen2017replica, SheikhCoolen2019} to arbitrary GLMs. In section 3 we calculate the quantitative relation between true and inferred association parameters in the overfitting regime, for potentially correlated covariates, and show how our results can be used to compute new estimators that are decontaminated for overfitting distortions (via bias removal, or MSE minimization). In section 4 we test our theoretical predictions 
via application to  linear, logistic and Cox regression,   recovering some known results as a test, and deriving  several new ones. We close with a discussion of present and future work. 
Most of the more technical calculations are relegated to appendices, to focus the main text on the key ideas and outcomes.
 In contrast to most analytical studies on overfitting in literature, our theory is not limited to linear models, to uncorrelated covariates, to scalar outputs,  or to models with output noise. 
Our present results enable one  to correct ML and MAP inferences in generalized linear regression models for overfitting distortions, and thereby extend the applicability of these popular regression methods into the hitherto forbidden regime $p\!=\! \order(N)$. 

\section{General theory of GLM regression}

\subsection{Definitions and notation}

In generalized linear regression models, the probability (density) of observing an outcome $s\in\Omega$ depends on the values of covariates $\bz\in\R^p$ via an expression of the form ${\rm Prob}(s|\bz)=p(s|\bbeta\cdot\!\bz/\!\sqrt{p},\theta)$, with $\theta$ representing any auxiliary parameters that are not coupled to covariates. The covariates appear only in an inner product with a vector $\bbeta$ of so-called association parameters. The outcome set $\Omega$ can be continuous, discrete, or a combination of both (for multi-valued outcomes), and the auxiliary parameters $\theta$ can  even be a function, as in the Cox model \cite{Cox}. We consider MAP inference without model mismatch, where upon observing a data set $\Data=\{(\bz_1,s_1),\ldots,(\bz_N,s_N)\}$ in which all samples $(\bz_i,s_i)$ are assumed to have been drawn randomly and independently from a distribution of the form $p(\bz,s)= p(\bz) p(s|\bbeta^\star\!\cdot\bz/\!\sqrt{p},\theta^\star)$, the inferred parameters $(\hat{\bbeta},\hat{\theta})$ are those that maximize the Bayesian posterior parameter probability 
  $p(\bbeta,\theta|\Data)$:
  \begin{eqnarray}
  p(\bbeta,\theta|\Data)&=& \frac{p(\Data|\bbeta,\theta) p(\bbeta,\theta)}{\int\!\rmd\bbeta^\prime \rmd\theta^\prime~p(\Data|\bbeta^\prime,\theta^\prime) p(\bbeta^\prime,\theta^\prime)}
  \nonumber
  \\
  &=& \frac{p(\bbeta,\theta)\prod_{i=1}^N p(s_i|\bbeta\!\cdot\!\bz_i/\sqrt{p},\theta)}{\int\!\rmd\bbeta^\prime \rmd\theta^\prime~ p(\bbeta^\prime,\theta^\prime)\prod_{i=1}^N p(s_i|\bbeta^\prime\!\cdot\!\bz_i/\sqrt{p},\theta^\prime)}.
  \end{eqnarray}
Hence, upon taking a logarithm and discarding an irrelevant  constant, 
\begin{eqnarray}
(\hat{\bbeta},\hat{\theta})_{\rm MAP}&=& {\rm argmax}_{\bbeta,\theta}\Big\{\sum_{i=1}^N \log p\Big(s_i|\frac{\bbeta\!\cdot\!\bz_i}{\sqrt{p}},\theta\Big)+\log p(\bbeta,\theta)\Big\}.
\label{eq:MAPdefn}
\end{eqnarray}
Choosing a regression model implies choosing a parametrization $p(s|\xi,\theta)$ and a prior $p(\bbeta,\theta)$. We recover ML from MAP regression by choosing the prior to be constant. Our convention to define GLMs in terms of $\bbeta\!\cdot\!\bz/\sqrt{p}$ as opposed to $\bbeta\!\cdot\!\bz$ ensures that even for $p\to\infty$ the components of $\bbeta$ will typically scale as $\beta_\mu=\order(1)$.
Following mainstream  literature, we will for simplicity choose nontrivial priors only for the coefficients $\bbeta$, where their inclusion is indeed most critical, so $p(\bbeta,\theta)\propto p(\bbeta)$.

For instance, the simplest GLM is 
linear regression, where one has outcomes from $\Omega=\R$, two auxiliary parameters  $\theta=(\beta_0,\sigma)$ with $\beta_0\in\R$ and $\sigma>0$, and 
\begin{eqnarray}
p(s|\xi,\beta_0,\sigma)&=&(2\pi\sigma^2)^{-\frac{1}{2}}\rme^{-\frac{1}{2}(s-\xi-\beta_0)^2/\sigma^2}\!.
\end{eqnarray}
In logistic regression, which can be seen  as a stochastic generalization of the binary perceptron \cite{Coolenbook},  we have  $\Omega=\{-1,1\}$, one auxiliary parameter  $\theta=\beta_0\in\R$, and 
\begin{eqnarray}
p(s|\xi,\beta_0)&=& \frac{1}{2}+\frac{1}{2}s\tanh(\xi\!+\!\beta_0)
=\frac{\rme^{s(\xi+\beta_0)}}{2\cosh(\xi\!+\!\beta_0)}.
\end{eqnarray}
In Cox regression \cite{Cox} without censoring we have $\Omega=[0,\infty)$ and a functional auxiliary parameter $\theta=\{\lambda(t)\}$, the base hazard rate, with
\begin{eqnarray}
p(s|\xi, \lambda)&=& \lambda(s)\exp\Big[\xi\!-\!\rme^\xi\!\int_0^s\!\!\rmd s^\prime~\lambda(s^\prime)\Big]
=-\frac{\partial}{\partial s}\exp\Big[\!-\!\rme^{\xi}\!\int_0^s\!\!\rmd s^\prime~\lambda(s^\prime)\Big].
\nonumber
\\[-1mm]&&
\end{eqnarray}
For Cox regression with censoring, $\Omega=[0,\infty)\times \{0,1\}$ (the outcome is a pair $(t,r)$ of an event time $t$ and a  binary label $r$ indicating whether the event was a primary one or censoring), with two functional auxiliary parameters $\theta=\{\lambda_0(t),\lambda_1(t)\}$ (the base rates of the primary and the censoring events), and 
\begin{eqnarray}
p(t,r|\xi, \lambda_0,\lambda_1)&=& \lambda_r(t)\exp\Big[\xi\delta_{r 1}\!-\!\int_0^t\!\!\rmd t^\prime~\lambda_0(t^\prime)-\rme^\xi\!\int_0^t\!\!\rmd t^\prime~\lambda_1(t^\prime)\Big].
\end{eqnarray}
In proportional hazards ordinal class regression with $C$ discrete possible outcomes we have $\Omega=\{1,2,\ldots,C\}$ and $\theta=(\lambda_2,\ldots,\lambda_C)\in\R^{C-1}$, with
\begin{eqnarray}
\hspace*{-5mm}
&& p(c|\xi,\lambda_2,\ldots,\lambda_C)= \Big(1\!-\!\tilde{p}(c|\xi,\lambda_c)\Big)\prod_{c^\prime>c}^{C+1} \tilde{p}(c^\prime|\xi,\lambda_{c^\prime})
\\
\hspace*{-5mm}
&& \tilde{p}(1|\xi,\lambda)=0,~~~\tilde{p}(C\!+\!1|\xi,\lambda)=1,~~~~~~
  1\!<c\leq  C\!:~~ \tilde{p}(c|\xi,\lambda)=   \rme^{- \lambda\exp(\xi)}.
  \nonumber
\end{eqnarray}

\subsection{The information-theoretic overfitting measure}

We follow closely the procedure in  \cite{coolen2017replica, SheikhCoolen2019}, which can be adapted to  arbitrary GLMs with only minimal change. We start from the observation that MAP regression for any model of the type $p(s|\bz,\btheta)$ (whether or not of the GLM form)  
  is equivalent to minimization over the model parameters $\btheta$ of the quantity 
\begin{eqnarray}
\Omega(\btheta|\Data)&=& D(\hat{p}_{\Data}|| p_{\btheta})-N^{-1}\log p(\btheta).
\end{eqnarray}
Here $\hat{p}_{\Data}$ is the empirical distribution of covariates and outcomes in the data,  $\hat{p}(s,\bz|\Data)=N^{-1}\sum_{i\leq N}\delta(s\!-\!s_i)\delta(\bz\!-\!\bz_i)$, 
$D(\hat{p}_{\Data}|| p_{\btheta})$ is the Kullback-Leibler distance
\begin{eqnarray}
D(\hat{p}_{\Data}||p_{\btheta})=\int\!\rmd\bz \rmd s ~\hat{p}(s,\bz|\Data)\log\Big(\frac{\hat{p}(s|\bz,\Data)}{p(s|\bz,\btheta)}
\Big),
\end{eqnarray}
and $p_{\btheta}$ is the assumed parametrized regression model, with covariate-conditioned outcome probabilities $p(s|\bz,\btheta)$.  
For discrete variables, delta functions are replaced by Kronecker delta symbols.\footnote{Note that this definition is very different from the Kullback-Leibler distance employed in e.g. \cite{Peng,Quintero}, which, in our present notation,  measures the deviation between the parameter posteriors $p(\btheta|\Data)$ computed before and after removal of a single data sample.}
Assuming that our data were indeed generated from a model of the assumed form, with (unknown) parameters $\btheta^\star\!$, a transparent overfitting measure can be defined as $E(\btheta^\star\!, \Data)={\rm min}_{\btheta}~ \Omega(\btheta|\Data)-\Omega(\btheta^\star|\Data)$, giving
\begin{eqnarray}
E(\btheta^\star\!,\Data)&=& \min_{\btheta}\Bigg\{
\!\frac{1}{N}\sum_{i=1}^N \log\Big(\frac{p(s_i|\bz_i,\btheta^\star)}{p(s_i|\bz_i,\btheta)}\Big)
+\frac{1}{N}\log\Big(\frac{p(\btheta^\star)}{p(\btheta)}\Big)\Bigg\}.
\label{eq:Hamiltonian}
\end{eqnarray}
Perfect regression would give
$E(\btheta^\star\!,\Data)=0$, finding $E(\btheta^\star\!,\Data)<0$ implies overfitting, and finding $E(\btheta^\star\!,\Data)>0$ implies underfitting. In GLMs with the MAP regression protocol  (\ref{eq:MAPdefn}) the parameters would be $\btheta=(\bbeta,\theta)$. 
Our calculations focus on evaluating the average of (\ref{eq:Hamiltonian}) over the possible realizations of the data set $\Data$, whose samples are drawn randomly from $p(s,\bz)=p(s|\bz,\btheta^\star)p(\bz)$ for some $p(\bz)$. The average is handled using the replica identity 
\begin{eqnarray}
\bra \log Z\ket=\lim_{n\to 0}n^{-1}\log\bra Z^n\ket, 
\label{eq:standard_replica}
\end{eqnarray}
and we write the minimization as the computation of the ground state energy density of a statistical mechanical system with degrees of freedom $\btheta\in\R^p$ and Hamiltonian 
\begin{eqnarray}
H(\btheta|\btheta^\star\!,\Data)&=& \sum_{i=1}^N \log\Big(\frac{p(s_i|\bz_i,\btheta^\star)}{p(s_i|\bz_i,\btheta)}\Big)+\log\Big(\frac{p(\btheta^\star)}{p(\btheta)}\Big).
\end{eqnarray}
  We can thus model MAP regression as the zero noise limit of a  stochastic minimization of  $H(\btheta|\btheta^\star\!,\Data)$ at inverse noise level $\gamma$, giving, with the help of (\ref{eq:standard_replica}), 
\begin{eqnarray}
\hspace*{-10mm}
\bra E(\btheta^\star\!,\Data)\ket_{\Data}&=& \lim_{\gamma\to \infty}E_\gamma(\btheta^\star),
\\
\hspace*{-10mm}
E_\gamma(\btheta^\star)
&=&- \frac{\partial}{\partial\gamma}\frac{1}{N}\Big\bra \log \int\!\rmd\btheta~\rme^{-\gamma  H(\btheta|\btheta^\star\!,\Data)}\Big\ket_{\!\Data}
\nonumber
\\
\hspace*{-10mm}
&=&  -  \lim_{n\to 0}\frac{\partial}{\partial\gamma}\frac{1}{Nn}\log \int\!\rmd\btheta^1\!\ldots\rmd\btheta^n  \Big\bra\rme^{-\gamma  \sum_{\alpha=1}^n H(\btheta^\alpha|\btheta^\star\!,\Data)}\Big\ket_{\!\Data}
\nonumber
\\
\hspace*{-10mm}
&=& -  \lim_{n\to 0}\frac{\partial}{\partial\gamma}\frac{1}{Nn}\log \int\!\rmd\btheta^1\!\ldots\rmd\btheta^n  
 \prod_{\alpha=1}^n 
\Big[\frac{p(\btheta^\alpha)}{p(\btheta^\star)}\Big]^\gamma
\nonumber
\\
\hspace*{-10mm}
&&\hspace*{5mm} \times
\Big\{
\int\!\rmd\bz \rmd s~p(\bz)p(s|\bz,\btheta^\star)
 \prod_{\alpha=1}^n
\Big[\frac{p(s|\bz,\btheta^\alpha)}{p(s|\bz,\btheta^\star)}\Big]^\gamma\Big\}^N.
\label{eq:E_gamma}
\end{eqnarray}
Integrals over outcome variables become summations when these variables are discrete, and integrals over functional parameters are interpreted as path integrals.
In the alternative limit $\gamma\to 1$ the quantity $E_\gamma(\btheta^\star)$ would involve the average over all data realizations of the Bayesian estimator,  $E_1(\btheta^\star)=\bra \int\!\rmd\btheta~p(\btheta|\Data)\Omega(\btheta|\Data)-\Omega(\btheta^\star|\Data)\ket_{\Data}$.

Application of expression (\ref{eq:E_gamma}) to generalized linear regression models implies setting $\btheta\to (\bbeta,\theta)$ and $p(s|\bz,\btheta)\to 
p(s|\bbeta\!\cdot\!\bz/\sqrt{p},\theta)$, so that we obtain
\begin{eqnarray}
\bra E(\bbeta^\star\!,\theta^\star\!,\Data)\ket_{\Data}&=& \lim_{\gamma\to \infty}E_\gamma(\bbeta^\star\!,\theta^\star),
\end{eqnarray}
with
\begin{eqnarray}
\hspace*{-10mm}
E_\gamma(\bbeta^\star,\theta^\star)
&=& -  \lim_{n\to 0}\frac{\partial}{\partial\gamma}\frac{1}{Nn}\log \int\!\rmd\theta^1\!\ldots\rmd\theta^n  
\int\!\rmd\bbeta^1\!\ldots\rmd\bbeta^n
 \prod_{\alpha=1}^n 
\Big[\frac{p(\bbeta^\alpha)}{p(\bbeta^\star)}\Big]^\gamma
\nonumber
\\
\hspace*{-10mm}
&&\hspace*{0mm} \times
\Big\{
\int\!\rmd\bz \rmd s~p(\bz)p(s|\bz,\bbeta^\star,\theta^\star)
 \prod_{\alpha=1}^n
\Big[\frac{p(s|\bz,\bbeta^\alpha,\theta^\alpha)}{p(s|\bz,\bbeta^\star,\theta^\star)}\Big]^\gamma\Big\}^N\!.~~
\label{eq:starting_point_of_replicas}
\end{eqnarray}

\subsection{Replica analysis in the regime $p=\order(N)$}

In \ref{app:SheikhCoolen} we compute (\ref{eq:starting_point_of_replicas})  for $N,p\to\infty$ with fixed ratio $\zeta=p/N$,  assuming $p(\bz)$ to be a zero-average distribution on $\R^p$, and for $L2$ priors $p(\bbeta)\propto\exp(-\eta\bbeta^2)$\footnote{Note that this latter choice would become $p(\bbeta)\propto\exp(-\eta p\bbeta^2)$ for the alternative convention where the association coefficients are not rescaled by $\sqrt{p}$, i.e. for GLMs written as $p(s|\bbeta\cdot\bz,\theta)$.}.
 We include this derivation, which follows  \cite{SheikhCoolen2019}, for completeness. 
The outcome of the regression  process is characterized by the values of a finite number of order parameters\footnote{The order parameters have been determined within the so-called replica-symmetric (RS) ansatz, which implies the assumption that the stochastic optimization process at finite $\gamma$ is ergodic.}, in terms of which one can quantify the relation between inferred and true regression coefficients.
The result of \ref{app:SheikhCoolen}  is:
\begin{eqnarray}
\hspace*{-20mm}
\lim\limits_{N \to \infty} E_{\gamma}(\bb^\star, \theta^\star) &=&
\int\! {\rm D}y_0\!\int\!\rmd s~ p(s |S\bra a\ket^{\frac{1}{2}}y_0, \theta^\star)
\log p(s | S\bra a\ket^{\frac{1}{2}}y_0, \theta^\star)  
-\zeta\eta  S^2  
 \nonumber\\
  \hspace*{-20mm}
&&
 \hspace*{-20mm}
+~\eta\zeta\Bigg\{
w^2\bra a\ket  
 \Big\langle \frac{a^2}{2 \eta \gamma\! +\! ga}\Big\rangle^{\!\!-2}
 \Big\langle  \frac{a^2}{(2 \eta \gamma\! +\! ga)^2}\Big\rangle
+
 \Big\langle\! \frac{1}{2 \eta \gamma\!+\!ga} \Big\rangle- f\Big\langle\frac{a}{(2 \eta \gamma\! +\! ga)^2} \Big\rangle 
\Bigg\}
  \nonumber  \\
 \hspace*{-20mm}
&& 
\hspace*{-31mm}
 -\! \int\!{\rm D}z {\rm D}y_0\!\int\!\!\rmd s~ p(s |S\bra a\ket^{\frac{1}{2}}y_0, \theta^\star)
\frac{  \int \!{\rm D} y~
p^\gamma(s | uy \!+\! wy_0\!+\!vz, \theta)\log p(s | uy \!+\! wy_0\!+\!vz, \theta) }
{  \int \!{\rm D} y~
p^\gamma(s| uy \!+\! wy_0\!+\!vz, \theta) },
\nonumber
\\[-1mm]
\hspace*{-20mm}
&&\end{eqnarray}
with the shorthand ${\rm D}z=(2\pi)^{-\frac{1}{2}}\rme^{-\frac{1}{2}z^2}\rmd z$. Brackets denote averages over the limit $p\to\infty$ of the eigenvalue spectrum $\varrho(a) $ of the $p\times p$ covariate correlation matrix $\bA$, with entries $A_{\mu\nu}=\int\!\rmd\bz~p(\bz)z_\mu z_\nu$. This result depends on the true association parameter vector $\bbeta^\star$ only via the asymptotic rescaled amplitude $S^2=\lim_{p\to\infty}p^{-1}\bbeta^{\star 2}$, assuming the components of $\bbeta^\star$ to have been drawn randomly from a symmetric distribution with finite second and fourth moments. 
Of the covariate covariance matrix $\bA$ we only require that its eigenvalue spectrum obeys 
 $\lim_{p\to\infty} \int\!\rmd a~\varrho(a)a \in \R$ and 
$\lim_{p\to\infty}p^{-1}\int\!\rmd a~\varrho(a)a^2 =0$.
The order parameters $(u,v,w,f,g,\theta)$ are determined by extremization of the following quantity, which acts as a free energy density:
\begin{eqnarray}
\Psi_{\rm RS}(u,v,w,f,g,\theta)&=& \Psi^A_{\rm RS}(u,v,w,f,g) - \Psi^B_{\rm RS}(u,v,w,\theta),
\end{eqnarray}
with
\begin{eqnarray}
\hspace*{-20mm}
\Psi^A_{\rm RS}(\ldots)&=&
-\frac{1}{2}\zeta (g\!+\!f) u^2-\frac{1}{2}\zeta g(v^2\!+\! w^2)
\nonumber
\\
\hspace*{-20mm}
&&
\hspace*{0mm}
+
 \frac{1}{2}\zeta\Big\{
w^2\bra a\ket
 \Big\bra \frac{a^2}{
2\gamma\eta\!+\!ga}\Big\ket^{\!-1}\!\!\!
+\Big\bra\log\Big(2\gamma\eta\!+\!ga\Big)\Big\ket
+
f\Big\bra  
\frac{a}{2\gamma\eta\!+\!ga}\Big\ket
\Big\},
\\
\hspace*{-20mm}
\Psi^B_{\rm RS}(\ldots)&=&
\int\!{\rm D}y_0{\rm D}z\!\int\!\rmd s~ p(s|S\bra a\ket^{\frac{1}{2}}y_0,\theta^\star)
\log \!
\int\!\rmD y~
p^\gamma (s|uy\!+\!
wy_0\!+\!vz,\theta).~
\label{eq:Psi_before_gamma_limit}
\end{eqnarray}
Here 
$\theta^\star$ are the true (unknown) auxiliary model parameters assumed to have been used to generate the observed data. 
The physical meaning of the two main order parameters, expressed in terms of the MAP-inferred parameters $\hat{\bbeta}$  and the true  parameters $\bbeta^\star$ responsible for the data, is
\begin{eqnarray}
v&=& \lim_{p\to\infty}
\Bigg\bra \frac{1}{\sqrt{p}}\Big\{\hat{\bbeta}\cdot\bA\hat{\bbeta}-\frac{\big(\bbeta^\star\!\cdot\bA\hat{\bbeta}\big)^2}{\bbeta^\star\!\cdot\bA\bbeta^\star}\Big\}^{\frac{1}{2}}\Bigg\ket_{\!\!\Data}
\label{eq:meaning_v}
\\
w&=&  \lim_{p\to\infty}\Bigg\bra \frac{1}{\sqrt{p}}\frac{\bbeta^\star\!\cdot\bA\hat{\bbeta}}{\sqrt{\bbeta^\star\!\cdot\bA\bbeta^\star}}
\Bigg\ket_{\!\!\Data}
\label{eq:meaning_w}
\end{eqnarray}
Perfect regression, i.e. $\hat{\bbeta}=\bbeta^\star$, would give $v=0$ and $w=\lim_{p\to\infty}\bra \sqrt{\bbeta^\star\!\cdot\bA\bbeta^\star/p}\ket_{\Data}$.

In the limit $\gamma\to\infty$ the maximization of the posterior becomes deterministic, and we recover the formulae describing MAP inference. 
In the precursor studies \cite{coolen2017replica, SheikhCoolen2019} it was found that the canonical scaling of the RS order parameters for large $\gamma$ is
\begin{eqnarray}
u=\tilde{u}/\sqrt{\gamma},~~~~~~v,w,\theta=\order(1),~~~~~~g=\tilde{g}\gamma,~~~~~~f=\tilde{f}\gamma^2.
\end{eqnarray}
Assuming this scaling to hold more generally gives
\begin{eqnarray}
\hspace*{-22mm}
\lim_{\gamma\to\infty}\frac{1}{\gamma}\Psi^A_{\rm RS}(\ldots)&=&
 \frac{1}{2}\zeta\Big\{\!
 w^2\bra a\ket
 \Big\bra \frac{a^2}{
2\eta\!+\!\tilde{g}a}\Big\ket^{\!-1}\!\!
+
\tilde{f}\Big[\Big\bra  
\frac{a}{2\eta\!+\!\tilde{g}a}\Big\ket
\!-\! \tilde{u}^2\Big]- \tilde{g}(v^2\!+\! w^2)
\Big\},
\label{eq:Psi_A_large_gamma}
\\
\hspace*{-22mm}
\lim_{\gamma\to\infty}\frac{1}{\gamma}\Psi^B_{\rm RS}(\ldots)&=&
 \int\!{\rm D}y_0{\rm D}z\int\!\rmd s~ p(s|S\bra a\ket^{\frac{1}{2}}y_0,\theta^\star)
 \nonumber
 \\
\hspace*{-22mm}
 &&\times
\lim_{\gamma\to\infty}\frac{1}{\gamma}
\log 
\int\!\rmd y~\rme^{-\frac{1}{2}y^2}
p^\gamma (s|\tilde{u}y/\sqrt{\gamma}\!+\!
wy_0\!+\!vz,\theta)
\nonumber
\\
\hspace*{-22mm}
&&\hspace*{-15mm} = 
\int\!{\rm D}y_0{\rm D}z\!\int\!\rmd s~ p(s|S\bra a\ket^{\frac{1}{2}}y_0,\theta^\star)
{\rm max}_{y\in\R}\Big[
\log p(s|\tilde{u}y\!+\!
wy_0\!+\!vz,\theta)
-\frac{1}{2} y^2\Big]
\nonumber
\\
\hspace*{-22mm}
&&\hspace*{-15mm} = 
\int\!{\rm D}y_0{\rm D}z\!\int\!\rmd s~ p(s|S\bra a\ket^{\frac{1}{2}}y_0,\theta^\star)
{\rm max}_{\xi\in\R}\Big[
\log p(s|\xi,\theta)
\!-\!\frac{(\xi\!-\!wy_0\!-\!vz)^2}{2\tilde{u}^2}\Big].
\nonumber
\\[-0mm]\hspace*{-22mm}&&
\label{eq:Psi_B_large_gamma}
\end{eqnarray}
We next abbreviate $\Xi_A(\tilde{f},\tilde{g},\tilde{u},v,w)=\lim_{\gamma\to\infty}\gamma^{-1}\Psi^A_{\rm RS}(\ldots)$ and 
$\Xi_B(\tilde{u},v,w,\theta)=\lim_{\gamma\to\infty}\gamma^{-1}\Psi^B_{\rm RS}(\ldots)$, and write the various averages as 
$\bra\!\bra \ldots \ket\!\ket=\int\!{\rm D}y_0{\rm D}z \ldots$ and $\bra f(s)\ket_s
=\int\!\rmd s~ p(s|S\bra a\ket^{\frac{1}{2}}y_0,\theta^\star)f(s)$ (with the integral over $s$ replaced by a sum if  $s$ is discrete). We also define 
\begin{eqnarray}
\xi(\mu,\sigma,s,\theta)&=& {\rm argmax}_{\xi\in\R}\Big[
\log p(s|\xi,\theta)
\!-\!\frac{1}{2} (\xi\!-\!\mu)^2/\sigma^2\Big].
\end{eqnarray}
A sufficient condition for $\xi(\mu,\sigma,s,\theta)$ to exists is that ${\rm argmax}_{\xi\in\R} p(s|\xi,\theta)$ exists for all $(s,\theta)$, which we have found to be true in all GLM models considered so far.  
Note that $ \xi(\mu,\sigma,s,\theta)$ is the solution of 
\begin{eqnarray}
\frac{\partial}{\partial\xi}\log p(s|\xi,\theta)=
(\xi\!-\!\mu)/\sigma^2.
\label{eq:xi_equation}
\end{eqnarray}
Hence we may write the model-independent part of the quantity to be extremized as
\begin{eqnarray}
\hspace*{-20mm}
\Xi_A(\tilde{f}\!,\tilde{g},\tilde{u},v,w)&=&
 \frac{1}{2}\zeta\Big\{
 w^2\bra a\ket
 \Big\bra \frac{a^2}{
2\eta\!+\!\tilde{g}a}\Big\ket^{\!\!-1}\!\!
\!+
\tilde{f}\Big[\Big\bra  
\frac{a}{2\eta\!+\!\tilde{g}a}\Big\ket
\!-\! \tilde{u}^2\Big]\!- \tilde{g}(v^2\!+\! w^2)
\Big\},~
\end{eqnarray}
and the model-dependent part as
\begin{eqnarray}
\hspace*{-15mm}
\Xi_B(\tilde{u},v,w,\theta)&=& 
\Big\bra\!\Big\bra\! \Big\bra 
{\rm max}_{\xi\in\R}\Big[
\log p(s|\xi,\theta)
\!-\!\frac{1}{2} (\xi\!-\!wy_0\!-\!vz)^2/\tilde{u}^2\Big]
\Big\ket_{\!s}\Big\ket\!\Big\ket
\nonumber
\\
\hspace*{-15mm}
&=& 
\Big\bra\!\Big\bra \!\Big\bra 
\Big[
\log p(s|\xi,\theta)
\!-\!\frac{(\xi\!-\!wy_0\!-\!vz)^2}{2\tilde{u}^2}\Big]\Big|_{\xi=\xi(wy_0+vz,\tilde{u},s,\theta)}\Big\ket_{\!s}\Big\ket\!\Big\ket.
\label{eq:Psi_B_large_gamma_compact}
\end{eqnarray}
The RS order parameter equations can then be written as
\begin{eqnarray}
\hspace*{-10mm}
&& \frac{\partial\Xi_A}{\partial\tilde{f}}=\frac{\partial\Xi_A}{\partial\tilde{g}}= \frac{\partial\Xi_B}{\partial\theta}=0,
\label{eq:RS_eqns_uncoupled}
\\
&& 
 \frac{\partial\Xi_A}{\partial\tilde{u}}=\frac{\partial\Xi_B}{\partial\tilde{u}},~~~~~~
\frac{\partial\Xi_A}{\partial v}=\frac{\partial\Xi_B}{\partial v},~~~~~~
\frac{\partial\Xi_A}{\partial w}=\frac{\partial\Xi_B}{\partial w}.
\label{eq:RS_eqns_coupled}
\end{eqnarray}
In \ref{app:further} we analyse and simplify these RS order parameter equations further,  and find that we can rewrite our closed MAP order parameter equation set as:
\begin{eqnarray}
\hspace*{-20mm}
\Big\bra  
\frac{a}{2\eta\!+\!\tilde{g}a}\Big\ket
&=& \tilde{u}^2,
\label{eq:ddf2=0}
\\
\hspace*{-20mm}
w^2\Big[
\bra a\ket
 \Big\bra \frac{a^2}{
2\eta\!+\!\tilde{g}a}\Big\ket^{\!\!-2}
\!
 \Big\bra \frac{a^3}{
(2\eta\!+\!\tilde{g}a)^2}\Big\ket
\!-\!1\Big]
\!-\!
\tilde{f}\Big\bra  
\frac{a^2}{(2\eta\!+\!\tilde{g}a)^2}\Big\ket
&=&v^2,
\label{eq:ddg2=0}
\\
\hspace*{-20mm}
\Big\bra\!\Big\bra\!\Big\bra 
 [\xi(wy_0\!+\!vz,\tilde{u},s,\theta)\!-\!wy_0\!-\!vz]^2\Big\ket_{\!s}\Big\ket\!\Big\ket
 &=& - \zeta
\tilde{f} \tilde{u}^4,
 \label{eq:ddu2=0}
\\[2mm]
\hspace*{-20mm}
 \Big\bra\!\Big\bra \!\Big\bra 
(\partial_1\xi)(wy_0\!+\!vz,\tilde{u},s,\theta)\Big\ket_{\!s}\Big\ket\!\Big\ket 
&=& 1- \zeta \tilde{g}\tilde{u}^2,
 \label{eq:ddv2=0}
\\[1mm]
\hspace*{-20mm}
 \Big\bra\!\Big\bra\!\Big\bra  \xi(wy_0\!+\!vz,\tilde{u},s,\theta) \frac{\partial \log p(s|S\bra a\ket^{\frac{1}{2}}y_0,\theta^\star)}{\partial y_0}\Big\ket_{\!s}\Big\ket\!\Big\ket
&=& \zeta w \tilde{u}^2
 \bra a\ket
 \Big\bra \frac{a^2}{
2\eta\!+\!\tilde{g}a}\Big\ket^{\!-1}\!\!,
 \label{eq:ddw2=0}
\\[1mm]
\hspace*{-20mm}
\Big\bra\!\Big\bra\! \Big\bra \frac{\partial \log p(s|\xi,\theta)}{\partial\theta}\Big|_{\xi=\xi(wy_0+vz,\tilde{u},s,\theta)}
\Big\ket_{\!s}\Big\ket\!\Big\ket&=& 0.
 \label{eq:ddtheta2=0}
\end{eqnarray}
The function  $\xi(\mu,\sigma,s,\theta)$, defined as the solution of (\ref{eq:xi_equation}), obeys 
 $\lim_{\sigma\to 0} \xi(\mu,\sigma,s,\theta)=\mu$. 
Its partial derivative $(\partial_1\xi)(\mu,\sigma,s,\theta)$ follows upon working out the partial derivative with respect to $\mu$ of (\ref{eq:xi_equation}), 
\begin{eqnarray}
0&=&\frac{1}{\sigma^2}+ \frac{\partial\xi}{\partial\mu}\frac{\partial}{\partial\xi}\Big[
\frac{\partial}{\partial\xi}\log p(s|\xi,\theta)+
\frac{\mu\!-\!\xi}{\sigma^2}\Big].
\end{eqnarray}
Hence
\begin{eqnarray}
\hspace*{-5mm}
(\partial_1\xi)(\mu,\sigma,s,\theta)&=&\Bigg[1-
\sigma^2\frac{\partial^2\log p(s|\xi,\theta)}{\partial\xi^2}\Bigg]^{-1}_{\xi=\xi(\mu,\sigma,s,\theta)}
\end{eqnarray}
and 
\begin{eqnarray}
\hspace*{-5mm}
(\partial_1\xi)(wy_0\!+\!vz,\tilde{u},s,\theta)&=&\Bigg[1-
\tilde{u}^2\frac{\partial^2\log p(s|\xi,\theta)}{\partial\xi^2}\Bigg]^{-1}_{\xi=\xi(wy_0+vz,\tilde{u},s,\theta)}.
\end{eqnarray}

\subsection{The limit $\zeta\to 0$ for ML regression}

For $\eta=0$ we revert back to ML regression. Here equations (\ref{eq:ddf2=0},\ref{eq:ddg2=0}) simplify to 
$\tilde{g}=1/\tilde{u}^2$ and $\tilde{f}=-v^2/\tilde{u}^4$, equation (\ref{eq:ddtheta2=0}) remains unaltered, and the three equations (\ref{eq:ddu2=0}, \ref{eq:ddv2=0}, \ref{eq:ddw2=0}) 
referring to extremization over $(\tilde{u},v,w)$ simplify to
\begin{eqnarray}
\Big\bra\!\Big\bra \!\Big\bra 
 [\xi(wy_0\!+\!vz,\tilde{u},s,\theta)\!-\!wy_0\!-\!vz]^2\Big\ket_{\!s}\Big\ket\!\Big\ket
 &=& \zeta v^2,
 \label{eq:ddu_ML=0}
\\[2mm]
 \Big\bra\!\Big\bra \!\Big\bra 
(\partial_1\xi)(wy_0\!+\!vz,\tilde{u},s,\theta)\Big\ket_{s}\Big\ket\!\Big\ket 
&=& 1- \zeta,
 \label{eq:ddv_ML=0}
\\[1mm]
 \Big\bra\!\Big\bra\!\Big\bra  \xi(wy_0\!+\!vz,\tilde{u},s,\theta) \frac{\partial \log p(s|S\bra a\ket^{\frac{1}{2}}y_0,\theta^\star)}{\partial y_0}\Big\ket_{\!s}\Big\ket\!\Big\ket
&=& \zeta w.
 \label{eq:ddw_ML=0}
\end{eqnarray}
As a test, let us consider the classical regime $\zeta\to 0$ where the number of samples is much larger than the number of covariates. 
We can show relatively easily, for any model $p(s|\xi,\theta)$, that in this limit the remaining ML equations are solved by the correct solution $(\tilde{u},v,w,\theta)=(0,0,S\bra a\ket^{\frac{1}{2}},\theta^\star)$, as one should expect. To see this we use
\begin{eqnarray}
\lim_{\tilde{u}\to 0}\xi(wy_0\!+\!vz,\tilde{u},s,\theta)&=& wy_0\!+\!vz,
\\
\lim_{\tilde{u}\to 0}(\partial_1\xi)(wy_0\!+\!vz,\tilde{u},s,\theta)&=&1.
\end{eqnarray}
Upon inserting  $(\tilde{u},v,w,\theta)=(0,0,S\bra a\ket^{\frac{1}{2}},\theta^\star)$
we find that (\ref{eq:ddu_ML=0}) and (\ref{eq:ddv_ML=0}) are trivially satisfied, whereas  (\ref{eq:ddtheta2=0})  and (\ref{eq:ddw_ML=0}) reduce, respectively,  to the trivial statements
\begin{eqnarray}
0&=& \Big\bra\!\Big\bra   \frac{\partial \log p(s| wy_0,\theta^\star)}{\partial \theta^\star}\Big\ket_{\!s}\Big\ket
 \nonumber
 \\&=& 
 \int\!{\rm D}y_0
\int\!\rmd s ~\frac{\partial }{\partial\theta^\star}p(s|S\bra a\ket^{\frac{1}{2}}y_0,\theta^\star)=0
\end{eqnarray}
and
\begin{eqnarray}
0&=&
 \Big\bra y_0\Big\bra  \frac{\partial \log p(s|wy_0,\theta^\star)}{\partial y_0}\Big\ket_{\!s}\Big\ket
 \nonumber
 \\&=& 
S\bra a\ket^{\frac{1}{2}} \int\!{\rm D}y_0~y_0\int\!\rmd s~\frac{\partial}{\partial y_0}  p(s|S\bra a\ket^{\frac{1}{2}}y_0,\theta^\star)
 =0.
\end{eqnarray}

\section{Link between true and inferred association parameters}
\label{app:the_link}

 \subsection{Replica evaluation of the joint distribution}

We can also calculate the (probabilistic) relation\footnote{Note that, although anticipated at the time, this was not yet done in  the previous studies \cite{coolen2017replica, SheikhCoolen2019}.}  between the true parameters $\beta_\mu^\star$ and the MAP-inferred values $\hat{\beta}_\mu$ in regression models of the GLM form. The relevant object to be computed in the case of MAP regression with prior $p(\bbeta)$ and data $\Data=\{(\bz_1,s_1),\ldots,(\bz_N,s_N)\}$ is the joint distribution
\begin{eqnarray}
\hspace*{-10mm}
\Prob(\beta,\beta^\star|\Data)&=& \lim_{\gamma\to\infty}
 \frac{1}{p}\sum_{\mu=1}^p \frac{\int\!\rmd\theta\rmd\bbeta~\rme^{\gamma \log p(\theta,\bbeta|\Data)}\delta(\beta-\beta_\mu)}
 {\int\!\rmd\theta\rmd\bbeta~\rme^{\gamma \log p(\theta,\bbeta|\Data)}}
 \delta(\beta^\star-\beta^\star_\mu),
 \label{eq:link_between_betas_start}
 \end{eqnarray}
 with the posterior parameter likelihood
 \begin{eqnarray}
 p(\theta,\bbeta|\Data)&=& \frac{p(\bbeta,\theta)\prod_{i=1}^N p(s_i|\bbeta\cdot\bz_i/\sqrt{p},\theta)}
 {\int\!\rmd\bbeta^\prime \rmd\theta^\prime~p(\bbeta^\prime,\theta^\prime)\prod_{i=1}^N p(s_i|\bbeta^\prime\!\cdot\bz_i/\sqrt{p},\theta^\prime)}.
 \label{eq:posterior}
 \end{eqnarray}
 The limit $\gamma\to\infty$ ensures that  the integrations in  (\ref{eq:link_between_betas_start}) are dominated by the parameter values where $p(\theta,\bbeta|\Data)$ is maximized. Hence the fraction in  (\ref{eq:link_between_betas_start}) reduces to $\delta(\beta-\hat{\beta}_\mu)$, where $\hat{\beta}_\mu$ is the MAP estimator of the parameter $\beta_\mu$, given dataset $\Data$. Note that for $\gamma\to 1$  expression (\ref{eq:link_between_betas_start}) would have reduced to the joint distribution of true coefficients and their Bayesian estimators, viz.
 \begin{eqnarray}
\Prob(\beta,\beta^\star|\Data)&=&
 \frac{1}{p}\sum_{\mu=1}^p \delta(\beta^\star\!-\beta^\star_\mu)\int\!\rmd\theta\rmd\bbeta~p(\theta,\bbeta|D)\delta(\beta-\beta_\mu).
 \end{eqnarray}

As always we are interested mainly in the typical form of  $\Prob(\beta,\beta^\star|\Data)$, so we average over the possible realizations of the data set $\Data$, assuming all samples $(\bz_i,s_i)$ to be drawn randomly and independently from $p(\bz)p(s|\bbeta^\star\!\!\cdot\bz/\!\sqrt{p},\theta^\star)$:
\begin{eqnarray}
\hspace*{-10mm}
\Prob(\beta,\beta^\star)&=& \bra \Prob(\beta,\beta^\star|\Data)\ket_{\Data}
\nonumber
\\
\hspace*{-10mm}
&=&
  \lim_{\gamma\to\infty}
 \frac{1}{p}\sum_{\mu=1}^p  \delta(\beta^\star\!-\beta^\star_\mu)~\Bigg\bra\frac{\int\!\rmd\theta\rmd\bbeta~\rme^{\gamma \log p(\theta,\bbeta|\Data)}\delta(\beta-\beta_\mu)}
 {\int\!\rmd\theta\rmd\bbeta~\rme^{\gamma \log p(\theta,\bbeta|\Data)}}
  \Bigg\ket_{\!\!\Data}.
\label{eq:joint_beta_stats_nearly}
 \end{eqnarray}
 Upon inserting (\ref{eq:posterior}) into  (\ref{eq:joint_beta_stats_nearly}) we obtain for this ${\Data}$-independent joint distribution:
 \begin{eqnarray}
 \hspace*{-10mm}
\Prob(\beta,\beta^\star)&=&
  \lim_{\gamma\to\infty}
 \frac{1}{p}\sum_{\mu=1}^p 
  \delta(\beta^\star\!-\beta^\star_\mu)~
  \nonumber
  \\[-1mm]
   \hspace*{-10mm}
  &&
  \hspace*{-5mm} \times\Bigg\bra\frac{\int\!\rmd\theta\rmd\bbeta~\Big[p(\bbeta)p(\theta)\prod_{i=1}^N p(s_i|\bbeta\cdot\bz_i/\sqrt{p},\theta)\Big]^\gamma \delta(\beta\!-\!\beta_\mu)}
 {\int\!\rmd\theta\rmd\bbeta~\Big[p(\bbeta)p(\theta)\prod_{i=1}^N p(s_i|\bbeta\cdot\bz_i/\sqrt{p},\theta)\Big]^\gamma}
  \Bigg\ket_{\!\!\Data}\!.
\label{eq:joint_beta_stats}
 \end{eqnarray}
 We can evaluate  (\ref{eq:joint_beta_stats}) using the following alternative form of the replica identity, which can be shown to be equivalent to the previous version $\bra \log Z\ket=\lim_{n\to 0}n^{-1}\log \bra Z^n\ket$,
  \begin{eqnarray}
  \hspace*{-15mm}
 \Bigg\bra\frac{\int\!\rmd x~w(x,y)f(x)}{\int\!\rmd x~w(x,y)}\Bigg\ket_{\!\!y}&=& \lim_{n\to 0}
 \Big\bra\Big[ \int\!\rmd x~w(x,y)f(x)\Big]\Big[ \int\!\rmd x~w(x,y)\Big]^{n-1} \Big\ket_{\!y}
 \nonumber
 \\[-1mm]
 &=& \lim_{n\to 0}
 \int\!\Big[\prod_{\alpha=1}^n \rmd x^\alpha\Big] f(x^1) \Big\bra \prod_{\alpha=1}^n w(x^\alpha,y)\Big\ket_{\!y}.
 \end{eqnarray}
 Application of this identity to (\ref{eq:joint_beta_stats}), with the choices $x\to (\theta,\bbeta)$, $y\to \Data$, $w(x,y)\to 
[p(\bbeta,\theta)\prod_{i=1}^N p(t_i|\bbeta\cdot\bz_i/\sqrt{p},\theta)]^\gamma$, and $f(x)\to \delta(\beta\!-\!\beta_\mu)$, 
followed by working out the definition of the average over the data sets $\Data$,
 gives
  \begin{eqnarray}
  \hspace*{-15mm}
\Prob(\beta,\beta^\star) &=& 
  \lim_{\gamma\to\infty} \lim_{n\to 0}
 \frac{1}{p}\sum_{\mu=1}^p \delta(\beta^\star\!-\beta_\mu^\star)
 \int\!\Big\{\prod_{\alpha=1}^n \rmd \theta^\alpha\rmd\bbeta^\alpha [p(\bbeta^\alpha)p(\theta^\alpha)]^\gamma \Big\}  
 ~\delta(\beta\!-\!\beta^1_\mu)\nonumber
 \\
   \hspace*{-15mm}
 &&\hspace*{-3mm}\times 
 \Big\{\int\!\rmd\bz \rmd s~p(\bz)p(s|\bbeta^\star\!\cdot\bz/\sqrt{p},\theta^\star)
  \prod_{\alpha=1}^n 
\Big[p(s|\bbeta^\alpha\!\cdot\bz/\sqrt{p},\theta^\alpha)\Big]^\gamma
 \Big\}^N.
 \label{eq:betas_starting_point}
 \end{eqnarray}
 In \ref{app:beta_stats} we calculate the limit $p,N\to\infty$ of  (\ref{eq:betas_starting_point}), for finite ratio $\zeta=p/N$. This builds on the replica calculation in \ref{app:SheikhCoolen}.  For the simplest case of uncorrelated and normalized covariates, i.e. $\bA=\one$, we find that
     \begin{eqnarray}
\lim_{N\to\infty}\Prob(\beta|\beta^\star) &=&
  \frac{1}{v\sqrt{2\pi}}\rme^{-\frac{1}{2}(\beta-w\beta^\star/S)^2/v^2}
\end{eqnarray}
This confirms what was suggested by simulation data and exploited in \cite{coolen2017replica}: if we plot inferred versus true association parameters in a plane, we will find for $L2$ priors and uncorrelated covariates a linear cloud with slope $w/S$ and zero-average Gaussian noise of width $v$. We have now proved  this analytically, for {\em any} generalized linear model. 
For the more tricky case of correlated covariates, i.e. arbitrary covariance matrices $\bA$ subject only to the spectral conditions $\lim_{p\to\infty}\bra a\ket\in \R$ and $\lim_{p\to\infty}p^{-1}\bra a^2\ket=0$ of \ref{app:self_averaging}, we find 
 \begin{eqnarray}
 \hspace*{-20mm}
\lim_{N\to\infty}\Prob(\beta,\beta^\star) &=& 
\lim_{p\to\infty} 
 \frac{1}{p}\sum_{\mu=1}^p 
 \frac{\delta(\beta^\star\!-\beta_\mu^0)}{\sqrt{2\pi |\tilde{f}|[ 
  (\tilde{g} \bA+2\eta\one)^{-1}\bA(\tilde{g} \bA+2\eta\one)^{-1}]_{\mu\mu}}}
  \label{eq:beta_relation_derived}
\\
 \hspace*{-20mm}
&&
\times
 \rme^{-\frac{1}{2}\Big[\beta+ \tilde{d}_0
 [ (\tilde{g}\one+2\eta\bA^{-1})^{-1}
 \bbeta^0]_\mu\Big]^2/|\tilde{f}|[ 
  (\tilde{g} \bA+2\eta\one)^{-1}\bA(\tilde{g} \bA+2\eta\one)^{-1}]_{\mu\mu}
}.
\nonumber
\end{eqnarray}
Once more the inferred vector $\hat{\bbeta}$ depends linearly on the true vector $\bbeta^\star$, supplemented with Gaussian noise. However, in the presence of covariate correlations, we obtain a scalar relation $\hat{\bbeta}=\kappa\bbeta^\star+{\it noise}$ typically only when $\eta=0$ (i.e. no regularizer, giving ML regression).

Expression (\ref{eq:beta_relation_derived}) is consistent with 
the more general propositions
\begin{eqnarray}
 \bra \hat{\bbeta}\ket&=&-\tilde{d}_0 [\tilde{g}\one+2\eta\bA^{-1}]^{-1}
 \bbeta^\star,
 \label{eq:average_beta}
 \\[1mm]
 \bra \hat{\beta}_\mu\hat{\beta}_\nu\ket-\bra \hat{\beta}_\mu\ket \bra \hat{\beta}_\nu\ket &=&|\tilde{f}|[ 
  (\tilde{g} \bA+2\eta\one)^{-1}\bA(\tilde{g} \bA+2\eta\one)^{-1}]_{\mu\nu}.
  \label{eq:covariance_beta}
 \end{eqnarray}
  Using expression (\ref{eq:found_d0}) for $d_0$, and $c_0=S\bra a\ket^{\frac{1}{2}}w$, we can  write (\ref{eq:average_beta}) also as
 \begin{eqnarray}
 \bra \hat{\bbeta}\ket&=& \frac{w\bra a\ket^{\frac{1}{2}}}{S} \Big\bra\frac{a^2}{2\eta\!+\!a\tilde{g}}\Big\ket^{\!-1} [\tilde{g}\one+2\eta\bA^{-1}]^{-1}
 \bbeta^\star.
 \label{eq:average_beta_again}
 \end{eqnarray}
When covariates are correlated, MAP regression with Gaussian priors sufficiently strong to have an impact will for finite $\zeta>0$ not just rescale the length of the inferred association vectors but also change its direction.  Only for small $\eta$ or weak correlations (or if by accident  $\bbeta^\star$ happens to be an eigenvector of $\bA$) will the relation between $\bra \hat{\beta}\ket$ and $\bbeta^\star$ reduce to scalar multiplication. This is reminiscent of what happens for linear regression, and we will explore the connection in more detail in a subsequent section. It is interesting that the form of the  above expressions is universal; GLM model dependencies enter only via the order parameters $(\tilde{d}_0,\tilde{f},\tilde{g})$.

 Since the vectors and matrices in (\ref{eq:average_beta},\ref{eq:covariance_beta}) have diverging dimensionality as $p\to\infty$, it is not possible to derive these propositions directly using the steepest descent route followed in deriving  (\ref{eq:beta_relation_derived}). Only for linear ML regression will we be able to confirm (\ref{eq:average_beta},\ref{eq:covariance_beta}) rigorously. 
However, in addition to simulation experiments with different models (described in a subsequent section), one can envisage several indirect mathematical tests of expressions (\ref{eq:average_beta},\ref{eq:covariance_beta}). First, they can be used to compute  the two order parameters $c_0$ and $C$, testing their consistency with our RS order parameter equations derived earlier. This gives, using \ref{app:self_averaging}, 
 \begin{eqnarray}
 \hspace*{-10mm}
 c_0&=& \lim_{p\to\infty}\frac{1}{p}\sum_{\mu\nu=1}^p \bra \hat{\beta}_\mu\ket A_{\mu\nu}\beta_\nu^\star
 \label{eq:check_c0}
 \\
 \hspace*{-10mm}
 &=& \frac{w\bra a\ket^{\frac{1}{2}}}{S} \Big\bra\frac{a^2}{2\eta\!+\!a\tilde{g}}\Big\ket^{\!-1}\! \!\lim_{p\to\infty}\frac{1}{p}  \sum_{\mu\nu=1}^p 
 \beta^\star_\mu [(\tilde{g}\bA\!+\!2\eta\one)^{-1}\bA^2]_{\mu\nu}\beta_\nu^\star
 ~=~ w\bra a\ket^{\frac{1}{2}}S,
 \nonumber
 \\
  \hspace*{-10mm}
 C&=& \lim_{p\to\infty}\frac{1}{p}\sum_{\mu\nu=1}^p \bra \hat{\beta}_\mu \hat{\beta}_\nu\ket A_{\mu\nu}
 \nonumber
 \\
  \hspace*{-10mm}
 &=&  \lim_{p\to\infty}\frac{1}{p}{\rm Tr} 
 \Big[|\tilde{f}|[ 
  (\tilde{g} \bA\!+\!2\eta\one)^{-1}\!\bA^2(\tilde{g} \bA\!+\!2\eta\one)^{-1}\Big]
  \nonumber
  \\
  \hspace*{-10mm}
  &&+ \frac{w^2\bra a\ket}{S^2} \Big\bra\frac{a^2}{2\eta\!+\!a\tilde{g}}\Big\ket^{\!-2} \!
  \lim_{p\to\infty}\frac{1}{p}\!\sum_{\mu\nu=1}^p
 [ (\tilde{g}\bA\!+\!2\eta\one)^{-1}\!\bA^3
   (\tilde{g}\bA\!+\!2\eta\one)^{-1}]_{\mu\nu}
 \beta^\star_\mu\beta^\star_\nu
 \nonumber
 \\
 \hspace*{-10mm}
 &=&|\tilde{f}|\Big\bra\frac{a^2}{(2\eta\!+\!a\tilde{g})^2}\Big\ket
+ w^2\bra a\ket\Big\bra\frac{a^2}{2\eta\!+\!a\tilde{g}}\Big\ket^{\!-2} 
\Big\bra\frac{a^3}{(2\eta\!+\!a\tilde{g})^2}\Big\ket.
\label{eq:check_C}
 \end{eqnarray}
 Clearly, (\ref{eq:check_c0}) is identical to the result of combining the first identity of (\ref{eq:inverse_transformation}) with the expression for $\tilde{S}$ in (\ref{eq:Stilde_to_S}). Similarly, upon combining the third identity of (\ref{eq:inverse_transformation})  with $\lim_{\gamma\to\infty}u=\lim_{\gamma\to\infty}\tilde{u}/\sqrt{\gamma}=0$, we find that (\ref{eq:check_C}) gives in the limit $\gamma\to\infty$:
 \begin{eqnarray}
 v^2+w^2=|\tilde{f}|\Big\bra\frac{a^2}{(2\eta\!+\!a\tilde{g})^2}\Big\ket
+ w^2\bra a\ket\Big\bra\frac{a^2}{2\eta\!+\!a\tilde{g}}\Big\ket^{\!-2} 
\Big\bra\frac{a^3}{(2\eta\!+\!a\tilde{g})^2}\Big\ket,
\end{eqnarray}
which, in combination with $\tilde{f}<0$, reproduces equation (\ref{eq:ddg2=0}). Hence one can compute the correct RS order parameter equations from (\ref{eq:covariance_beta},\ref{eq:average_beta_again}).
 Secondly, 
in the ML limit $\eta\to 0$, where we know that $|\tilde{f}|=v^2\tilde{g}^2$, the formulae are seen to simplify as follows, confirming {\em en passant}  an ansatz made in \cite{SheikhCoolen2019}:
 \begin{eqnarray}
 \bra \hat{\bbeta}\ket= (w/S\bra a\ket^{\frac{1}{2}})  \bbeta^\star,~~~~~~
 \bra \hat{\beta}_\mu^2\ket\!-\!\bra \hat{\beta}_\mu\ket^2=v^2
 (\bA^{-1})_{\mu\mu}.
 \end{eqnarray}
As a third test we can also verify from (\ref{eq:average_beta},\ref{eq:covariance_beta}) our earlier results for uncorrelated and normalized covariates. Substitution of the appropriate values $\bA=\one$, $\tilde{f}=-v^2/\tilde{u}^4$, $2\eta+\tilde{g}=\tilde{u}^{-2}$, and $\tilde{d}_0=-w/S\tilde{u}^2$ into (\ref{eq:average_beta},\ref{eq:covariance_beta}) gives indeed the correct expressions
 \begin{eqnarray}
 \bra \hat{\bbeta}\ket= (w/S) \bbeta^\star,~~~~~~
 \bra \hat{\beta}_\mu^2\ket-\bra \hat{\beta}_\mu\ket^2=
 v^2.
 \end{eqnarray}
 
 \subsection{Correction of association parameters for overfitting effects}

To work out the replica order parameters and all associated theoretical predictions in practice, we first need to estimate the true covariate correlation matrix $\bA$ from the available covariate samples $\{\bz_1,\ldots,\bz_N\}$ (which is a standard statistical problem in portfolio theory), and the value of $S$ (which controls the amplitude of the unknown vector $\bbeta^\star$). The latter can be 
found for sufficiently large $p$ by evaluation of  $p^{-1}\hat{\bbeta}\cdot\bA\hat{\bbeta}$, using the outcome $\hat{\bbeta}$ of MAP/ML inference on the given data, in combination with equation (\ref{eq:check_C}).  One similarly uses the MAP/ML inferred auxiliary parameters $\hat{\theta}$ together with their associated order parameter equations that express the link between $\hat{\theta}$ and $\theta^\star$  to eliminate the need to know $\theta^\star$.  Once $\bA$, $S$, $\theta^\star$,  and the solution of our RS equations are known, expressions  (\ref{eq:average_beta},\ref{eq:covariance_beta}) allow us to construct alternative estimators from the MAP estimator $\hat{\bbeta}$ of the association parameters, decontaminated from the distorting effects of overfitting. 
To compactify notation we first define two $p\times p$ matrices $\bG$ and $\bXi$:
\begin{eqnarray}
\bG&=& |\tilde{d}_0|[\tilde{g}\one\!+\!2\eta \bA^{-1}]^{-1},
\\
\bXi&=& |\tilde{f}|(\tilde{g}\bA\!+\!2\eta\one)^{-1}\bA(\tilde{g}\bA\!+\!2\eta\one)^{-1},
\end{eqnarray}
with which  (\ref{eq:average_beta},\ref{eq:covariance_beta})  become
\begin{eqnarray}
\bra \hat{\bbeta}\ket=\bG\bbeta^\star,~~~~~~
\bra \hat{\beta}_\mu\hat{\beta}_\nu\ket-\bra \hat{\beta}_\mu\ket\bra \hat{\beta}_\nu\ket=\Xi_{\mu\nu}.
\end{eqnarray}
Both $\bG$ and $\bXi$ are symmetric matrices, which commute,  and $\bXi=|\tilde{f}|\tilde{d}_0^{-2}\bG\bA^{-1}\bG$. 
We will limit ourselves to linear correction protocols of the form $\hat{\bbeta}\to \hat{\bbeta}^\star\!=\bF\hat{\bbeta}$, where the correction matrix $\bF$ is restricted to be non-stochastic.
One could a priori envisage several natural criteria for determining  $\bF$, dependent upon the desired properties of the new estimator $\hat{\bbeta}^\star$, such as:
\begin{enumerate}
\item Removal of the inference bias, i.e.   $\bra \hat{\bbeta}^\star\ket=\bbeta^\star $. 
\item Minimization of the expected MSE (mean squared error)  $\sum_{\mu=1}^p \bra (\hat{\beta}_\mu^\star\!-\beta_\mu^\star)^2\ket$. 
\item Minimization of the expected generalization error. 
\end{enumerate}
In \ref{app:generalization_error} we show that, somewhat counterintuitively, minimization of the generalization error can  lead to nonsensical results (an excessive bias and a hyperconfident outcome prediction model), and should therefore not be used. 
 We will next compute the correction matrices and corresponding new estimators for the more reliable criteria (i) and (ii) in explicit form. 
 
 Criterion (i), removal of inference bias, is immediately seen to require choosing $\bF_{\rm opt}=\bG^{-1}$, giving the new and unbiased estimator
 \begin{eqnarray}
 \hat{\bbeta}^\star 
&=& |\tilde{d}_0|^{-1}  [\tilde{g}\one+2\eta\bA^{-1}]\hat{\bbeta}.
\label{eq:Correction_debias}
 \end{eqnarray}
Its variance is
 \begin{eqnarray}
  \bra  \hat{\beta}^{\star 2}_\mu \ket-\bra  \hat{\beta}^\star_\mu\ket^2&=&
( |\tilde{f}| /\tilde{d}^2_0) 
(\bA^{-1})_{\mu\mu}.
\end{eqnarray}

Next we work out criterion (ii) for  large $p$, assuming the various matrices to obey the conditions of \ref{app:self_averaging}, so that we may use expressions such as $p^{-1}\sum_{\mu\nu\leq p}\beta_\mu^\star M_{\mu\nu}\beta_\nu^\star=(S^2/p){\rm Tr}\bM+{\it o}(1)$. The objective function to be minimized over $\bF$ then becomes, after a rescaling by $p$ to ensure that it is $\order(1)$ as $p\to\infty$:
\begin{eqnarray}
\hspace*{-15mm}
\Omega(\bF)&=& \frac{1}{p} \bra (\bF\hat{\bbeta}-\bbeta^\star)^2\ket
\nonumber
\\
\hspace*{-15mm}
&=&  \frac{1}{p}\sum_{\mu} \Big(\sum_{\nu\rho} F_{\mu\nu}F_{\mu\rho}\bra \hat{\beta}_\nu\hat{\beta}_\rho\ket+(\beta^\star_\mu)^2-2\beta^\star_\mu
\sum_\nu F_{\mu\nu}\bra \hat{\beta}_\nu\ket\Big)
\nonumber
\\
\hspace*{-15mm}
&=&
 \frac{1}{p} \sum_{\mu\nu\rho} F_{\mu\nu}F_{\mu\rho}
 [\Xi_{\nu\rho}+(\bG\bbeta^\star)_\nu(\bG\bbeta^\star)_{\rho}]+S^2 - \frac{2}{p}\sum_{\mu\nu\rho}\beta^\star_\mu
 F_{\mu\nu}G_{\nu\rho}\beta_\rho^\star+{\it o}(1)
  \nonumber
\\
\hspace*{-15mm}
&=&
 \frac{1}{p}{\rm Tr}(\bF\bXi\bF^\dag)+ \frac{S^2}{p}
 \sum_{\mu\nu\rho\lambda} F_{\mu\nu}F_{\mu\rho}G_{\nu\lambda} G_{\rho\lambda}
 +S^2 - \frac{2S^2}{p}\sum_{\mu\nu}
F_{\mu\nu}G_{\nu\mu}+{\it o}(1)
 \nonumber
\\
\hspace*{-15mm}
&=&
\Omega_1(\bF)+\Omega_2(\bF)+{\it o}(1).
\label{eq:Omega}
\end{eqnarray}
with
\begin{eqnarray}
\hspace*{-10mm}
\Omega_1(\bF)= \frac{1}{p}{\rm Tr}(\bF\bXi\bF^\dag),~~~~~~
 \Omega_2(\bF)=\frac{S^2}{p}{\rm Tr}[(\bF\bG\!-\!\one)^\dag(\bF\bG\!-\!\one)].
 \label{eq:Omega_terms}
\end{eqnarray}
Removal of the inference bias gives $\Omega_2(\bF)=0$ (achieved for $\bF=\bG^{-1}$, following the previous criterion (i)), and removal of the inference noise gives $\Omega_1(\bF)=0$ (achieved for $\bF=\bnull$, or for any $\bF$ if $\bXi=\bnull$, i.e. if the MAP inference is already noise-free). Hence  we can interpret $\Omega_1(\bF)$ as the error contribution from the noise, and $\Omega_2(\bF)$ as the error contribution from the bias. In criterion (i) we minimized $\Omega_2(\bF)$ and this would generally increase $\Omega_1(\bF)$.  
Minimizing (\ref{eq:Omega}) requires balancing the two error sources.  This is  the  bias-variance trade-off in inference \cite{Bishop,CasellaBerger}. However, since $\Omega(\bF)$ is quadratic in $\bF$ we can find the location of the overall minimum in explicit form:  
\begin{eqnarray}
\bF_{\rm opt}&=&(\bXi/S^2\!+\!\bG^2)^{-1}\bG,
\label{eq:Correction_mse}
\\
\Omega_{\rm min}&=&   \frac{1}{p}{\rm Tr}(S^{-2}\one\!+\!\bXi^{-1}\bG^{2})^{-1}.
\end{eqnarray}

%%%%%%%%%%%%%%%%%%%%%%%%%%%%%%%%%%%%%%%%%%%%%%%%%%%%%%%%%%%%

\section{Applications to specific regression models}

We now apply   the  generic replica symmetric MAP order parameter equations (\ref{eq:ddf2=0})--(\ref{eq:ddtheta2=0}), where 
 $\xi(\mu,\sigma,s,\theta)$  represents the solution of (\ref{eq:xi_equation}), to different regression models of the GLM family.  We test the predictions of the theory for MAP and ML regression against measurements of simulations with different outcome types and  models, and with synthetic data.  In all cases we will for simplicity choose the covariate distribution $p(\bz)$ to be Gaussian, with zero average but potentially correlated components $\{z_\mu\}$. For the $p\times p$ covariance matrix $\bA$ with entries $A_{\mu\nu}=\bra z_\mu z_\nu\ket$ we will make the following choice, with $\epsilon\in[0,1]$:
 \begin{eqnarray}
\begin{array}{lll} 
A_{\mu\mu}&\!\!\!=\!\!\!&1
\\[1mm]
A_{\mu,\mu+1}&\!\!\!=\!\!\!&A_{\mu+1,\mu}~=~\epsilon
\\[1mm]
 A_{\mu\nu}&\!\!\!=\!\!\!&0~~~{\rm for~all~other~entries}.
 \end{array}
\label{eq:chosen_A}
\end{eqnarray}
This describes pairwise correlated covariates. The matrix (\ref{eq:chosen_A})  obeys the conditions in \ref{app:self_averaging}, and is trivially diagonalised to give $\varrho(a)=\frac{1}{2}\delta(a\!-\!1\!+\!\epsilon)+\frac{1}{2}\delta(a\!-\!1\!-\!\epsilon)$, enabling precise tests of the predictions of the theory.

\subsection{Linear regression}

{\em Replica equations for MAP linear regression.}
The simplest  case of a GLM corresponds to linear regression, where the outcomes of ML and MAP regression can in principle be computed in explicit form. It therefore serves as the simplest test for our general equations.  
In linear regression we have $\theta=(\beta_0,\Sigma)\in\R^2$ and 
\begin{eqnarray}
p(s|
\xi,\theta)=(2\pi\Sigma^2)^{-\frac{1}{2}}\rme^{-\frac{1}{2}(s-\xi-\beta_0)^2/\Sigma^2}.
\label{eq:linear_case}
\end{eqnarray}
Here we find that 
\begin{eqnarray}
\xi(\mu,\sigma,s,\theta)&=& \frac{\mu\Sigma^2+
\sigma^2 (s-\beta_0)}{\Sigma^2+\sigma^2}.
\end{eqnarray}
Hence 
$(\partial_1\xi)(\mu,\sigma,s,\theta)=\Sigma^2/(\sigma^2\!+\!\Sigma^2)$, and 
upon working out the relevant derivatives of $p(s|
\xi,\theta)$, we find the following closed set of MAP order equations:
\begin{eqnarray}
\hspace*{-15mm}
\Big\bra  
\frac{a}{2\eta\!+\!\tilde{g}a}\Big\ket
&=& \tilde{u}^2,
\label{eq:ddf_lin=0}
\\
\hspace*{-15mm}
w^2\Big[
\bra a\ket
 \Big\bra \frac{a^2}{
2\eta\!+\!\tilde{g}a}\Big\ket^{\!-2}
 \Big\bra \frac{a^3}{
(2\eta\!+\!\tilde{g}a)^2}\Big\ket
\!-\!1\Big]
-
\tilde{f}\Big\bra  
\frac{a^2}{(2\eta\!+\!\tilde{g}a)^2}\Big\ket
&=&v^2,
\label{eq:ddg_lin=0}
\\
\hspace*{-15mm}
\int\!{\rm D}t ~\Big\bra\!\Big\bra \bra 
 (\beta_0^\star\!-\!\beta_0\! +\!(S\bra a\ket^{\frac{1}{2}}\!-\!w)y_0\!+\!\Sigma^\star t\!-\!vz)^2\ket_{s}\Big\ket\!\Big\ket
 &=& - \!\zeta
\tilde{f}(\Sigma^2\!+\! \tilde{u}^2)^2,
 \label{eq:ddu_lin=0}
\\[1mm]
\hspace*{-15mm}
\frac{1}{\tilde{u}^2+\Sigma^2} 
&=&\zeta \tilde{g},
 \label{eq:ddv_lin=0}
\\[1mm]
\hspace*{-15mm}
  \frac{S}{\Sigma^2+\tilde{u}^2}
&=& \frac{\zeta w 
 \bra a\ket^{\frac{1}{2}}}
{ \Big\bra \frac{a^2}{
2\eta+\tilde{g}a}\Big\ket},
 \label{eq:ddw_lin=0}
\\[1mm]
\hspace*{-15mm}
\beta_0^\star-\beta_0&=& 0,
 \label{eq:ddbeta0_lin=0}
 \\[1mm]
 \hspace*{-15mm}
\int\!{\rm D}t~ \Big\bra\!\Big\bra \bra 
[\beta_0^\star\!-\!\beta_0\!+\!(S\bra a\ket^{\frac{1}{2}}\!-\!w)y_0\!+\!\Sigma^\star t\!-\!vz]^2\Big\ket_{\!s}\Big\ket\!\Big\ket
&=&\frac{(\Sigma^2+\tilde{u}^2)^2}{\Sigma^2}.
 \label{eq:ddSigma_lin=0}
\end{eqnarray}
Thus one always has $\beta_0=\beta_0^\star$, and the other equations can be compactified to
\begin{eqnarray}
&&
\Big\bra  
\frac{a}{2\eta\!+\!\tilde{g}a}\Big\ket
= \tilde{u}^2, ~~~~~~
1/\Sigma^2
 = - \zeta
\tilde{f}, ~~~~~~
\frac{1}{\tilde{u}^2+\Sigma^2} 
=\zeta \tilde{g},
\\
&&
w^2\Big[
\bra a\ket
 \Big\bra \frac{a^2}{
2\eta\!+\!\tilde{g}a}\Big\ket^{\!-2}
 \Big\bra \frac{a^3}{
(2\eta\!+\!\tilde{g}a)^2}\Big\ket
-1\Big]
-
\tilde{f}\Big\bra  
\frac{a^2}{(2\eta\!+\!\tilde{g}a)^2}\Big\ket
=v^2,
\\[2mm]
&&
  \frac{S}{\Sigma^2+\tilde{u}^2}
= \zeta w 
 \bra a\ket^{\frac{1}{2}}
 \Big\bra \frac{a^2}{
2\eta\!+\!\tilde{g}a}\Big\ket^{-1},
\\[1mm]
&&
 (S\bra a\ket^{\frac{1}{2}}\!-\!w)^2 \!+\!v^2\!+\!\Sigma^{\star 2}
= (\Sigma^2+\tilde{u}^2)^2/\Sigma^2.
\end{eqnarray}
Via substitutions one can reduce these coupled equations to a single nonlinear equation for $\tilde{g}$, the numerical solution of which then generates the other order parameters $(v,w,\tilde{f},\tilde{u},\Sigma)$. With the short-hand $\alpha_{k\ell}=\bra a^k/(2\eta\!+\!\tilde{g}a)^\ell\ket$ this equation takes the form
\begin{eqnarray}
\tilde{g}^{-1}&=& \zeta(1\!-\!\zeta\tilde{g}\alpha_{11})\big[S^2(\alpha_{10}\!-\!2\tilde{g}\alpha_{21}\!+\!\tilde{g}^2 \alpha_{32})\!+\!\Sigma^{\star 2}\big]+\zeta\tilde{g}\alpha_{22}.
\end{eqnarray}
Similarly, using the above formulae we can also simplify the predictions (\ref{eq:average_beta},\ref{eq:covariance_beta})  to
\begin{eqnarray}
\bra \hat{\bbeta}\ket &=& (\one+\frac{2\eta}{\tilde{g}}\bA^{-1})^{-1}\bbeta^\star,
\\
\bra\hat{\beta}^2_\mu\ket-\bra \hat{\beta}_\mu\ket^2&=& \frac{1}{\zeta\Sigma^2\tilde{g}^2}[(\one\!+\!\frac{2\eta}{\tilde{g}}\bA^{-1})\bA^{-1}(\one\!+\!\frac{2\eta}{\tilde{g}}\bA^{-1})]_{\mu\mu}.
\end{eqnarray}
For uncorrelated covariates, i.e. $\bA=\one$, these results are consistent with the well-known asymptotic behaviour of linear estimators with large random measurement matrices \cite{FA1,FA2,FA3,FA4}.
Setting $\eta=0$ brings us from MAP regression to ML regression. Here we find that the above equations reduce after some simple manipulations to
\begin{eqnarray}
&&
\hspace*{-10mm}
w=S\bra a\ket^{\frac{1}{2}},~~~~~~\Sigma=\Sigma^\star\sqrt{1\!-\!\zeta},~~~~~~\tilde{u}=\Sigma^\star\sqrt{\zeta},~~~~~~v=\Sigma^\star\sqrt{\frac{\zeta}{1\!-\!\zeta}},
\label{eq:linear_ML_replica}
\\[-1mm]
&&
\hspace*{-10mm}
\bra \hat{\bbeta}\ket = \bbeta^\star,~~~~~~~~
\bra\hat{\beta}^2_\mu\ket-\bra \hat{\beta}_\mu\ket^2= \frac{\zeta}{1\!-\!\zeta}(\Sigma^{\star})^2 (\bA^{-1})_{\mu\mu}.
\end{eqnarray}
Thus also the association parameters will on average be inferred correctly in ML, but there will be increasing overfitting induced noise (diverging at the transition point $\zeta\!=\!1$), and under-estimation of the true uncertainty $\Sigma^\star$ in the outcome predictions. 
\vsp

\noindent{\em Direct solution.}
For linear regression we can go beyond testing the replica predictions against numerical simulations, since the regression problem allows for exact solution. The parameter to be inferred are $\bbeta$ and $\Sigma$, whose MAP estimators are
\begin{eqnarray}
(\hat{\bbeta},\hat{\Sigma})&=&  {\rm argmin}_{\bbeta,\Sigma}\Big\{
\frac{1}{2\Sigma^2}\sum_{i=1}^N (s^i\!-\!\frac{\bbeta\cdot\bz_i}{\sqrt{p}})^2+N\log\Sigma
+\eta\bbeta^2\Big\}.
\end{eqnarray}
This minimization results in the following coupled equations, with the empirical $p\times p$ covariance matrix $\hat{\bA}$ with entries $\hat{A}_{\mu\nu}=N^{-1}\sum_{i\leq N} z_{i\mu}z_{i\nu}$:
\begin{eqnarray}
\hat{\Sigma}^2&=& \frac{1}{N}\sum_{i=1}^N (s_i-\frac{\hat{\bbeta}\cdot\bz_i}{\sqrt{p}})^2,
\\[-1mm]
\hat{\beta}_\mu&=&\sum_{\nu}\Big(2\eta\zeta\hat{\Sigma}^2\one+\hat{\bA}\Big)^{\!-1}_{\!\mu\nu}
\frac{\zeta}
{\sqrt{p}}\sum_{i=1}^N z_{i\nu} s_i.
\end{eqnarray}
The direct solution is formulated in terms of the empirical covariate covariance matrix $\hat{\bA}$, whereas the replica analysis involves the true population covariance matrix $\bA$. To understand the connection between the two descriptions, we need to compute disorder-averaged quantities from the above equations. 
To do this, we assume, as in the replica analysis, that the data are generated by a linear model of the type (\ref{eq:linear_case}), with unknown parameters $(\bbeta^\star,\Sigma^\star)$. Hence  $s_i=\bbeta^\star\!\cdot\bz_i/\sqrt{p}+\Sigma^\star\xi_i$, in which all $\xi$ are i.i.d. random variables, drawn from $p(\xi)=(2\pi)^{-\frac{1}{2}}\rme^{-\frac{1}{2}\xi^2}$. We will show below that in ML regression $\hat{\Sigma}$ is  self-averaging for $p\to\infty$, so that for large $p$ we can evaluate the distribution of inferred association parameters, averaged over all possible realizations of the data, i.e. over all $\{\xi_i,\bz_i\}$:
\begin{eqnarray}
\hspace*{-0mm}
P(\hat{\bbeta})&=& 
\Big\bra\!\Big\bra \delta\Big[\hat{\bbeta}-(2\eta\zeta\hat{\Sigma^2}\one+\hat{\bA})^{-1}\Big(\hat{\bA}\bbeta^\star+\frac{\zeta \Sigma^\star}
{\sqrt{p}}\sum_{i=1}^N \bz_i \xi_i\Big)\Big]\Big\ket\!\Big\ket_{\{\xi,\bz\}}
\nonumber
\\
\hspace*{-0mm}
&=&\Big\bra \int\!\frac{\rmd\bx}{(2\pi)^d}~\rme^{\rmi\bx\cdot(\bbeta-\hat{\bG}\bbeta^\star)
-\frac{1}{2} \zeta (\Sigma^\star)^2\bx\cdot \hat{\bG}\!\hat{\bA}^{-1}\hat{\bG}\bx}
\Big\ket_{\{\bz\}}
\nonumber
\\
\hspace*{-0mm}
&=&\int\!\rmd\hat{\bA} ~W(\hat{\bA})~
{\mathcal N}(\hat{\bbeta}|\hat{\bG}\bbeta^\star,2\zeta\Sigma^{\star 2}\bG
\hat{\bA}^{-1}\bG),
\label{eq:linear_betastats_1}
\end{eqnarray}
with $\hat{\bG}=(\one\!+\!2\eta\zeta\hat{\Sigma^2}\hat{\bA}^{-1})^{-1}$, and
\begin{eqnarray}
{\mathcal N}(\bbeta|\bmu,\bSigma)&=&
\frac{\rme^{-\frac{1}{2}(\bbeta-\bmu)\cdot\bSigma^{-1}
(\bbeta-\bmu)}}
{\sqrt{(2\pi)^d{\rm Det}\bSigma}},
\\
P(\hat{\bA})&=& \Big\bra \prod_{\mu\nu=1}^p \delta\Big[\hat{A}_{\mu\nu}-\frac{1}{N}\sum_{i=1}^N z_{i\mu} z_{i\nu} \Big]\Big\ket_{\{\bz\}}.
\label{eq:Ahat_measure}
\end{eqnarray}
Thus $P(\hat{\bbeta})$ is an average of Gaussian distributions, each weighted by the measure $P(\hat{\bA})$ of empirical covariate covariance matrices.  The integral in (\ref{eq:linear_betastats_1})  is still defined over all $p\times p$ matrices $\hat{\bA}$.

In \ref{app:towards_Wishart} we evaluate (\ref{eq:Ahat_measure}) and 
(\ref{eq:linear_betastats_1})
further for  the choice $p(\bz)=[(2\pi)^{-p}{\rm Det}\bA]^{\frac{1}{2}}\rme^{-\frac{1}{2}\bz\cdot\bA\bz}$,  and show that here $P(\hat{\bbeta})$ 
can be written as the following integral over the space $\Omega_p$ of symmetric positive definite matrices, involving 
the Wishart distribution $W(\hat{\bA})$ \cite{Wishart} with $N$ degrees of freedom:
\begin{eqnarray}
P(\hat{\bbeta})&=& 
\int_{\Omega_p}\!\rmd\hat{\bA} ~W(\hat{\bA})
~{\cal N}(\hat{\bbeta}|\hat{\bG}\bbeta^\star,2\zeta\Sigma^{\star 2}\hat{\bG}
\hat{\bA}^{-1}\hat{\bG}),
\label{eq:linear_betastats_2}
\end{eqnarray}
with 
\begin{eqnarray}
W(\hat{\bA})&=& \frac{\rme^{-\frac{1}{2}N{\rm Tr}(\hat{\bA}\bA^{-1})}({\rm Det}\hat{\bA})^{\frac{1}{2}(N-p-1)}}{{\cal Z}(\bA)}
\\[1mm]
{\cal Z}(\bA) &=& 
\int_{\Omega_p}\!\!\rmd\hat{\bA} ~
\rme^{-\frac{1}{2}N{\rm Tr}(\hat{\bA}\bA^{-1})}({\rm Det}\hat{\bA})^{\frac{1}{2}(N-p-1)}
\nonumber
\\[-1mm]
&=& \Big(\frac{2}{N}\Big)^{\!Np/2}
 \pi^{\frac{1}{4}p(p-1)}({\rm Det}\bA)^{\frac{1}{2}N} \prod_{j=\frac{1}{2}(N-p+1)}^{\frac{1}{2}N} \Gamma(j).
\end{eqnarray}
From the properties of the Wishart distribution follow average and variance of the entries of $\hat{\bA}$, which confirm, as expected, that $\hat{A}_{\mu\nu}=A_{\mu\nu}+\order(N^{-\frac{1}{2}})$:
\begin{eqnarray}
\bra \hat{A}_{\mu\nu}\ket = A_{\mu\nu},~~~~~~
\bra \hat{A}^2_{\mu\nu}\ket-\bra \hat{A}_{\mu\nu}\ket^2= N^{-1}A_{\mu\nu}\!+\!N^{-2}A_{\mu\mu}A_{\nu\nu}.
\end{eqnarray}

The integral in (\ref{eq:linear_betastats_2}) is still nontrivial, so we now focus on the case of linear ML regression and take the limit $\eta\to 0$, where our previous result simplifies considerably. We define the $p\times p$ matrix with entries $C_{\mu\nu}=p^{-1}(\hat{\beta}_\mu\!-\!\beta^\star_\mu)(\hat{\beta}_\nu\!-\!\beta^\star_\nu)$. This leads to
\begin{eqnarray}
\hspace*{-15mm}
P(\hat{\bbeta})&=& 
\int_{\Omega_p}\!\!\frac{\rmd\hat{\bA} ~
\rme^{-\frac{1}{2}N{\rm Tr}(\hat{\bA}\bA^{-1})}({\rm Det}\hat{\bA})^{\frac{1}{2}(N-p-1)}}{{\cal Z}(\bA)}
~{\cal N}(\bbeta|\bbeta^\star\!,\zeta (\Sigma^\star)^2\hat{\bA}^{-1})
\nonumber
\\
\hspace*{-15mm}
&=& \frac{[2\pi \zeta (\Sigma^\star)^2]^{-p/2}}{{\cal Z}(\bA)}
\int_{\Omega_p}\!\rmd\hat{\bA} ~
\rme^{-\frac{1}{2}N{\rm Tr}[\hat{\bA}(\bA^{-1}\!+\bC/(\Sigma^\star)^2)]}({\rm Det}\hat{\bA})^{\frac{1}{2}(N-p)}.
~~~~
\end{eqnarray}
This is again an integral of the Wishart form that can be evaluated analytically, now with $N+1$ degrees of freedom. 
Thus we get
\begin{eqnarray}
\hspace*{-15mm}
P(\hat{\bbeta})
&=& \frac{N^{Np/2-\frac{1}{2}p(N+1)} \Gamma(\frac{1}{2}(N\!+\!1))\sqrt{{\rm Det}\bA}}{[\pi \zeta (\Sigma^\star)^2]^{p/2}   \Gamma(\frac{1}{2}(N\!-\!p\!+\!1))}
[{\rm Det}(\one\!+\!\bA\bC/(\Sigma^\star)^2)]^{-\frac{N+1}{2}}.
\end{eqnarray}
Finally we use the identity 
\begin{eqnarray}
{\rm Det}(\one\!+\!\bA\bC/(\Sigma^\star)^2)&=& 1+\frac{(\hat{\bbeta}\!-\!\bbeta^\star)\cdot\bA(\hat{\bbeta}\!-\!\bbeta^\star)}{p (\Sigma^\star)^{2}}
\end{eqnarray}
to show that $P(\hat{\bbeta})$ is for any $(p,N)$ a  multivariate student's $t$-distribution with $N\!-\!p\!+\!1$ degrees of freedom:
\begin{eqnarray}
\hspace*{-15mm}
P(\hat{\bbeta})
&=&\pi^{-p/2} \frac{\Gamma(\frac{1}{2}(N\!+\!1))}{\Gamma(\frac{1}{2}(N\!-\!p\!+\!1))}
 \frac{\sqrt{{\rm Det}\bA}}{[p (\Sigma^\star)^2]^{p/2}  }
\Big[ 1\!+\!\frac{(\hat{\bbeta}\!-\!\bbeta^\star)\cdot\bA(\hat{\bbeta}\!-\!\bbeta^\star)}{p (\Sigma^\star)^{2}}\Big]^{-\frac{N+1}{2}}.
\label{eq:Exact_ML}
\end{eqnarray}
Equivalently we can write $\hat{\bbeta}=\bbeta^\star+\bA^{\frac{1}{2}}\bx$, 
where
\begin{eqnarray}
P(\bx)
&=&\frac{\Gamma(\frac{1}{2}(N\!+\!1))}{\Gamma(\frac{1}{2}(N\!-\!p\!+\!1))}
 \frac{\pi^{-p/2} }{[p (\Sigma^\star)^2]^{p/2}  }
\Big[ 1\!+\!\frac{\bx^2}{p (\Sigma^\star)^{2}}\Big]^{-\frac{N+1}{2}}
\label{eq:Px}
\end{eqnarray}
 Mean and covariance matrix of(\ref{eq:Exact_ML}) are in the limit $N,p\to\infty$, with $p/N=\zeta$ fixed, exactly as predicted by the replica theory, since
\begin{eqnarray}
\bra \hat{\beta}_\mu\ket&=& \beta^\star_\mu,
\label{eq:t_moment_1}
\\
\bra \hat{\beta}_\mu\hat{\beta}_\nu\ket- \bra \hat{\beta}_\mu\ket\bra \hat{\beta}_\nu\ket&=& 
 \frac{\zeta (\Sigma^\star)^2}{1\!-\!\zeta}(\bA^{-1})_{\mu\nu}+\order(\frac{1}{N}).
 \label{eq:t_moment_2}
\end{eqnarray}
Along the same lines one can also compute higher order moments of $P(\hat{\bbeta})$, giving results such as  
\begin{eqnarray}
\Big\bra \Big[\frac{1}{p}(\hat{\bbeta}\!-\!\bbeta^\star)^2\Big]^2\Big\ket-\Big\bra\frac{1}{p}(\hat{\bbeta}\!-\!\bbeta^\star)^2\Big\ket^2=\order(p^{-1})
\end{eqnarray}
Although 
(\ref{eq:Exact_ML}) is itself not a Gaussian distribution, for the marginal distribution of any finite set of components of $\hat{\bbeta}$  it predicts  Gaussian statistics in the limit $p,N\to\infty$ with fixed ratio $\zeta=p/N$. 
It is a general property of the multivariate student's $t$-distribution that all its marginals also obey multivariate student's $t$-distributions \cite{Nadarajah}. Let us define the set of indices corresponding to non-marginalized components of $\hat{\bbeta}$ as ${\cal S}\subset\{1,\ldots,p\}$, and write these components as $\tilde{\bbeta}=\{\beta_\mu,\mu\!\in\! {\cal S}\}$.  We also define an $|{\cal S}|\!\times\!|{\cal S}|$ matrix $\tilde{\bA}$, defined by the property that $(\tilde{\bA}^{-1})_{\mu\nu}=(\bA^{-1})_{\mu\nu}$ for all $(\mu,\nu)\in|{\cal S}|^2$. 
Then the marginal distribution for $\tilde{\bbeta}$ is  \cite{Nadarajah}:
\begin{eqnarray}
P(\tilde{\bbeta}) &\propto &
\Big[1\!+\!\frac{(\tilde{\bbeta}-\tilde{\bbeta}^\star)\cdot\tilde{\bA}(\tilde{\bbeta}-\tilde{\bbeta}^\star)}{p(\Sigma^\star)^2}\Big]^{-\frac{1}{2}(N-p+1+|S|)}.
\end{eqnarray}
For $|{\cal S}|$ is finite and $p,N\!\to\!\infty$ with $p/N=\zeta$ fixed, we can expand this and find
\begin{eqnarray}
P(\tilde{\bbeta}) &\propto & \rme^{-\frac{1}{2}\frac{1-\zeta}{\zeta(\Sigma^\star)^2}
(\tilde{\bbeta}-\tilde{\bbeta}^\star)\cdot\tilde{\bA}(\tilde{\bbeta}-\tilde{\bbeta}^\star)+\order(p^{-1})}.
\end{eqnarray}
So in the relevant limit, exact evaluation of $P(\hat{\bbeta})$ gives for linear ML regression a Gaussian distribution for the marginals (if $|{\cal S}|$ is finite), with, upon using $(\tilde{\bA}^{-1})_{\mu\nu}=({\bA}^{-1})_{\mu\nu}$ and in accordance with (\ref{eq:t_moment_1}, \ref{eq:t_moment_2}),
\begin{eqnarray}
&&\hspace*{-10mm}
\mu,\nu\in {\cal S}:~~~~
 \bra \hat{\beta}_\mu\ket=\beta_\mu^\star,~~~~ \bra \hat{\beta}_\mu\hat{\beta}_\nu\ket-\bra \hat{\beta}_\mu\ket\bra\hat{\beta}_\nu\ket=\frac{\zeta(\Sigma^\star)^2}{1\!-\!\zeta}(\bA^{-1})_{\mu\nu}.
 \end{eqnarray}
 We conclude from the above analysis that, in those cases where exact evaluation enables direct comparison with the predictions of the replica theory (i.e. for linear ML regression), there is full agreement between the two, and that the two propositions (\ref{eq:average_beta},\ref{eq:covariance_beta}) hold. Going beyond ML to do the same test for linear MAP regression requires  evaluation of the integral in (\ref{eq:linear_betastats_2}), which we have so far been unable to do.\vsp

The direct calculation of the statistics of the inferred noise parameter $\hat{\Sigma}$ in ML linear regression also confirms the replica prediction. After some simple manipulations one finds that $\hat{\Sigma}^2$ can be written in terms of the data as
\begin{eqnarray}
\hat{\Sigma}^2&=&
\frac{\Sigma^{\star 2}}{N}\sum_{ij=1}^N \xi_i\Big\{
\delta_{ij}
 - \frac{1}{N}
 \bz_i\cdot
\hat{\bA}^{-1}\bz_j 
\Big\}\xi_j.
\end{eqnarray}
We define the $p\!\times\! N$ matrix $\bZ$ with entries $Z_{\mu i}=z_{i\mu}/\sqrt{N}$. The characteriztic function of the distribution $P(\hat{\Sigma}^2)$ of $\hat{\Sigma}^2$ over the realizations of the outcome noise $\bxi$ is 
\begin{eqnarray}
\phi(k)&=&\int\!{\rm D}\bxi~ \rme^{\rmi k \frac{\Sigma^{\star 2}}{N}\bxi\cdot[
\one
 - \bZ^\dag
\hat{\bA}^{-1}\bZ 
]\bxi}
\nonumber
\\
&=&
\Big[
{\rm Det} 
\Big(
(1\! -\frac{2\rmi k}{N} \Sigma^{\star 2})\one +\frac{2\rmi k}{N} \Sigma^{\star 2} \bZ^\dag\hat{\bA}^{-1}\bZ 
\Big)\Big]^{-\frac{1}{2}}.
\end{eqnarray}
We note that 
$\bZ\bZ^\dag\!=\!\hat{\bA}$, from which it follows in turn that 
$(\bZ^\dag\hat{\bA}^{-1}\bZ )^2=
\bZ^\dag\hat{\bA}^{-1}\bZ $. Hence  $\bZ^\dag\hat{\bA}^{-1}\bZ $ is a projection matrix, with eigenvalues 0 and 1. Moreover, since ${\rm Tr}(\bZ^\dag\hat{\bA}^{-1}\bZ )=p$,  we know in fact that it has precisely $\zeta N$ eigenvalues 1 and $(1\!-\!\zeta)N$ eigenvalues 0. 
Hence, for any realization of the covariates we have
\begin{eqnarray}
{\rm Det} 
\Big[(1\! -\!\frac{2\rmi k \Sigma^{\star 2}}{N})\one \!+\!\frac{2\rmi k \Sigma^{\star 2} }{N}\bZ^\dag\hat{\bA}^{-1}\bZ 
\Big)\Big]=(1\! -\!\frac{2\rmi k \Sigma^{\star 2}}{N})^{(1-\zeta)N}\!,~~
\end{eqnarray}
so that
\begin{eqnarray}
\phi(k)=(1\! -\frac{2\rmi k \Sigma^{\star 2}}{N})^{-\frac{1}{2}(1-\zeta)N}.
\end{eqnarray}
We recognize that this is the characteriztic function of the gamma distribution, with average $(1\!-\!\zeta)\Sigma^{\star 2}$ and width $\Sigma^{\star 2}\sqrt{2(1\!-\!\zeta)/N}$. Hence $\hat{\Sigma}$ obeys the gamma distribution and is self-averaging with respect to the realization of the data for $N\to\infty$, and $\lim_{N\to\infty}\hat{\Sigma}=\Sigma^{\star}\sqrt{1\!-\!\zeta}$,  confirming the prediction in (\ref{eq:linear_ML_replica})  of the replica theory. 
\vsp

\unitlength=0.337mm
\begin{figure}[t]
\hspace*{-8mm}
\begin{picture}(300,318)

\put(25,305){\small  $\eta=0.01$, $\epsilon\!=\!0$, }\put(25,291){\small  $\Sigma^\star\!=\!0.1$}
\put(125,305){\small $\eta=0.01$, $\epsilon\!=\!0.75$,}\put(125,291){\small $\Sigma^\star\!=\!0.1$}
\put(225,305){\small $\eta=0.01$, $\epsilon\!=\!0$, }\put(225,291){\small $\Sigma^\star\!=\!0.5$}
\put(325,305){\small $\eta=0.1$, $\epsilon\!=\!0.75$, }\put(325,291){\small $\Sigma^\star\!=\!0.5$}

\put(3,233){\small $\hat{\beta}_\mu$}
\put(60,175){\small $\beta^\star_\mu$}\put(160,175){\small $\beta^\star_\mu$}
\put(260,175){\small $\beta^\star_\mu$}\put(360,175){\small $\beta^\star_\mu$}
\put(28,265){\small $\zeta\!=\!0.5$}\put(128,265){\small $\zeta\!=\!0.5$}
\put(228,265){\small $\zeta\!=\!0.5$}\put(328,265){\small $\zeta\!=\!0.5$}
\put(0,185){\includegraphics[width=135\unitlength]{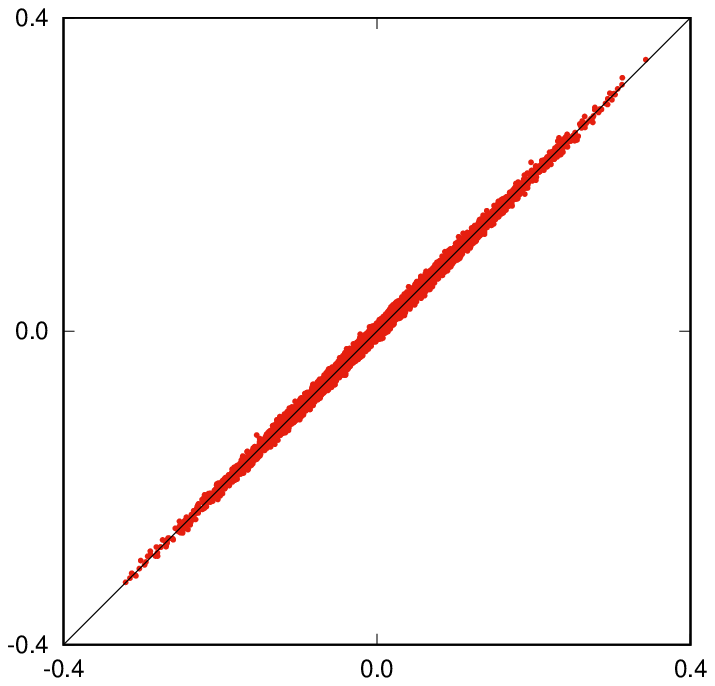}}
\put(100,185){\includegraphics[width=135\unitlength]{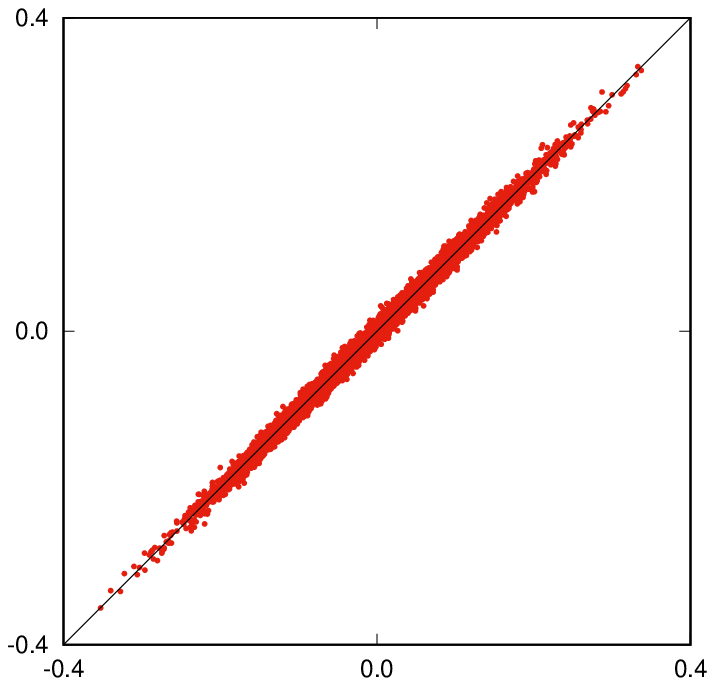}}
\put(200,185){\includegraphics[width=135\unitlength]{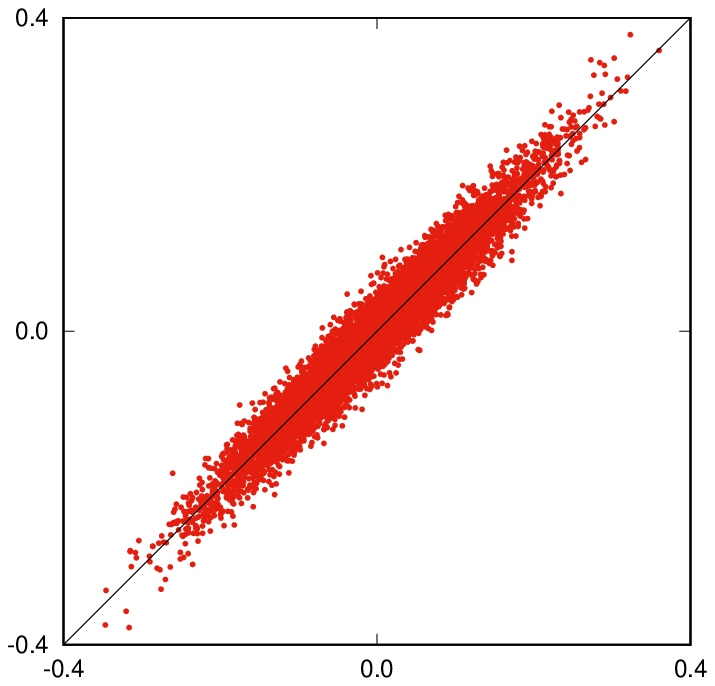}}
\put(300,185){\includegraphics[width=135\unitlength]{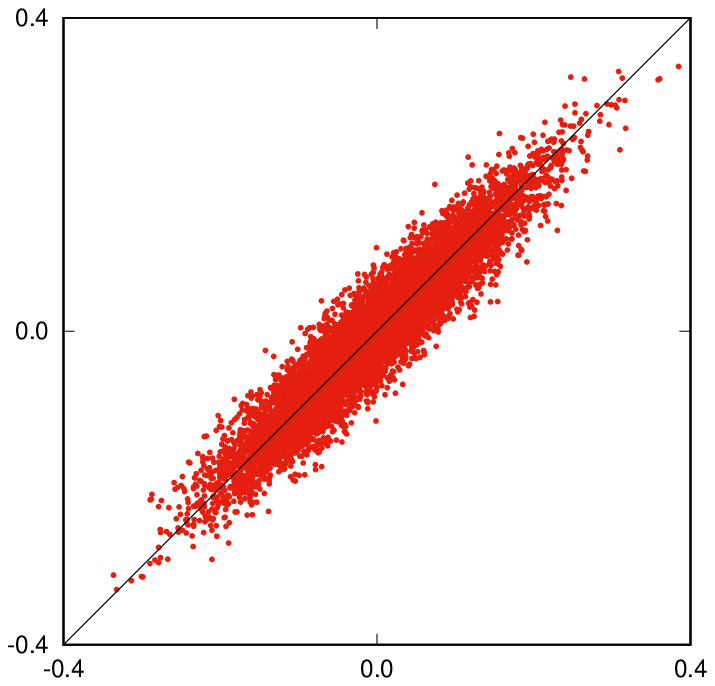}}

\put(2,120){$w$}
\put(0,70){\includegraphics[width=135\unitlength]{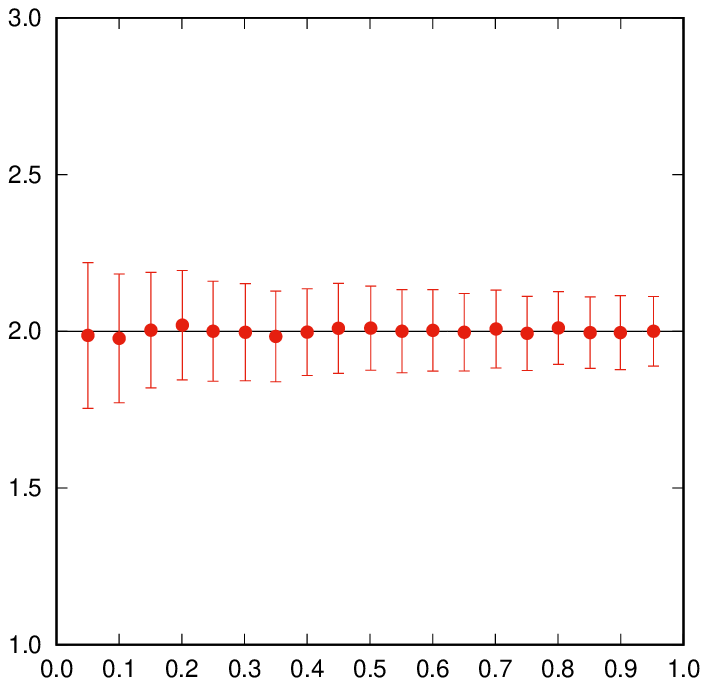}}
%\put(63,95){\small $\zeta$}
\put(100,70){\includegraphics[width=135\unitlength]{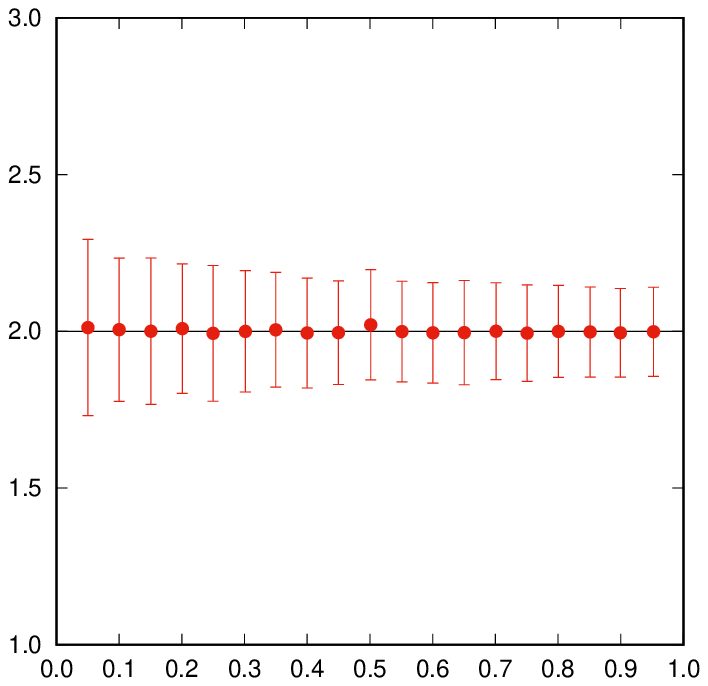}}
%\put(163,95){\small $\zeta$}
\put(200,70){\includegraphics[width=135\unitlength]{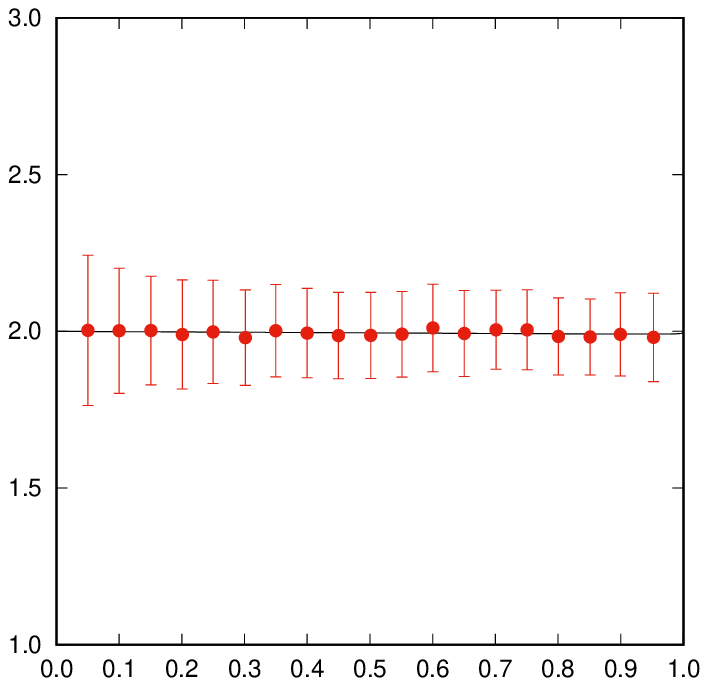}}
%\put(263,95){\small $\zeta$}
\put(300,70){\includegraphics[width=135\unitlength]{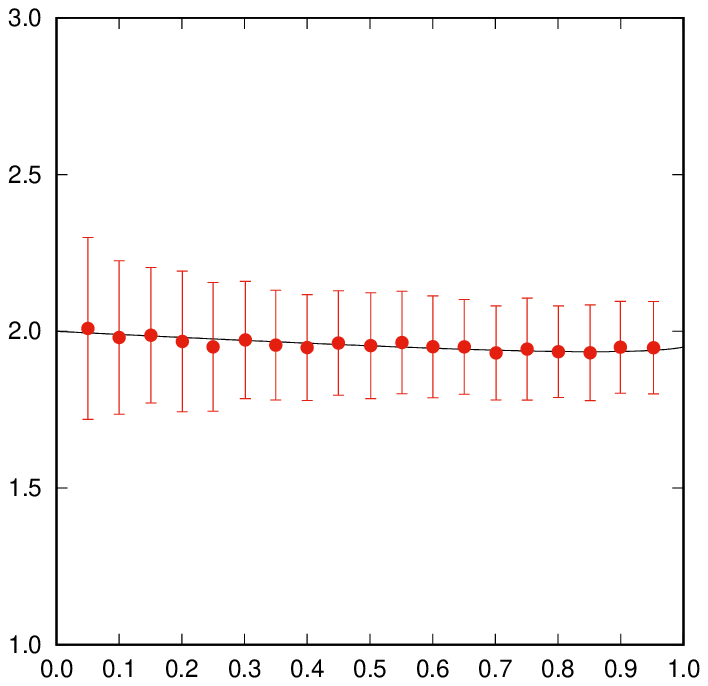}}
%\put(363,95){\small $\zeta$}

\put(2,20){$v$}
\put(0,-30){\includegraphics[width=135\unitlength]{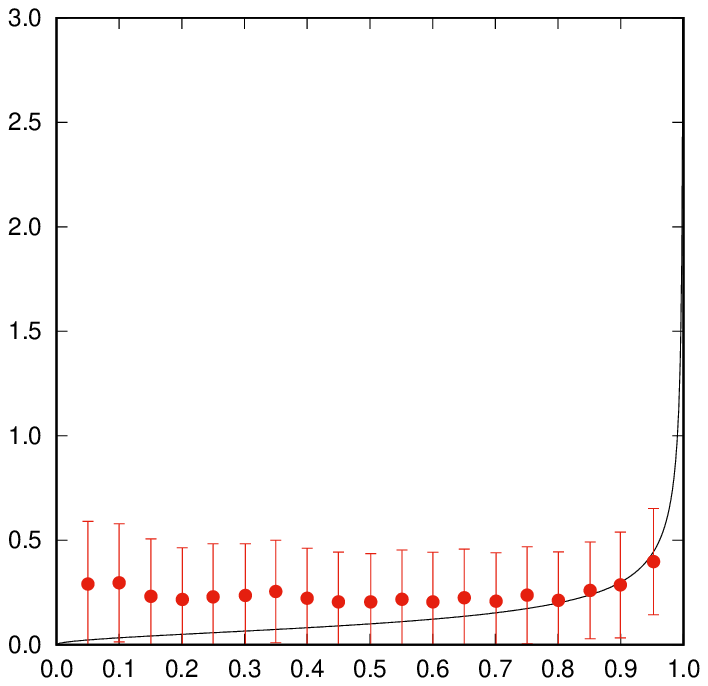}}
%\put(63,-40){\small $\zeta$}
\put(100,-30){\includegraphics[width=135\unitlength]{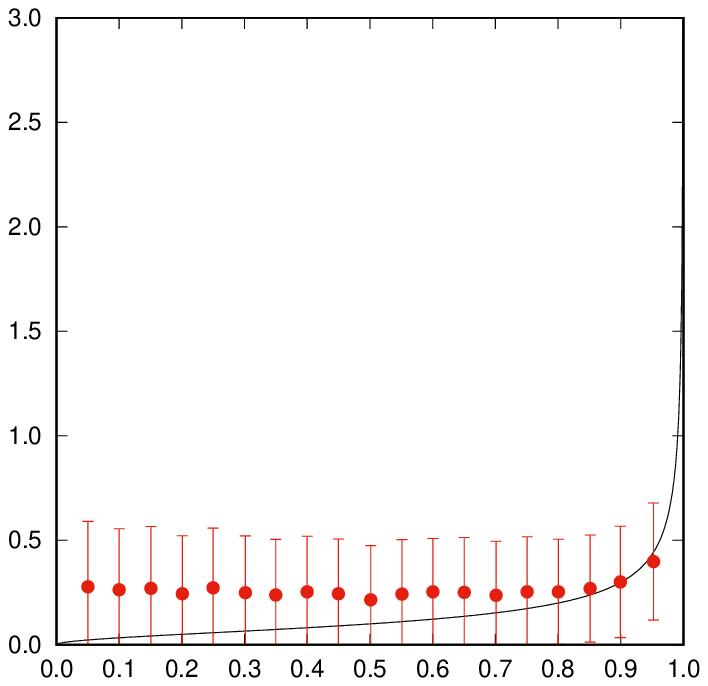}}
%\put(163,-40){\small $\zeta$}
\put(200,-30){\includegraphics[width=135\unitlength]{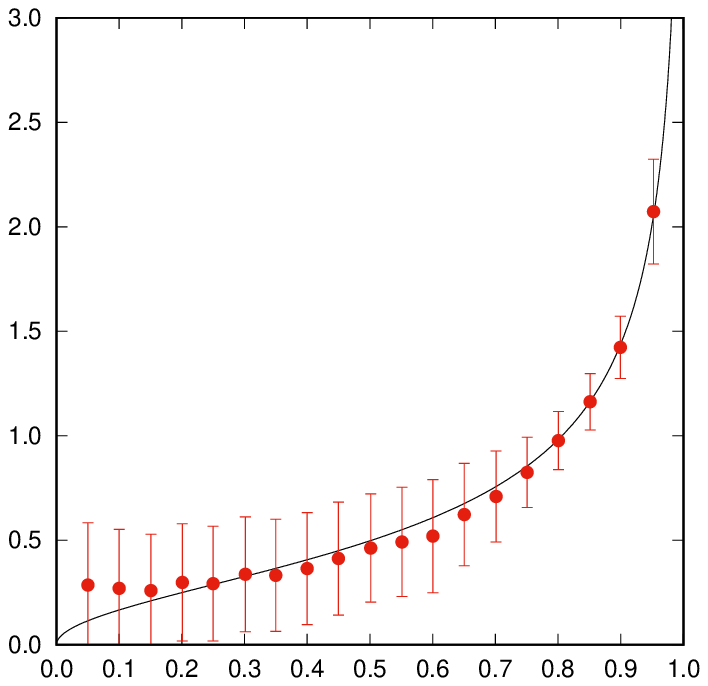}}
%\put(263,-40){\small $\zeta$}
\put(300,-30){\includegraphics[width=135\unitlength]{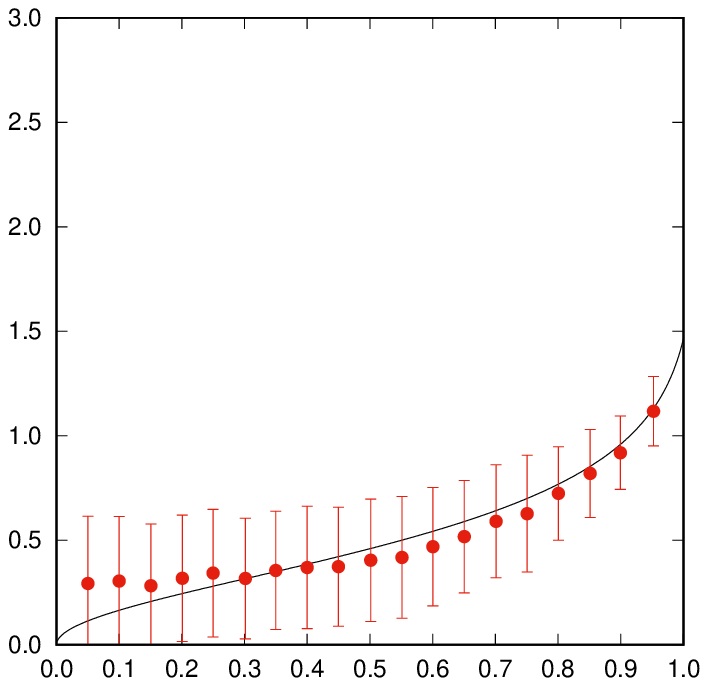}}
%\put(363,-40){\small $\zeta$}

\put(2,-80){$\Sigma$}
\put(0,-130){\includegraphics[width=135\unitlength]{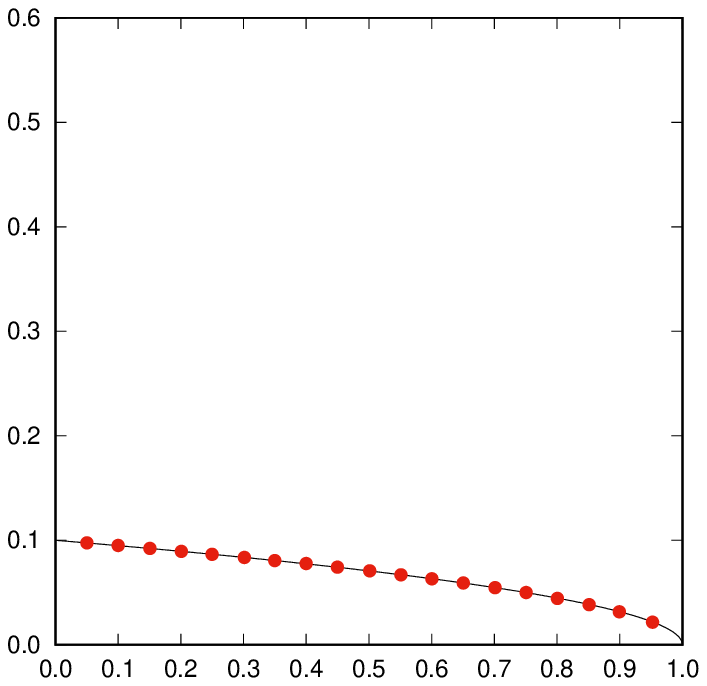}}
\put(63,-140){\small $\zeta$}
\put(100,-130){\includegraphics[width=135\unitlength]{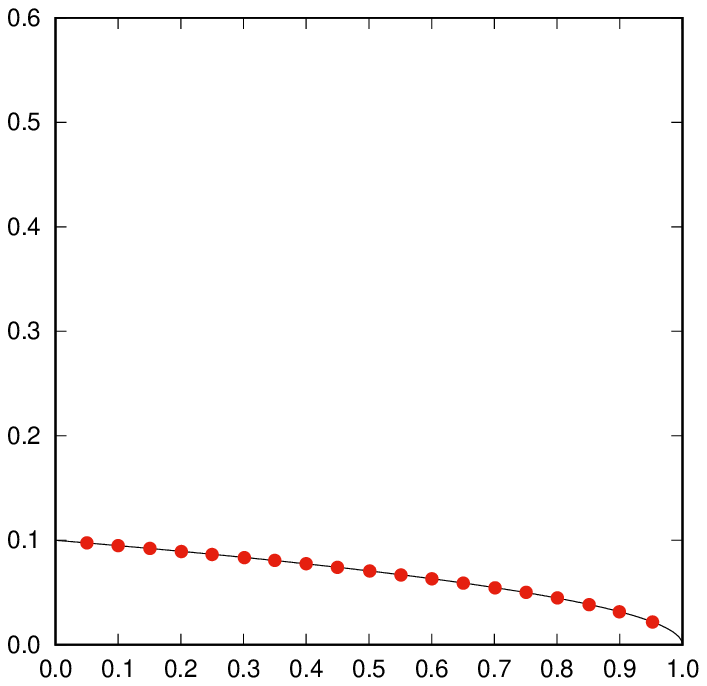}}
\put(163,-140){\small $\zeta$}
\put(200,-130){\includegraphics[width=135\unitlength]{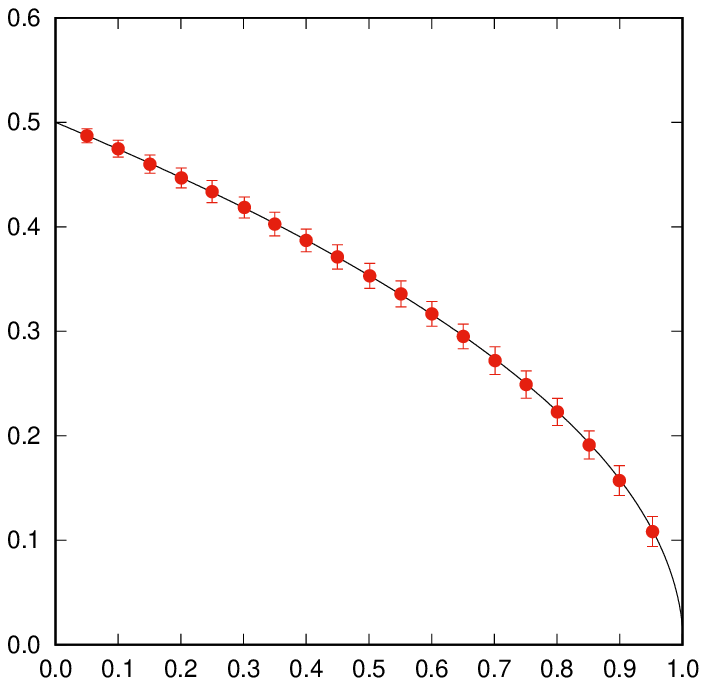}}
\put(263,-140){\small $\zeta$}
\put(300,-130){\includegraphics[width=135\unitlength]{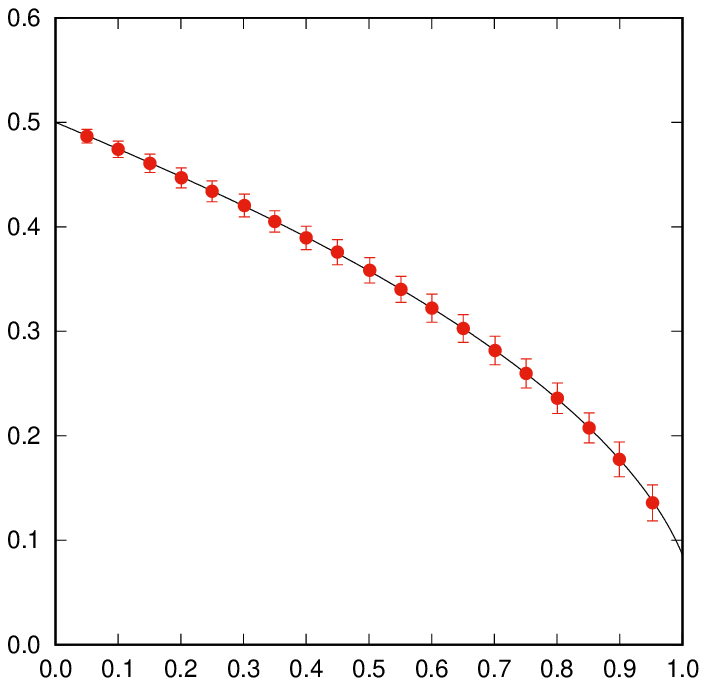}}
\put(363,-140){\small $\zeta$}

\end{picture}
\vspace*{55mm}

\caption{\small Results of linear MAP regression simulations with $Np=400,000$ and $\beta_0^\star=0$, for different combinations of $\eta$ (regularizer), $\epsilon$ (covariate correlations), and $\Sigma^\star$ (true noise strength).  In all cases $S=2$. Top row: inferred versus true association parameters for $\zeta=0.5$. 
Second and third row: order parameters $w$ and $v$ plotted versus $\zeta$. Bottom row: inferred noise strength $\Sigma$ versus $\zeta$. Each simulation data point represents average and standard deviation computed over 400 synthetic data sets and regressions. Solid curves: theoretical predictions obtained by solving the RS equations.   }
\label{fig:MAP_linear_vw}
\end{figure}

\unitlength=0.337mm
\begin{figure}[t]
\hspace*{-8mm}
\begin{picture}(300,233)

\put(25,219){\small  $\eta=0.01$, $\epsilon\!=\!0$, }\put(25,205){\small  $\Sigma^\star\!=\!0.1$}
\put(125,219){\small $\eta=0.01$, $\epsilon\!=\!0.75$,}\put(125,205){\small $\Sigma^\star\!=\!0.1$}
\put(225,219){\small $\eta=0.01$, $\epsilon\!=\!0$, }\put(225,205){\small $\Sigma^\star\!=\!0.5$}
\put(325,219){\small $\eta=0.1$, $\epsilon\!=\!0.75$, }\put(325,205){\small $\Sigma^\star\!=\!0.5$}

\put(-12,150){\em\small slope}
\put(0,100){\includegraphics[width=135\unitlength]{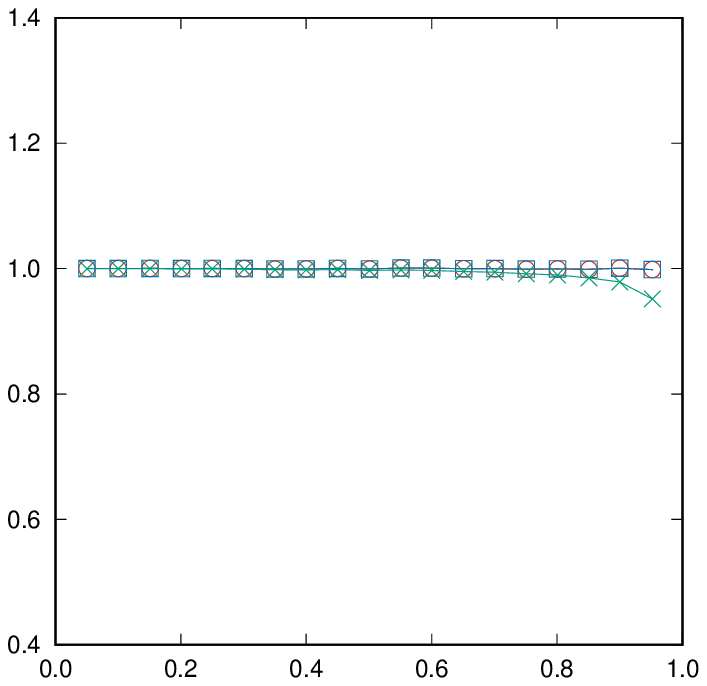}}
%\put(63,95){\small $\zeta$}
\put(100,100){\includegraphics[width=135\unitlength]{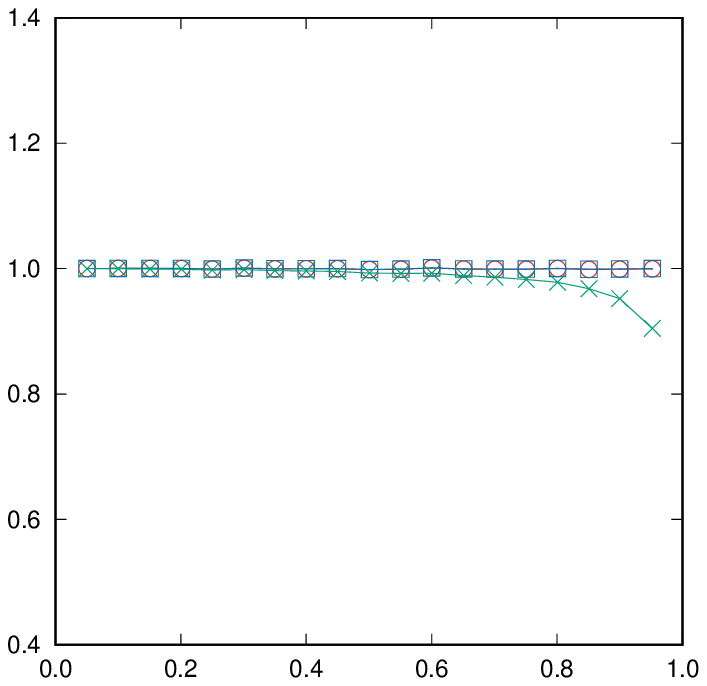}}
%\put(163,95){\small $\zeta$}
\put(200,100){\includegraphics[width=135\unitlength]{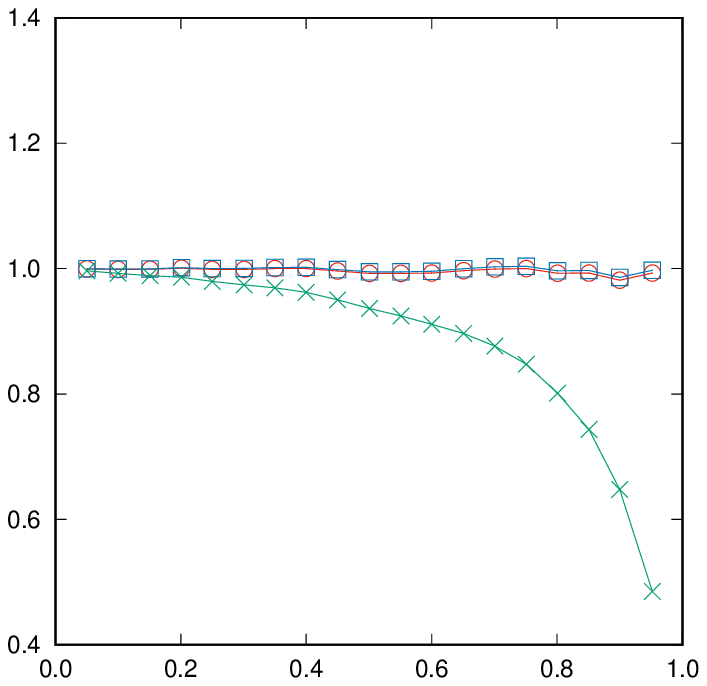}}
%\put(263,95){\small $\zeta$}
\put(300,100){\includegraphics[width=135\unitlength]{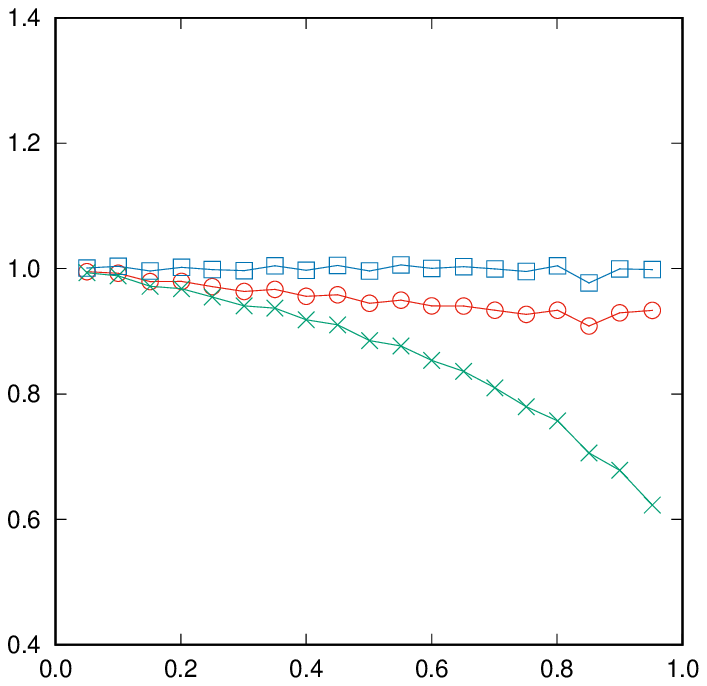}}
%\put(363,95){\small $\zeta$}

\put(-20,50){\em\small $\sqrt{\it MSE}$}
\put(0,0){\includegraphics[width=135\unitlength]{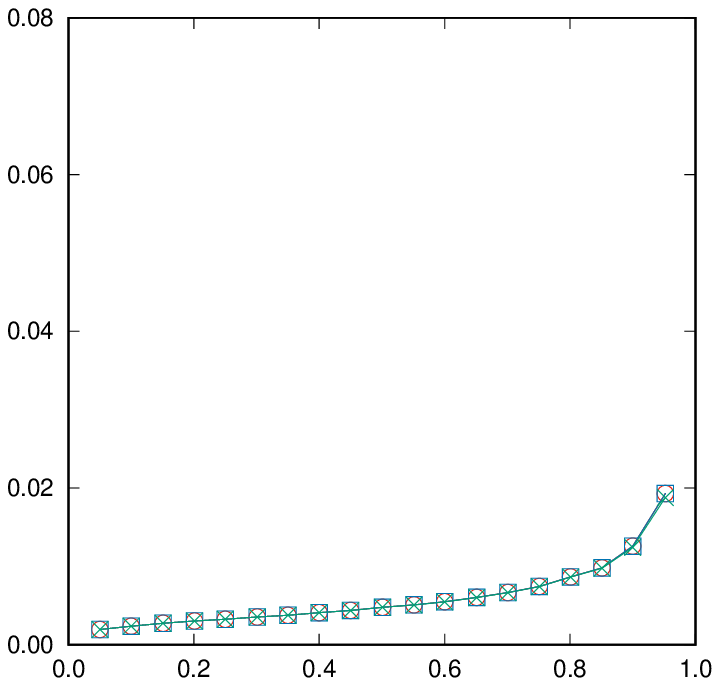}}
\put(63,-10){\small $\zeta$}
\put(100,0){\includegraphics[width=135\unitlength]{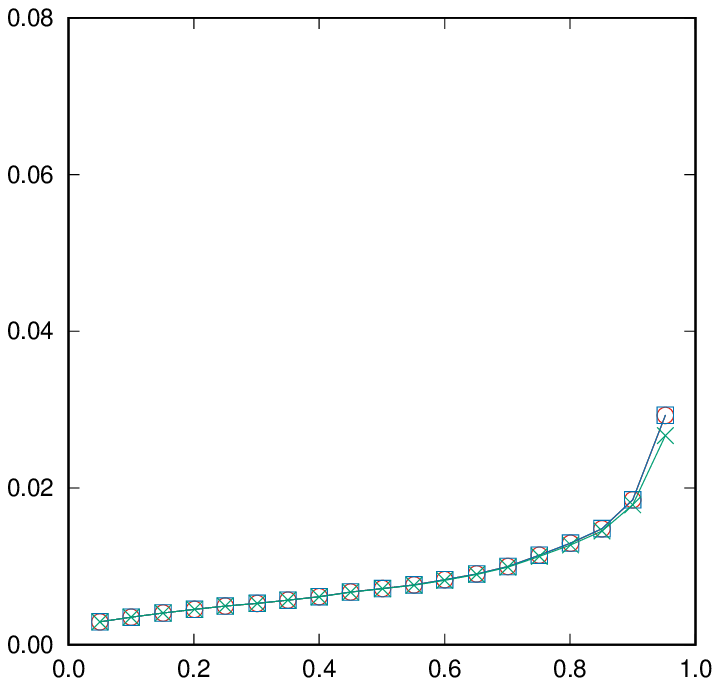}}
\put(163,-10){\small $\zeta$}
\put(200,0){\includegraphics[width=135\unitlength]{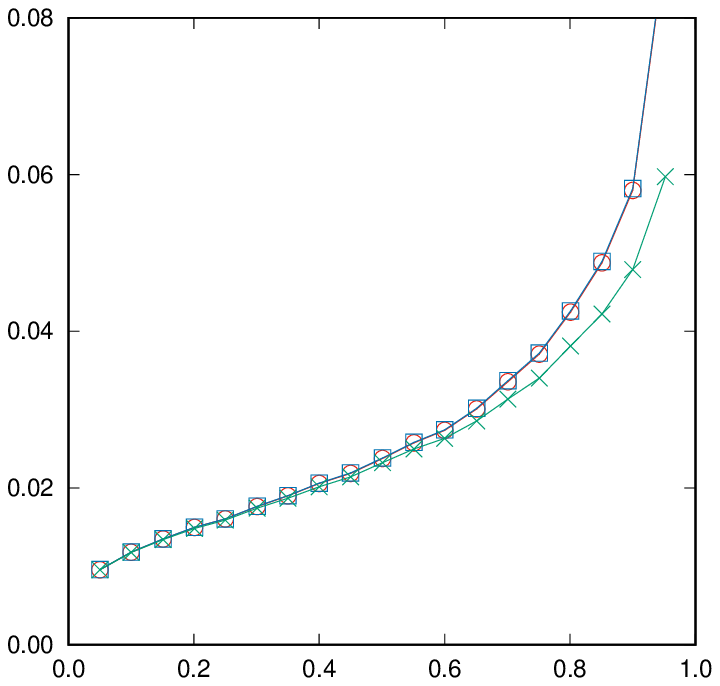}}
\put(263,-10){\small $\zeta$}
\put(300,0){\includegraphics[width=135\unitlength]{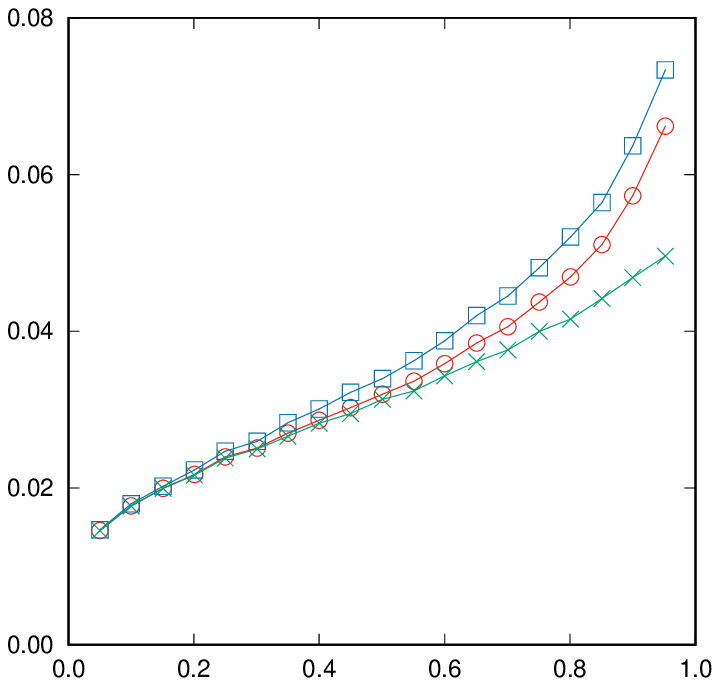}}
\put(363,-10){\small $\zeta$}

\end{picture}
\vspace*{5mm}

\caption{\small Tests of the correction protocols for MAP estimators, applied to the data of the previous figure. 
Red circles: the MAP estimator $\hat{\beta}_\mu$. Blue squares: the corrected estimator (\ref{eq:Correction_debias}), aimed at removing inference bias. Green crosses: the corrected estimator (\ref{eq:Correction_mse}), aimed at minimizing the MSE. 
The top row show as a function of $\zeta$ the slopes of the relation between the three estimators  and the true values 
$\beta_\mu^\star$ of association parameters (the slopes of the data clouds in the top row of the previous figure, computed via least squares analysis); this slope would be unity for unbiased estimators. 
The bottom row shows the values of $\sqrt{\rm MSE}$, where ${\rm MSE}=p^{-1}\sum_{\mu\leq p}(\hat{\beta}_\mu\!-\!\beta_\mu^\star)^2$.  
 }
\label{fig:MAP_linear_correction}
\end{figure}

\noindent{\em Numerical simulations of MAP linear regression with correlated covariates.}
The result of solving numerically the RS order parameter equations in the presence of covariate correlations of the type (\ref{eq:chosen_A}) is shown in Figure \ref{fig:MAP_linear_vw}, where we plot the resulting values of the order parameters $v$ and $w$ and the inferred noise strength $\Sigma$ together with regression simulation data (for synthetic Gaussian covariates), as  functions of the ratio $\zeta=p/N$. In these experiments we chose $\beta_0=\beta_0^\star=0$, for simplicity. Once more we observe excellent agreement between theory and simulation. 
 In the top row we also plot for each parameter combination the MAP-inferred parameters $\hat{\beta}_\mu$ versus the corresponding true association strengths $\beta_\mu^\star$, for pooled data from 20 regressions. In the two columns on the right we see that the width of the data cloud (top picture) reflects inference noise in the case where $\epsilon=0$ (no covariate correlations), with a larger $v$, whereas  for $\epsilon=0.75$ the inference noise is reduced, so that there the wider cloud reflects the correlation-induced bias. 

These data also enable us to  tests our two protocols (\ref{eq:Correction_debias},\ref{eq:Correction_mse}) for correcting the MAP estimator of the association parameters for the distortions caused by overfitting.
 See Figure \ref{fig:MAP_linear_correction}. 
 We plot the slopes of the data clouds of estimators versus true parameter values, as shown for $\zeta=0.5$ in the top row of Figure \ref{fig:MAP_linear_vw}, including the  MAP estimator (red circles), the minimum bias estimator  (\ref{eq:Correction_debias}) (blue squares), and the minimum MSE estimator (\ref{eq:Correction_mse}) (green crosses). 
 For linear regression with small regularizers (i.e. small $\eta$) the slopes of the data clouds for the MAP estimator are very close to unity, the inference bias is weak, and hence the red and blue data points coincide (debiasing makes no difference).  For stronger correlations and stronger regularization (column on the right), this is  no longer true.   Similarly, the MSE values of the minimum MSE estimator (\ref{eq:Correction_mse}) (green) are as predicted indeed always identical to or below those of the other two estimators.

\subsection{Logistic regression}

{\em Equations for MAP logistic regression.}
In logistic regression we have 
$s\in\{-1,1\}$ (alternatively one could define $s\in\{0,1\}$; with our present choice the equations will be somewhat more compact), with $\theta=\beta_0\in\R$, and 
\begin{eqnarray}\label{eq:logistic_regression_model}
p(s|\xi,\beta_0)&=&\frac{\rme^{s(\xi+\beta_0)}}{2\cosh(\xi\!+\!\beta_0)}.
\end{eqnarray}
Hence $\partial_\xi\log p(s|\xi,\beta_0)= s-\tanh(\xi\!+\!\beta_0)$ and 
$\partial_\xi^2\log p(s|\xi,\beta_0)=\tanh^2(\xi\!+\!\beta_0)-1$.
We will now compute the various model-dependent building blocks of our general order parameter equations (\ref{eq:ddf2=0}--\ref{eq:ddtheta2=0}). 
The function $\xi(\mu,\sigma,s,\beta_0)$ is the solution of 
\begin{eqnarray}
s-\tanh(\xi\!+\!\beta_0)=(\xi-\mu)/\sigma^2.
\end{eqnarray}
We switch from $\xi$ to the new variable $x=s(\xi+\beta_0)$, so $sx-\beta_0=\xi$. Now
\begin{eqnarray}
1-[x-s(\beta_0+\mu)]/\sigma^2=\tanh(x).
\end{eqnarray}
We next define $\tilde{x}(\mu,\sigma)$  as the solution of the following 
 transcendental equation, whose solution is  unique since the right-hand side increases monotonically from $-1$ to $1$, and the left-hand side decreases monotonically from $+\infty$ to $-\infty$:
\begin{eqnarray}
\tilde{x}(\mu,\sigma):&~~~& {\rm solution~of}~~\tanh(x)=1-(x\!-\!\mu)/\sigma^2.
\end{eqnarray}
Graphical inspection shows that $\tilde{x}(\mu,\sigma)$ increases monotonically with both $\mu$ and $\sigma\geq 0$, and that $\mu=\tilde{x}(\mu,0)\leq \tilde{x}(\mu,\sigma)\leq \tilde{x}(\mu,\infty)=\infty$. 
In terms of $\tilde{x}$ we may write
\begin{eqnarray}
\xi(\mu,\sigma,s,\beta_0)=s\tilde{x}(s(\beta_0\!+\!\mu),\sigma)-\beta_0.
\end{eqnarray}
We can now work out the relevant derivatives required in our equations:
\begin{eqnarray}
(\partial_1\xi)(\mu,\sigma,s,\beta_0)&=& \Big[1+\sigma^2[1\!-\!\tanh^2(\xi\!+\!\beta_0)]\Big]^{-1},
\\
\frac{\partial}{\partial\beta_0}\log p(s|\xi,\beta_0)&=&s-\tanh(\xi\!+\!\beta_0),
\\
\frac{\partial}{\partial y_0}\log p(s|S\bra a\ket^{\frac{1}{2}}y_0\!+\!\beta_0^\star)&=& 
 S\bra a\ket^{\frac{1}{2}}\Big[s-\tanh(S\bra a\ket^{\frac{1}{2}}y_0\!+\!\beta^\star_0)\Big].
\end{eqnarray}
Upon substituting the  above model-specific  expressions for logistic regression into our general RS order parameter equations (\ref{eq:ddf2=0}--\ref{eq:ddtheta2=0}), we obtain
\begin{eqnarray}
\hspace*{-20mm}
\Big\bra  
\frac{a}{2\eta\!+\!\tilde{g}a}\Big\ket
&=& \tilde{u}^2,
\label{eq:logistic_MAP_1}
\\
\hspace*{-20mm}
w^2\Big[
\bra a\ket
 \Big\bra \frac{a^2}{
2\eta\!+\!\tilde{g}a}\Big\ket^{-2}
 \Big\bra \frac{a^3}{
(2\eta\!+\!\tilde{g}a)^2}\Big\ket
-1\Big]
-
\tilde{f}\Big\bra  
\frac{a^2}{(2\eta\!+\!\tilde{g}a)^2}\Big\ket
&=&v^2,
\label{eq:logistic_MAP_2}
\\[1mm]
\hspace*{-20mm}
\Big\bra\!\Big\bra \bra 
 [s\tilde{x}(s(\beta_0\!+\!wy_0\!+\!vz),\tilde{u})\!-\!\beta_0\!-\!wy_0\!-\!vz]^2\ket_{s}\Big\ket\!\Big\ket
 &=& - \zeta
\tilde{f} \tilde{u}^4,
\label{eq:logistic_MAP_3}
\\[1mm]
\hspace*{-20mm}
 \Big\bra\!\Big\bra \bra 
  \Big[1+\tilde{u}^2[1\!-\!\tanh^2(s\tilde{x}(s(\beta_0\!+\!wy_0\!+\!vz),\tilde{u}))]\Big]^{-1}
 \ket_{s}\ket\!\Big\ket 
&=& 1- \zeta \tilde{g}\tilde{u}^2,
\label{eq:logistic_MAP_4}
\\[1mm]
\hspace*{-20mm}
 \Big\bra\!\Big\bra\Big\bra  
 \Big[
 s\tilde{x}(s(\beta_0\!+\!wy_0\!+\!vz),\tilde{u})\!-\!\beta_0
 \Big]
 \Big[s\!-\!\tanh(S\bra a\ket^{\frac{1}{2}}y_0\!+\!\beta^\star_0)\Big]
 \Big\ket_{\!s}\big\ket\!\Big\ket
&=& \frac{\zeta w \tilde{u}^2
 \bra a\ket^{\frac{1}{2}}}{S  \Big\bra \frac{a^2}{
2\eta+\tilde{g}a}\Big\ket},
\label{eq:logistic_MAP_5}
\\[1mm]
\hspace*{-20mm}
\Big\bra\!\Big\bra \Big\bra 
s-\tanh(s\tilde{x}(s(\beta_0\!+\!wy_0\!+\!vz),\tilde{u}))
\Big\ket_{\!s}\Big\ket\!\Big\ket&=& 0.
\label{eq:logistic_MAP_6}
\end{eqnarray}
\vsp

\noindent{\em Equations for ML logistic regression.}
For $\eta=0$ we revert from MAP to ML regression. Here we find the usual model-independent simplifications $\tilde{g}=1/\tilde{u}^2$ and $\tilde{f}=-v^2/\tilde{u}^4$, the covariate correlations (if present) drop out of the theory, and the remaining equations simplify to 
\begin{eqnarray}
\hspace*{-15mm}
\Big\bra\!\Big\bra \bra 
 [s\tilde{x}(s(\beta_0\!+\!wy_0\!+\!vz),\tilde{u})\!-\!\beta_0\!-\!wy_0\!-\!vz]^2\ket_{s}\Big\ket\!\Big\ket
 &=& \zeta
v^2,
\label{eq:ML_logistic_1}
\\[1mm]
\hspace*{-15mm}
 \Big\bra\!\Big\bra \bra 
  \Big[1+\tilde{u}^2[1\!-\!\tanh^2(\tilde{x}(s(\beta_0\!+\!wy_0\!+\!vz),\tilde{u}))]\Big]^{-1}
 \ket_{s}\ket\!\Big\ket 
&=& 1- \zeta,
\label{eq:ML_logistic_2}
\\[1mm]
\hspace*{-15mm}
 \Big\bra\!\Big\bra\Big\bra  
 \Big[
 s\tilde{x}(s(\beta_0\!+\!wy_0\!+\!vz),\tilde{u})\!-\!\beta_0
 \Big]
 \Big[s\!-\!\tanh(S\bra a\ket^{\frac{1}{2}}y_0\!+\!\beta^\star_0)\Big]
 \Big\ket_{\!s}\big\ket\!\Big\ket
&=& \frac{\zeta w}{
S \bra a\ket^{\frac{1}{2}}},
\label{eq:ML_logistic_3}
\\[1mm]
\hspace*{-15mm}
\Big\bra\!\Big\bra \Big\bra 
s-\tanh(s\tilde{x}(s(\beta_0\!+\!wy_0\!+\!vz),\tilde{u}))
\Big\ket_{\!s}\Big\ket\!\Big\ket&=& 0.
\label{eq:ML_logistic_4}
\end{eqnarray}
For numerical evaluation it is helpful to write (\ref{eq:ML_logistic_4})  in an alternative form, exploiting the equation that defines $\tilde{x}(\mu,\sigma)$: $\tanh(\tilde{x}(\mu,\sigma))=1-[\tilde{x}(\mu,\sigma)\!-\!\mu]/\sigma^2$. We see that
\begin{eqnarray}
\hspace*{-10mm}
\tanh(s\tilde{x}(s(\beta_0\!+\!wy_0\!+\!vz),\tilde{u}))&=& s\tanh(\tilde{x}(s(\beta_0\!+\!wy_0\!+\!vz),\tilde{u}))
\nonumber
\\
\hspace*{-10mm}
&&\hspace*{-25mm} =~ s-\frac{1}{\tilde{u}^2}\Big[s\tilde{x}(s(\beta_0\!+\!wy_0\!+\!vz),\tilde{u})\!-\!(\beta_0\!+\!wy_0\!+\!vz)\Big].
\label{eq:rewrite_beta0_logistic}
\end{eqnarray}
This enables us to write (\ref{eq:ML_logistic_4})  as
\begin{eqnarray}
\beta_0&=&
\Big\bra\!\Big\bra \Big\bra 
s\tilde{x}(s(\beta_0\!+\!wy_0\!+\!vz),\tilde{u})
\Big\ket_{\!s}\Big\ket\!\Big\ket.
\end{eqnarray}
Upon finally writing in explicit form all the averages, and after some simple rewriting, we obtain four equations that can be solved numerically via fixed-point iteration:
\begin{eqnarray}
\hspace*{-20mm}
\zeta v^2\!
&=&\! \int\!{\rm D}y_0{\rm D}z \Bigg\{
\frac{1}{2}\Big[1\!+\!\tanh(S\bra a\ket^{\frac{1}{2}}y_0\!+\!\beta_0^\star)\Big]
  \Big[\tilde{x}(\beta_0\!+\!wy_0\!+\!vz,\tilde{u})\!-\!(\beta_0\!+\!wy_0\!+\!vz)\Big]^2
  \nonumber
  \\
  \hspace*{-20mm}
  &&
 \hspace*{0mm} +
  \frac{1}{2}\Big[1\!-\!\tanh(S\bra a\ket^{\frac{1}{2}}y_0\!+\!\beta_0^\star)\Big]
  \Big[\tilde{x}(-(\beta_0\!+\!wy_0\!+\!vz),\tilde{u})\!+\!(\beta_0\!+\!wy_0\!+\!vz)\Big]^2
 \Bigg\},
 \nonumber
 \\[-0mm]
   \hspace*{-20mm}
  &&
\label{eq:ML_logistic_final_1}
\\[1mm]
\hspace*{-20mm}
\zeta &=&\!
\int\!{\rm D}y_0{\rm D}z
\Bigg\{
\frac{1}{2}\Big[1\!+\!\tanh(S\bra a\ket^{\frac{1}{2}}y_0\!+\!\beta_0^\star)\Big]
 \frac{\tilde{u}^2[1\!-\!\tanh^2(\tilde{x}(\beta_0\!+\!wy_0\!+\!vz,\tilde{u}))]}
 {1\!+\!\tilde{u}^2[1\!-\!\tanh^2(\tilde{x}(\beta_0\!+\!wy_0\!+\!vz,\tilde{u}))]}
 \nonumber
 \\
 \hspace*{-20mm}
 && \hspace*{0mm}
 +\frac{1}{2}\Big[1\!-\!\tanh(S\bra a\ket^{\frac{1}{2}}y_0\!+\!\beta_0^\star)\Big]
 \frac{\tilde{u}^2[1\!-\!\tanh^2(\tilde{x}(-(\beta_0\!+\!wy_0\!+\!vz),\tilde{u}))]}
 {1\!+\!\tilde{u}^2[1\!-\!\tanh^2(\tilde{x}(-(\beta_0\!+\!wy_0\!+\!vz),\tilde{u}))]}
 \Bigg\},
 \nonumber
 \\[-0mm]
   \hspace*{-20mm}
  &&
\label{eq:ML_logistic_final_2}
\\[1mm]
\hspace*{-20mm}
w
&=&
\frac{S \bra a\ket^{\frac{1}{2}}}{2\zeta}\int\!{\rm D}y_0{\rm D}z
\Big[1\!-\!\tanh^2(S\bra a\ket^{\frac{1}{2}}y_0\!+\!\beta_0^\star)\Big]
\nonumber
\\
\hspace*{-20mm}&&\hspace*{20mm}\times
\Big[
 \tilde{x}(\beta_0\!+\!wy_0\!+\!vz,\tilde{u})+
 \tilde{x}(-(\beta_0\!+\!wy_0\!+\!vz),\tilde{u})
 \Big],
\label{eq:ML_logistic_final_3}
\\[1mm]
\hspace*{-20mm}
\beta_0&=&\int\!{\rm D}y_0{\rm D}z
\Bigg\{
\frac{1}{2}\Big[1\!+\!\tanh(S\bra a\ket^{\frac{1}{2}}y_0\!+\!\beta_0^\star)\Big]
\tilde{x}(\beta_0\!+\!wy_0\!+\!vz,\tilde{u})
\nonumber
\\[-1mm]
\hspace*{-20mm}
&& \hspace*{20mm}
-
\frac{1}{2}\Big[1\!-\!\tanh(S\bra a\ket^{\frac{1}{2}}y_0\!+\!\beta_0^\star)\Big]
\tilde{x}(-(\beta_0\!+\!wy_0\!+\!vz),\tilde{u})
\Bigg\}.
\label{eq:ML_logistic_final_4}
\end{eqnarray}
\vsp

\begin{figure}[t]
\unitlength=0.32mm
\hspace*{-11mm}
\begin{picture}(400,175)
\put(0,0){\includegraphics[width=240\unitlength]{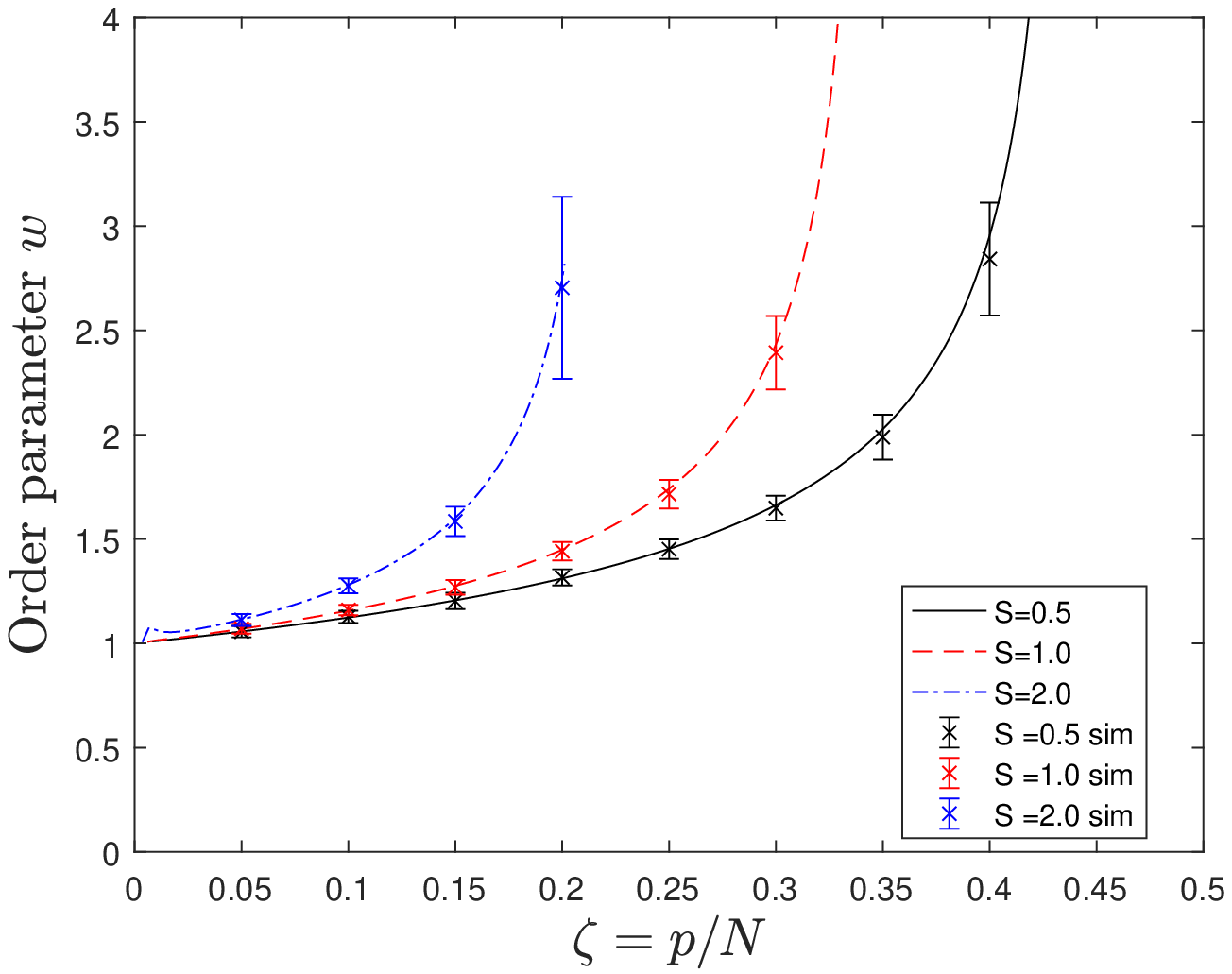}}
\put(220,0){\includegraphics[width=240\unitlength]{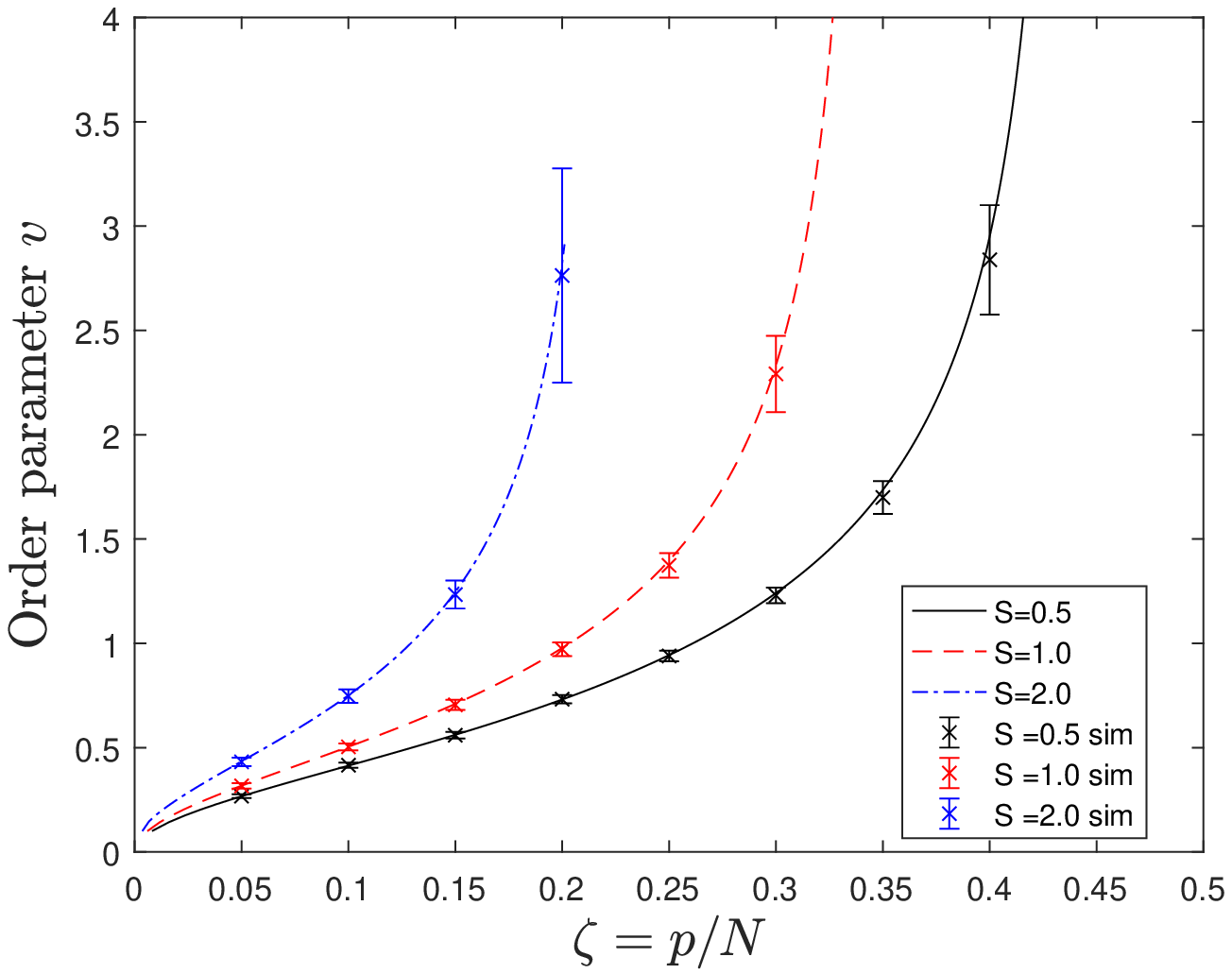}}
%\put(117,-8){\small\em $\zeta$}
%\put(277,-8){\small\em $\zeta$}
%\put(32,85){\small $w$}
%\put(192,85){\small $v$}

\end{picture}
\vspace*{0mm}

\caption{\small Regression simulations (markers) versus theoretical predictions (lines) for logistic ML regression with $Np=400,000$, $\bA=\one$, and $\beta_0^\star=0$. Left: order parameter $w$ versus $\zeta$, for different values of $S$. Right: order parameter $v$ versus $\zeta$, for different values of $S$. Each simulation data point represents average and standard deviation computed over 400 synthetic data sets and regressions.  
}
\label{fig:ML_logistic_vw}
\end{figure}

\unitlength=0.337mm
\begin{figure}[t]
\hspace*{-8mm}
\begin{picture}(300,318)

\put(25,300){\small $\eta\!=\!\frac{1}{10}$, $\epsilon\!=\!\frac{1}{2}$, $S\!=\!1$}
\put(125,300){\small $\eta\!=\!\frac{1}{20}$, $\epsilon\!=\!0$, $S\!=\!1$}
\put(225,300){\small $\eta\!=\!\frac{1}{20}$, $\epsilon\!=\!\frac{3}{4}$, $S\!=\!1$}
\put(325,300){\small $\eta\!=\!\frac{1}{20}$, $\epsilon\!=\!\frac{3}{4}$, $S\!=\!2$}

\put(3,233){\small $\hat{\beta}_\mu$}
\put(60,175){\small $\beta^\star_\mu$}\put(160,175){\small $\beta^\star_\mu$}
\put(260,175){\small $\beta^\star_\mu$}\put(360,175){\small $\beta^\star_\mu$}
\put(28,265){\small $\zeta\!=\!0.5$}\put(128,265){\small $\zeta\!=\!0.5$}
\put(228,265){\small $\zeta\!=\!0.5$}\put(328,265){\small $\zeta\!=\!0.5$}
\put(0,185){\includegraphics[width=135\unitlength]{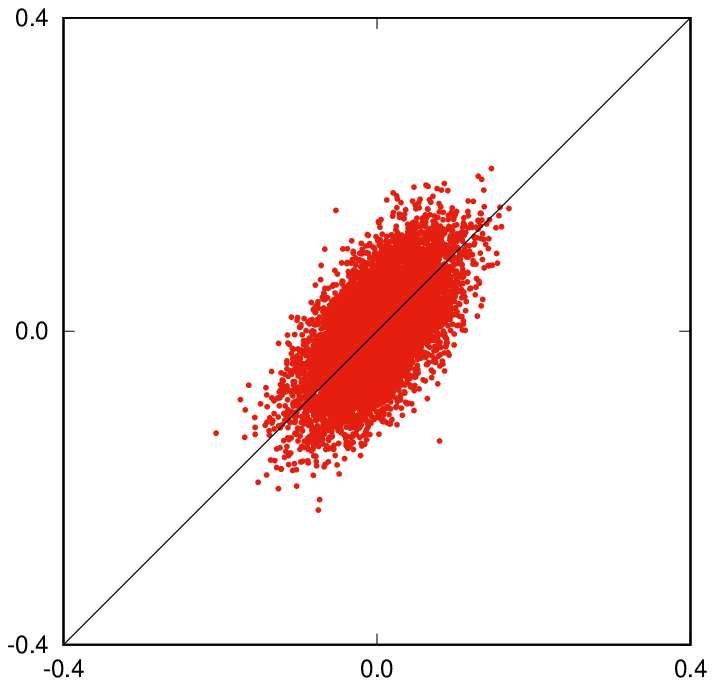}}
%\put(63,95){\small $\zeta$}
\put(100,185){\includegraphics[width=135\unitlength]{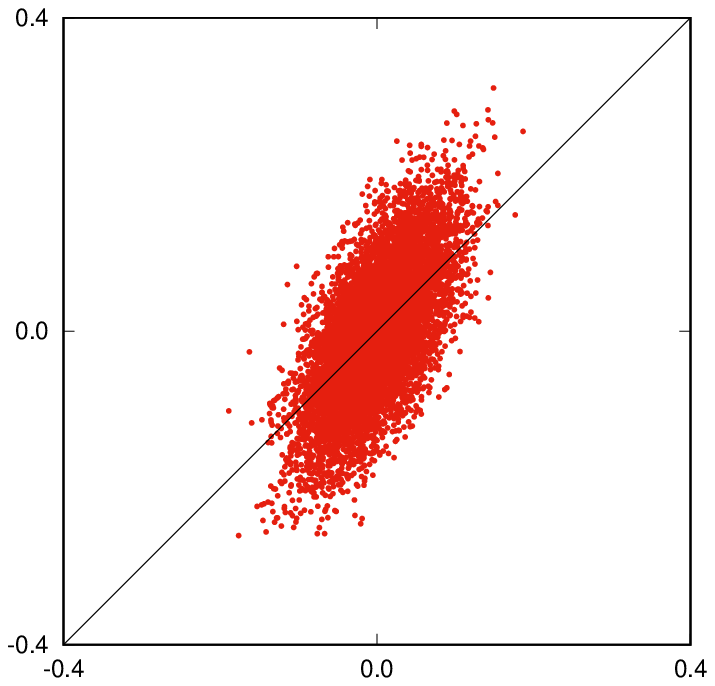}}
%\put(163,95){\small $\zeta$}
\put(200,185){\includegraphics[width=135\unitlength]{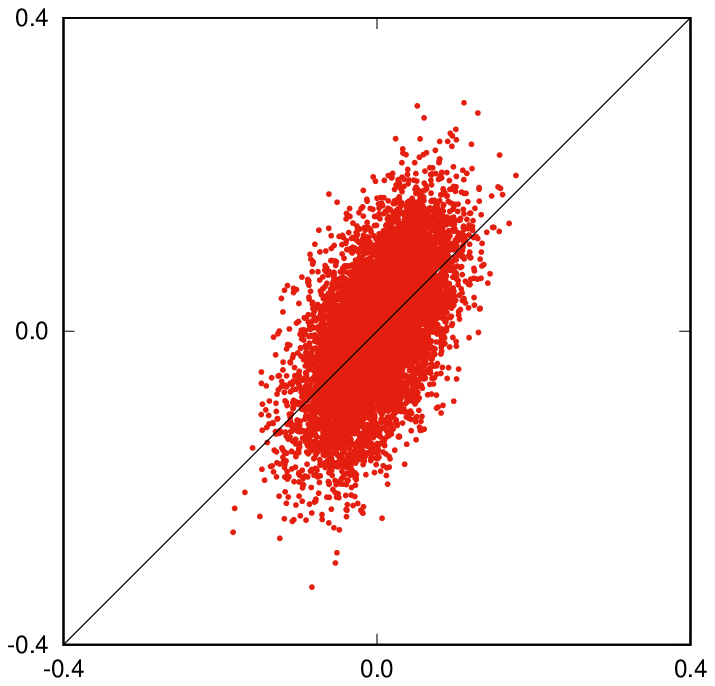}}
%\put(263,95){\small $\zeta$}
\put(300,185){\includegraphics[width=135\unitlength]{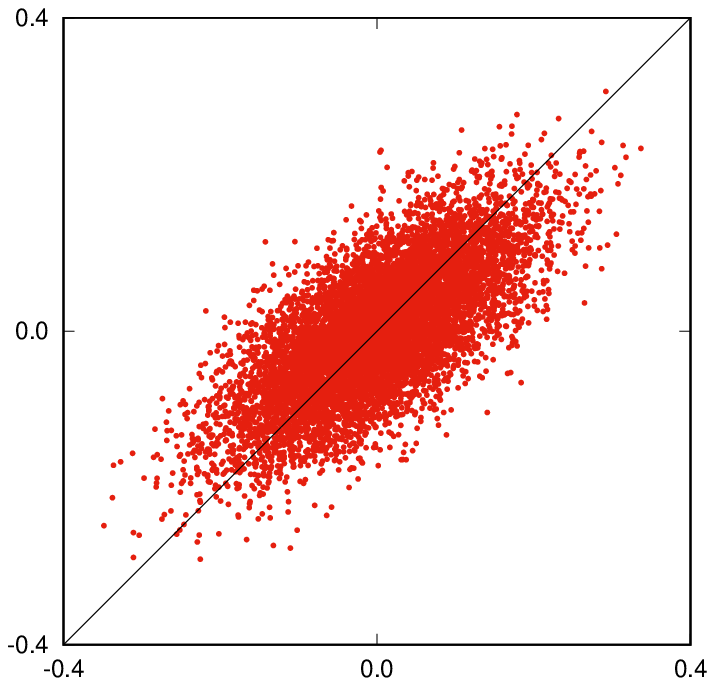}}
%\put(363,95){\small $\zeta$}

\put(2,120){$w$}
\put(0,70){\includegraphics[width=135\unitlength]{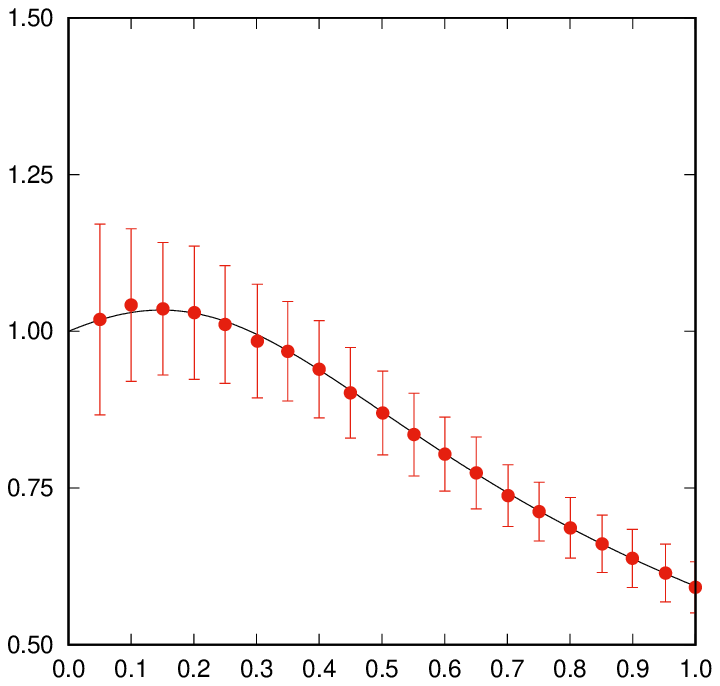}}
%\put(63,95){\small $\zeta$}
\put(100,70){\includegraphics[width=135\unitlength]{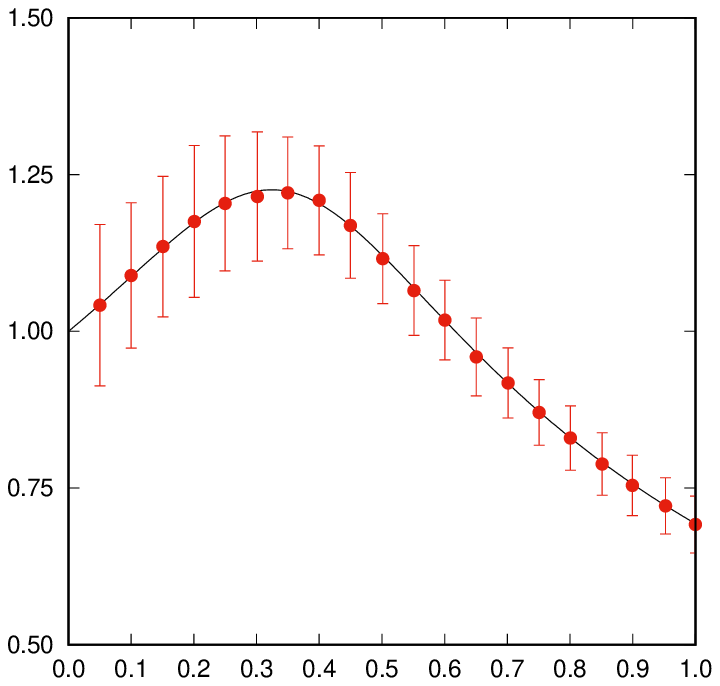}}
%\put(163,95){\small $\zeta$}
\put(200,70){\includegraphics[width=135\unitlength]{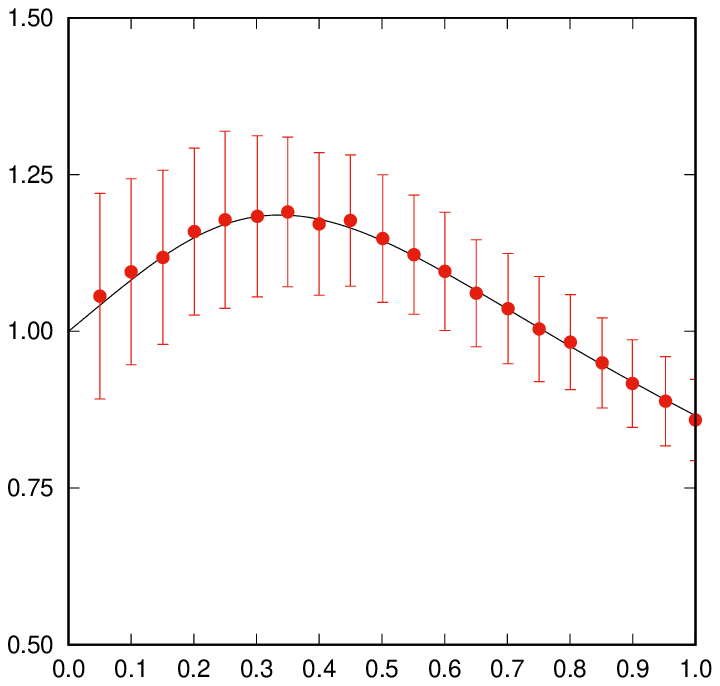}}
%\put(263,95){\small $\zeta$}
\put(300,70){\includegraphics[width=135\unitlength]{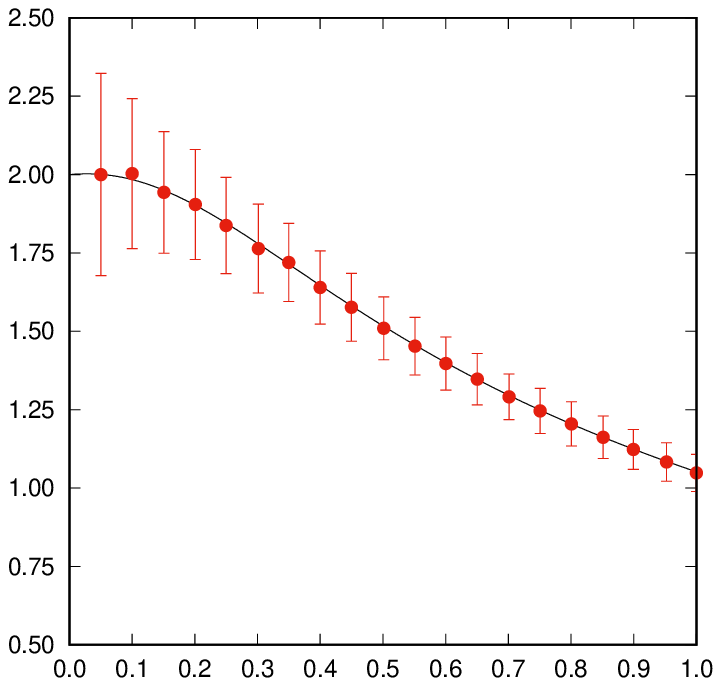}}
%\put(363,95){\small $\zeta$}

\put(2,20){$v$}
\put(0,-30){\includegraphics[width=135\unitlength]{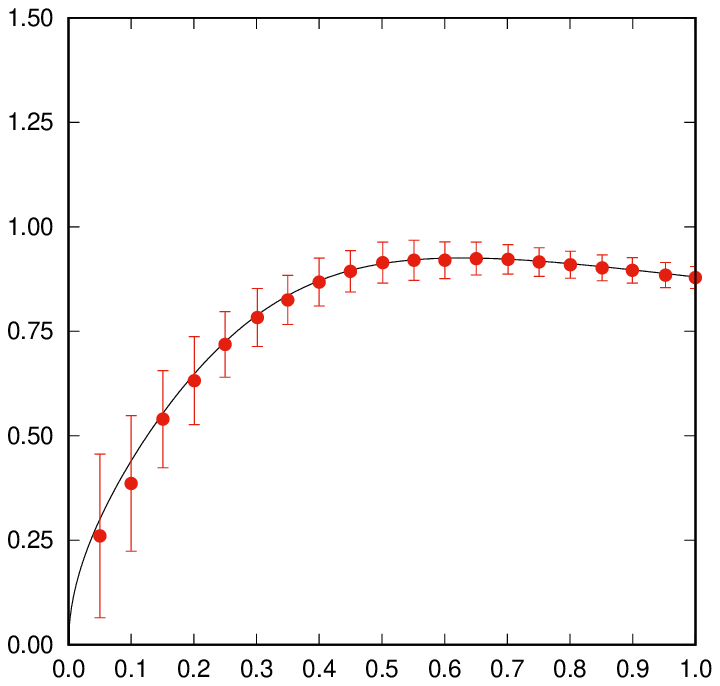}}
\put(63,-40){\small $\zeta$}
\put(100,-30){\includegraphics[width=135\unitlength]{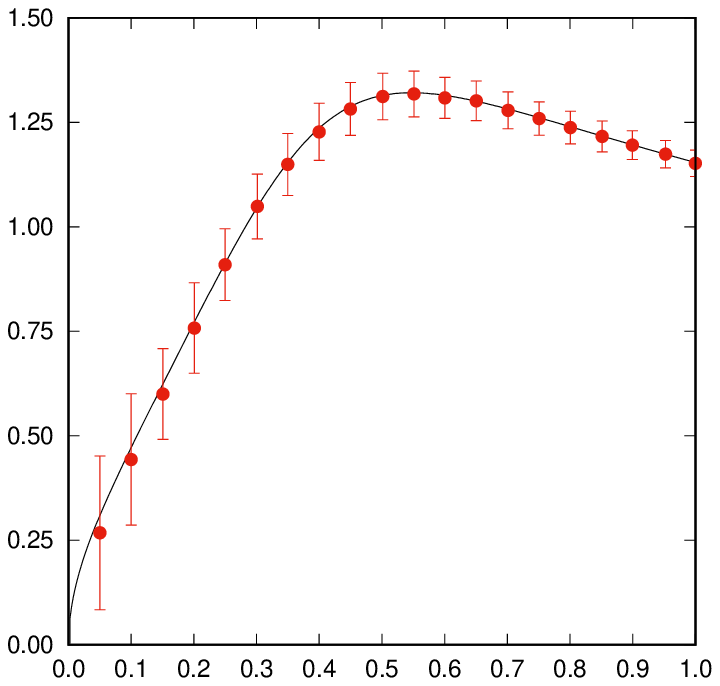}}
\put(163,-40){\small $\zeta$}
\put(200,-30){\includegraphics[width=135\unitlength]{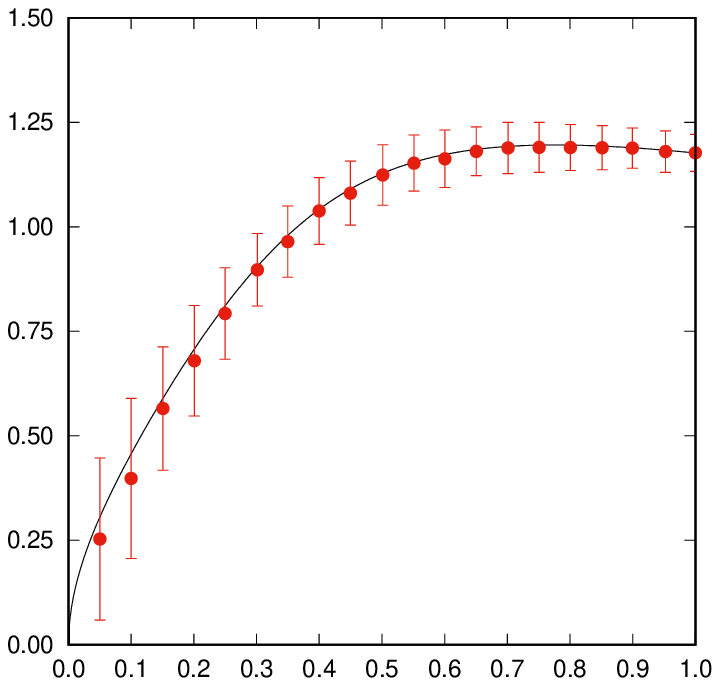}}
\put(263,-40){\small $\zeta$}
\put(300,-30){\includegraphics[width=135\unitlength]{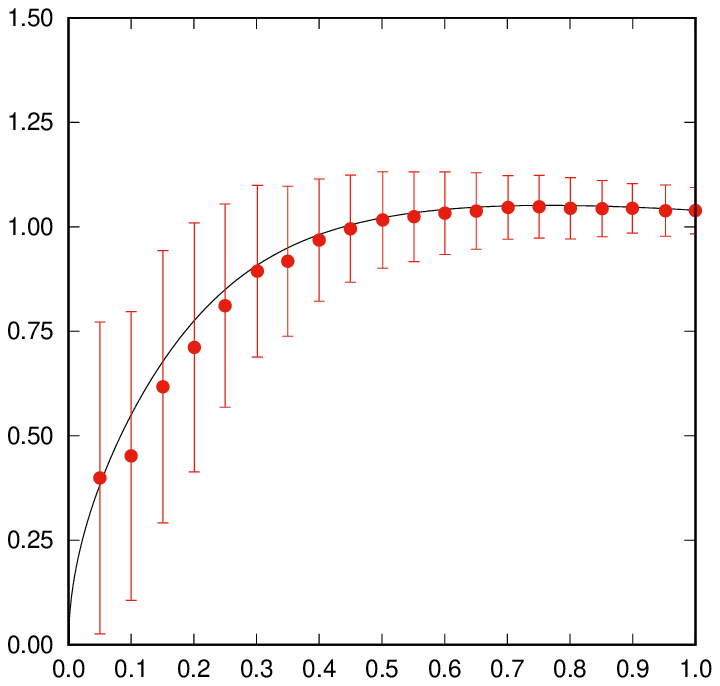}}
\put(363,-40){\small $\zeta$}

\end{picture}
\vspace*{15mm}

\caption{\small Results of  logistic MAP regression simulations with $Np=400,000$ and $\beta_0^\star=0$, for different combinations of $\eta$ (regularizer strength), $\epsilon$ (covariate correlations), and $S$ (true association strengths).  Top row: inferred versus true association parameters for the ratio $\zeta=p/N=0.5$. 
Middle and lower row: order parameters $w$ and $v$ plotted versus $\zeta$. Each simulation data point represents average and standard deviation computed over 400 synthetic data sets and regressions. Solid curves give the theoretical predictions obtained by solving the RS equations.   }
\label{fig:MAP_logistic_vw}
\end{figure}

\unitlength=0.337mm
\begin{figure}[t]
\hspace*{-8mm}
\begin{picture}(300,220)

\put(25,208){\small $\eta\!=\!\frac{1}{10}$, $\epsilon\!=\!\frac{1}{2}$, $S\!=\!1$}
\put(125,208){\small $\eta\!=\!\frac{1}{20}$, $\epsilon\!=\!0$, $S\!=\!1$}
\put(225,208){\small $\eta\!=\!\frac{1}{20}$, $\epsilon\!=\!\frac{3}{4}$, $S\!=\!1$}
\put(325,208){\small $\eta\!=\!\frac{1}{20}$, $\epsilon\!=\!\frac{3}{4}$, $S\!=\!2$}

\put(-12,150){\em\small slope}
\put(0,100){\includegraphics[width=135\unitlength]{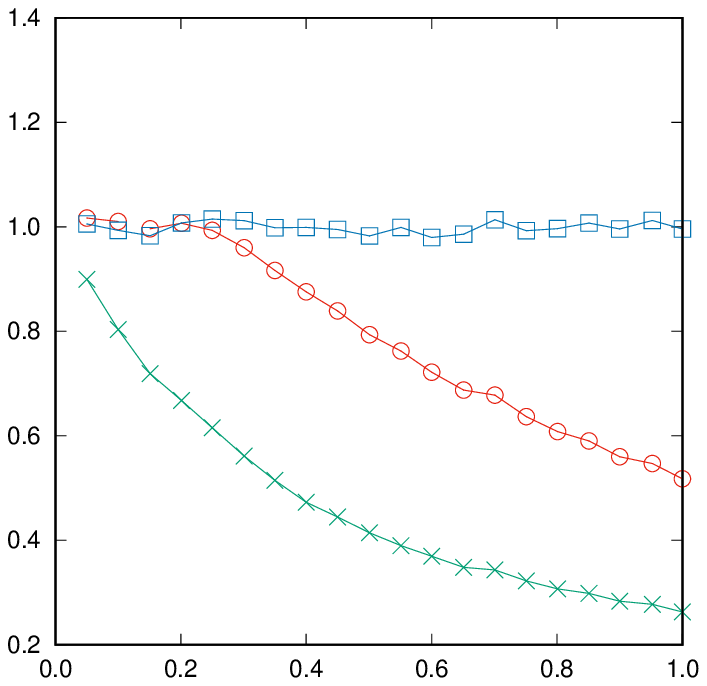}}
%\put(63,95){\small $\zeta$}
\put(100,100){\includegraphics[width=135\unitlength]{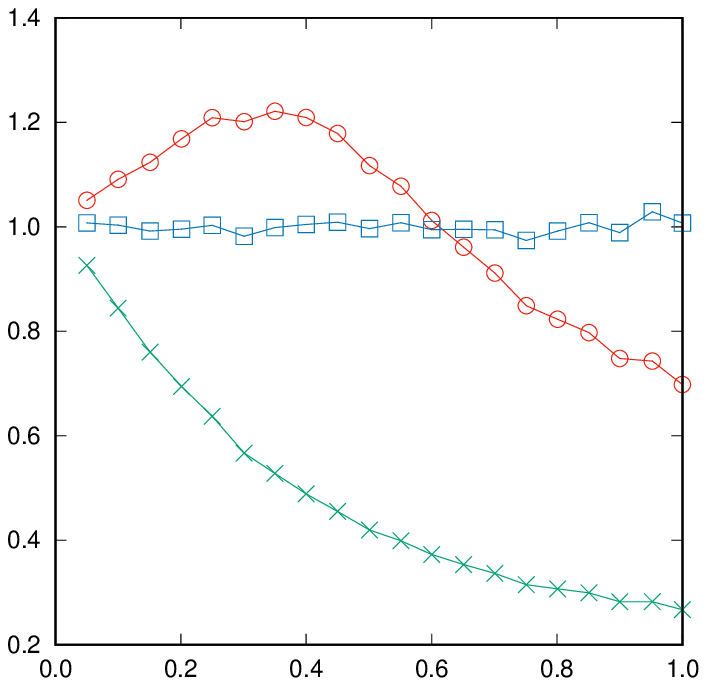}}
%\put(163,95){\small $\zeta$}
\put(200,100){\includegraphics[width=135\unitlength]{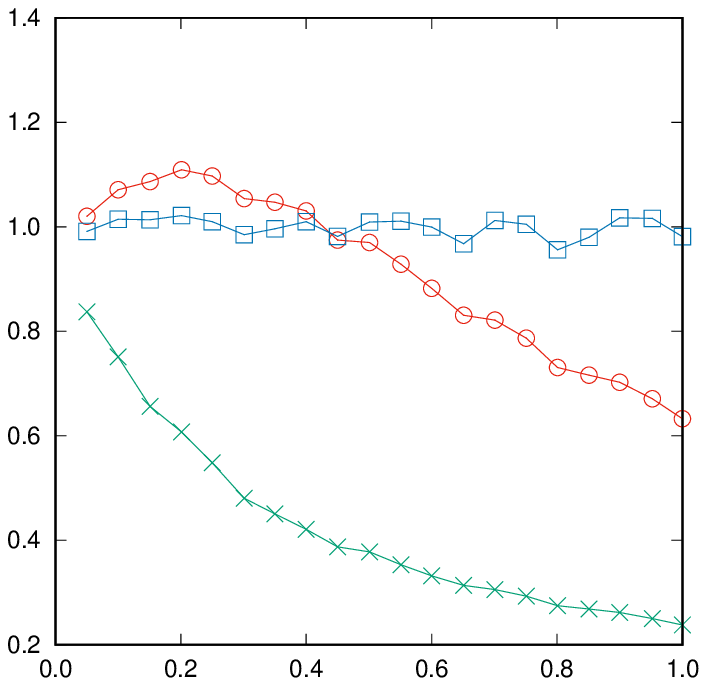}}
%\put(263,95){\small $\zeta$}
\put(300,100){\includegraphics[width=135\unitlength]{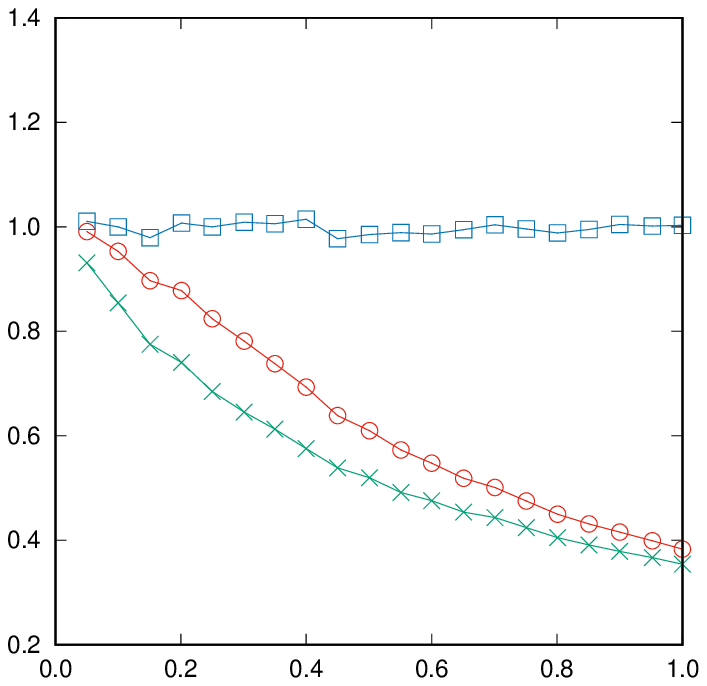}}
%\put(363,95){\small $\zeta$}

\put(-20,50){\em\small $\sqrt{\it MSE}$}
\put(0,0){\includegraphics[width=135\unitlength]{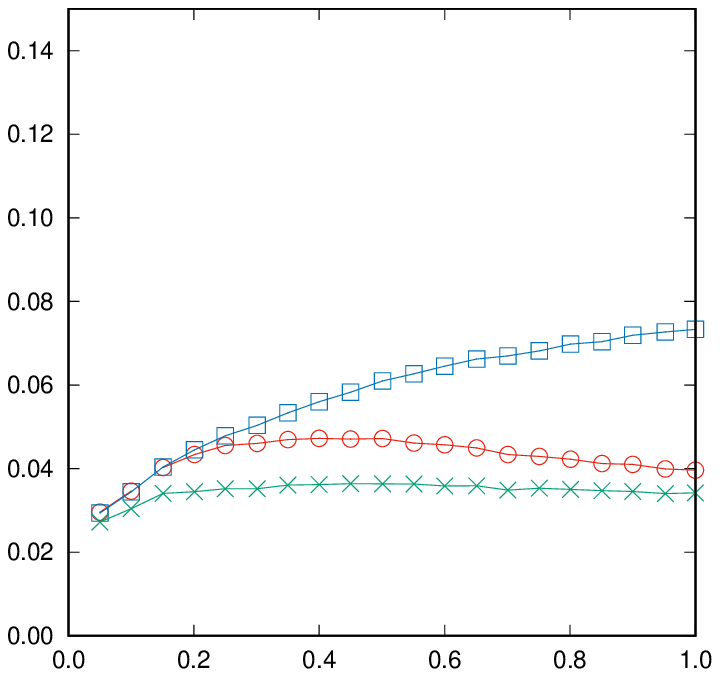}}
\put(63,-10){\small $\zeta$}
\put(100,0){\includegraphics[width=135\unitlength]{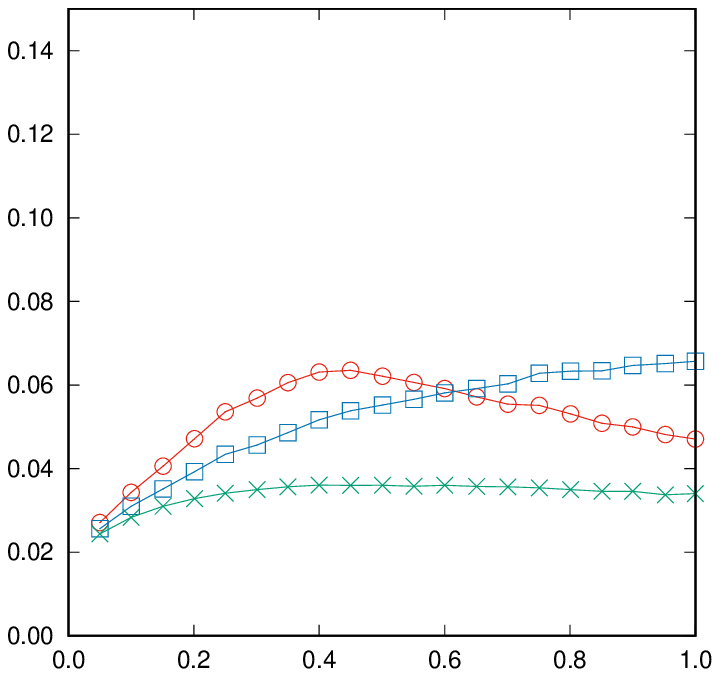}}
\put(163,-10){\small $\zeta$}
\put(200,0){\includegraphics[width=135\unitlength]{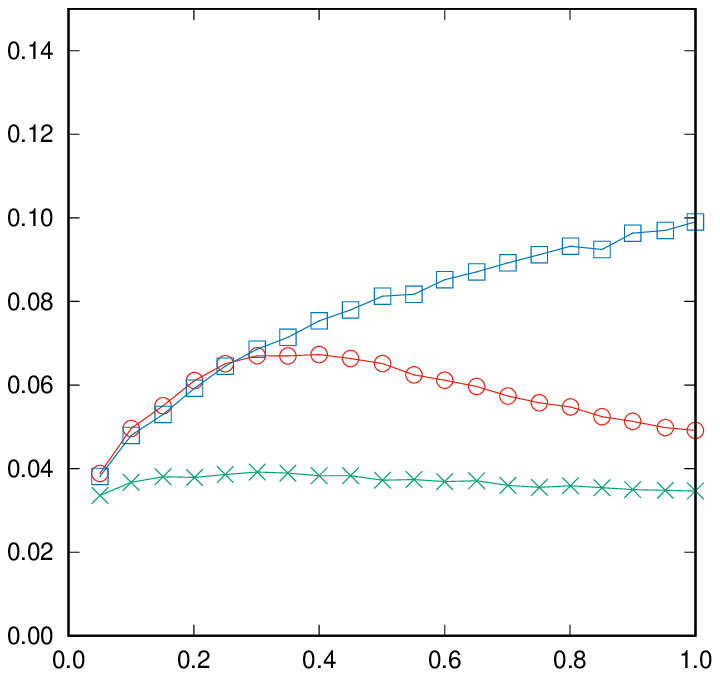}}
\put(263,-10){\small $\zeta$}
\put(300,0){\includegraphics[width=135\unitlength]{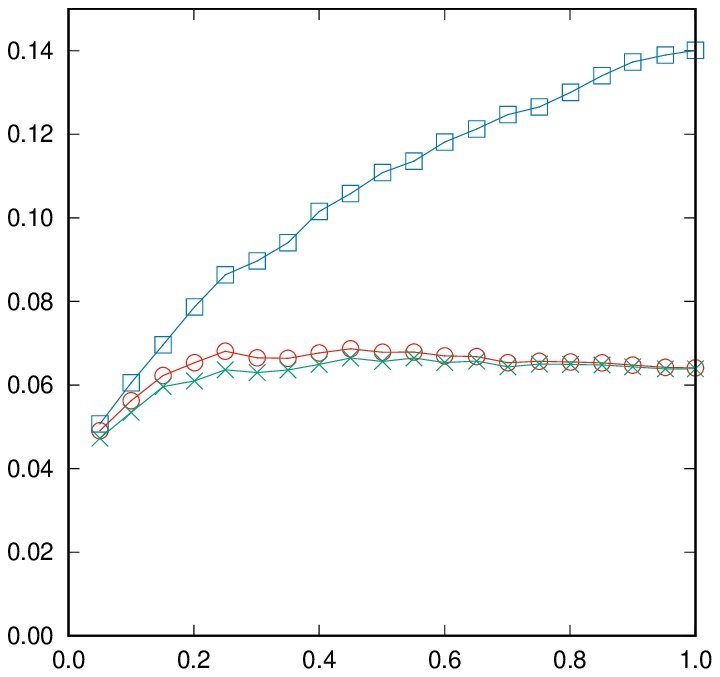}}
\put(363,-10){\small $\zeta$}

\end{picture}
\vspace*{5mm}

\caption{\small Tests of the correction protocols for MAP estimators, applied to the data of the previous figure. 
Red circles: the MAP estimator $\hat{\beta}_\mu$. Blue squares: the corrected estimator (\ref{eq:Correction_debias}), aimed at removing inference bias . Green crosses: the corrected estimator (\ref{eq:Correction_mse}), aimed at minimizing the MSE. 
The top row show as a function of $\zeta$ the slopes of the relation between the three estimators  and the true values 
$\beta_\mu^\star$ of association parameters (the slopes of the data clouds in the top row of the previous figure, computes via least squares analysis); this slope would be unity for unbiased estimators. 
The bottom row shows the values of $\sqrt{\rm MSE}$, where ${\rm MSE}=p^{-1}\sum_{\mu\leq p}(\hat{\beta}_\mu\!-\!\beta_\mu^\star)^2$.  
 }
\label{fig:MAP_logistic_correction}
\end{figure}

\begin{figure}[t]
\unitlength=0.42mm
\hspace*{30mm}
\begin{picture}(400,130)

\put(70,110){\small $\eta\!=\!0$}
\put(105,85){\small $\eta\!=\!0.025$}
\put(107,33){\small $\eta\!=\!0.05$}

\put(9,65){\small $\hat{\beta}_0$}
\put(0,0){\includegraphics[width=180\unitlength]{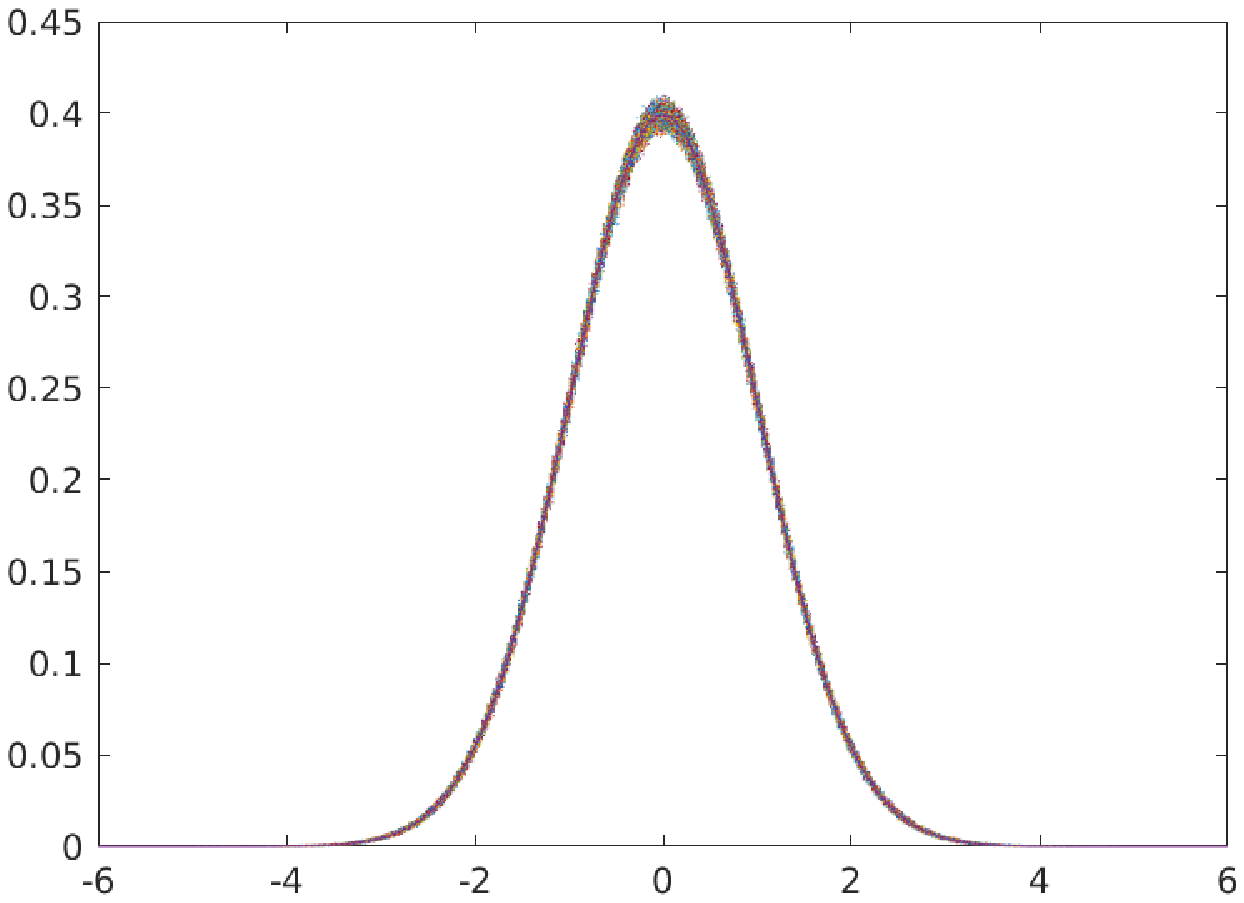}}
\put(91,-4){\small $\hat{\beta}^\prime$}
\put(-12,70){\small $P(\hat{\beta}^\prime)$}
\end{picture}
\vspace*{-2mm}

\caption{Histograms of rescaled values of the MAP estimators, each defined as $\hat{\beta}^\prime_\mu=
\big[\hat{\beta}_\mu+ \tilde{d}_0
 [ (\tilde{g}\one+2\eta\bA^{-1})^{-1}
 \bbeta^\star]_\mu\big]/\sqrt{|\tilde{f}|[ 
  (\tilde{g} \bA+2\eta\one)^{-1}\bA(\tilde{g} \bA+2\eta\one)^{-1}]_{\mu\mu}}$, computed 
for $10^6$  independent regularized logistic regression experiments with correlated covariates ($\epsilon=0.75$, $\eta=0.01$, $S=1$, $p=500$ and $N=1000$).  The resulting 500 histograms are plotted together in the present figure, one for each value of $\mu$. According to the theoretical prediction   (\ref{eq:beta_relation_derived}), these histograms should all asymptotically become zero average and unit average distributions. This is indeed seen to be the case.}

 \label{fig:logistic_gaussian}
 
\end{figure}

\noindent{\em ML logistic regression for data with zero offset.}
Further simplifications arise when we have $\beta_0^\star=0$. By symmetry of the Gaussian averages we now immediately obtain $\beta_0=0$ in (\ref{eq:ML_logistic_final_4}), which leaves us with just three coupled equations to be solved numerically. Upon using wherever possible the symmetry of the Gaussian averages these final equations 
take the relatively simple form:
\begin{eqnarray}
\hspace*{-15mm}
\zeta v^2
&=&\int\!{\rm D}y_0{\rm D}z ~
\Big[1\!+\!\tanh(S\bra a\ket^{\frac{1}{2}}y_0)\Big]
  \Big[\tilde{x}(wy_0\!+\!vz,\tilde{u})\!-\!(wy_0\!+\!vz)\Big]^2,
\label{eq:ML_logistic_final_betazero_1}
\\[1mm]
\hspace*{-15mm}
\zeta &=&
\int\!{\rm D}y_0{\rm D}z~
\Big[1\!+\!\tanh(S\bra a\ket^{\frac{1}{2}}y_0)\Big]
 \frac{\tilde{u}^2[1\!-\!\tanh^2(\tilde{x}(wy_0\!+\!vz,\tilde{u}))]}
 {1\!+\!\tilde{u}^2[1\!-\!\tanh^2(\tilde{x}(wy_0\!+\!vz,\tilde{u}))]},
\label{eq:ML_logistic_final_betazero_2}
\\[1mm]
\hspace*{-15mm}
\zeta w
&=& S \bra a\ket^{\frac{1}{2}}
\int\!{\rm D}y_0{\rm D}z~
\Big[1\!-\!\tanh^2(S\bra a\ket^{\frac{1}{2}}y_0)\Big]
 \tilde{x}(wy_0\!+\!vz,\tilde{u}).
\label{eq:ML_logistic_final_betazero_3}
\end{eqnarray}
In Figure \ref{fig:ML_logistic_vw} we plot the resulting values of the order parameters $v$ and $w$, whose physical meaning is given in (\ref{eq:meaning_v},\ref{eq:meaning_w}), as functions of $\zeta=p/N$, together with the corresponding results of  regression simulations on synthetic data with Gaussian covariates.  
The agreement between theory and simulations is very good.
\vsp

\noindent{\em MAP logistic regression with correlated covariates.}
The result of solving numerically the MAP equations (\ref{eq:logistic_MAP_1}-\ref{eq:logistic_MAP_6}) in the presence of covariate correlations of the type (\ref{eq:chosen_A}) is shown in Figure \ref{fig:MAP_logistic_vw}, where we plot the resulting values of the order parameters $v$ and $w$ together with regression simulation data (for synthetic Gaussian covariates) as  functions of the ratio $\zeta=p/N$. In these experiments we chose $\beta_0=\beta_0^\star=0$; we will address the  intercept parameter below. Once more we observe excellent agreement between theory and simulation. 
 In the top row we also plot for each parameter combination the MAP-inferred parameters $\hat{\beta}_\mu$ versus the corresponding true association strengths $\beta_\mu^\star$, for pooled data from 20 regressions and $\zeta=0.5$.

Again we can also for logistic regression test our two protocols (\ref{eq:Correction_debias},\ref{eq:Correction_mse}) for correcting the MAP estimator of the association parameters for the distortions caused by overfitting.
 See Figure \ref{fig:MAP_logistic_correction}. 
 The slopes of the data clouds of estimators versus true parameter values, as shown for $\zeta=0.5$ in the top row of Figure \ref{fig:MAP_logistic_vw}, are indeed typically away from unity (implying inference bias), both for the MAP estimator (red circles) and the minimum MSE estimator (\ref{eq:Correction_mse}) (green crosses). For the debiased estimator  (\ref{eq:Correction_debias}) (blue squares), in contrast, the slope is indeed unity, indicating that bias has been removed successfully.  Similarly, the MSE values of the minimum MSE estimator (\ref{eq:Correction_mse}) (green) are as predicted indeed always below those of the other two estimators. 
 
 In order to test prediction   (\ref{eq:beta_relation_derived}) for the distribution of inferred regression parameters we next generated $10^6$ data sets, each with $p=500$ and $N=1000$ (so $\zeta=0.5$),  with Gaussian covariates that are pairwise correlated according to (\ref{eq:chosen_A}) and $\epsilon=0.75$. The true association parameters were drawn as i.i.d. Gaussian random  variables with amplitude $S=1$. After carrying our regularized logistic regression with $\eta=0.05$, we carried out on each of the resulting MAP estimators $\{\hat{\beta}_\mu\}$ of each dataset the specific linear transformation that according to  (\ref{eq:beta_relation_derived})  should transform these into zero-average and unit variance Gaussian random variables (using the order parameters computed from the theory). Upon creating for each value of $\mu$ a histogram of the rescaled estimators $\hat{\beta}_\mu^\prime$, we obtain 500 histograms which according to theory should all collapse asymptotically to a zero average unit variance Gaussian. The result is shown in Figure \ref{fig:logistic_gaussian}.  This figure confirms that, even for the modest values of $p$ and $N$ used, the predicted Gaussian statistics of the estimators with the predicted values of average and width  given in (\ref{eq:beta_relation_derived}) are indeed correct. 
\vsp

\noindent{\em Intercept parameter for imbalanced class sizes.}
Having training data with vastly different outcome class sizes leads to the minority outcome rarely being predicted \cite{wallace2011class} in logistic regression. As this imbalance increases, especially in the overfitting regime the intercept term $\beta_0$ in parametrized models diverges  \cite{owen2007infinitely}, and  all new samples are  assigned the majority outcome. 
Medical data often exhibit large imbalances between numbers of diseased and healthy samples, with  the clinically important decision relying on identifying the rare cases correctly. Similarly, in financial fraud detection there may be millions of legitimate transactions against a handful of fraudulent ones, and we seek to identify the minority class. 
Existing methods to mitigate the effect of class imbalance have focused on data pre-processing \cite{chawla2002smote,drummond2003c4} or incorporating a cost function into the classification algorithm \cite{wallace2011class}. While these methods are useful to the practitioner, theoretical explanations are limited \cite{owen2007infinitely,sei2014infinitely}. Our present theory enables us to investigate class imbalance effects analytically.

\begin{figure}[t]
\unitlength=0.42mm
\hspace*{24mm}
\begin{picture}(400,130)

\put(70,103){\small $\eta\!=\!0$}
\put(135,80){\small $\eta\!=\!0.025$}
\put(137,30){\small $\eta\!=\!0.05$}

\put(15,65){\small $\hat{\beta}_0$}
\put(0,0){\includegraphics[width=221\unitlength,height=120\unitlength]{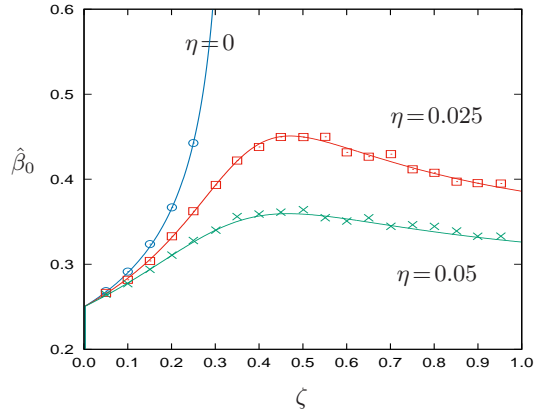}}
\put(105,-9){\small $\zeta$}
\end{picture}
\vspace*{2mm}

\caption{Predicted values of the offset parameter $\beta_0$ are drawn as solid curves, for $\A = \one$ (uncorrelated covariates), $S=1$, and three regularization strengths ($\eta=0$: blue circles; $\eta=0.025$: red squares; $\eta=0.05$: green crosses). Full circles give the average values of $\hat{\beta}_0$  found in MAP regression, for 400 simulations with Gaussian covariates and $NP=400,000$.  Standard deviations are not shown in order to reduce visual clutter, but range between 0.003 for small $\zeta$ and 0.1 for large $\zeta$).   The true offset used in generating the data was $\beta_0^\star=0.25$, representing an average class imbalance of $42:58$ according to (\ref{eq:imbalance}).  }

 \label{fig:intercept}
 
\end{figure}

The outcome class imbalance in logistic regression data is measured by $m= N^{-1} \sum_{i=1}^N s_i \in [-1,1]$. Averaging over the data in  (\ref{eq:logistic_regression_model}) gives the expectation value
\begin{eqnarray}
\label{eq:imbalance}
\bra m \ket  &= &\int \!\rmd \z ~ p(\z)  \tanh\big(\beta_0^\star \!+\!\bb^\star\!\!\cdot \z/\sqrt{p} \big).
\end{eqnarray}
Figure \ref{fig:intercept} shows our theoretical predictions for the inferred  order parameter $\beta_0$ in ML and MAP logistic regression, together with the result of regression simulations. For any given level of outcome imbalance, controlled by $\beta_0^\star$, the bias in the inferred intercept increases with $\zz$. Regularization mitigates this effect, leading to the possibility of correcting inferred class imbalances, in spite of the  regularization being applied only to the coefficients $\{\beta_{\mu}\}_{\mu=1}^p$,  not to  $\beta_0$ itself. This situation is reminiscent of  Cox's survival analysis model \cite{Cox}, where the inferred hazard rate (given by Breslow's estimator \cite{breslow1972contribution})  can be expressed in terms  of the inferred regression coefficients, and thereby inherits their inference bias.  In the $\eta>0$ case, we find the intercept inflation differs from  that of the association parameters, due to their rescaling with $\sqrt{p}$.
Again the agreement between theory and experiment is very satisfactory.

\subsection{Regularized Cox regression}

{\em Equations for MAP Cox regression.}
In Cox regression without censoring\footnote{Also the more complicated case of Cox regression with censoring falls within the scope of our present GLM equations, provided the censoring events are non-informative (as with end-of-trial censoring). For informative censoring, such as censoring caused by nontrivial competing risks, we first need to generalize the theory further to models in which outcome probabilities involve multiple linear combinations of covariates. This should be straightforward, but will be the subject of a future study. } we have 
$s=t\in[0,\infty)$, $\theta=\{\lambda(t)\}$ (the so-called base hazard rate, a nonnegative function on the time interval $[0,\infty)$ such that $\Lambda(t)=\int_0^t\!\rmd t^\prime~\lambda(t^\prime)$ diverges for $t\to\infty$), and 
\begin{eqnarray}
p(t|\xi,\lambda)&=& \lambda(t)\rme^{\xi-\exp(\xi)\Lambda(t)},\\
\log p(t|\xi,\lambda)&=&  \log\lambda(t)+\xi-\rme^\xi\Lambda(t).
\end{eqnarray}
Hence 
\begin{eqnarray}
\frac{\partial}{\partial\xi}\log p(t|\xi,\lambda)= 1-\rme^\xi\Lambda(t),
~~~~~~
\frac{\partial^2}{\partial\xi^2}\log p(t|\xi,\lambda)=-\rme^\xi\Lambda(t).
\end{eqnarray}
The analysis of overfitting in MAP Cox regression with arbitrary covariate covariance matrices $\bA$ was first carried out in \cite{SheikhCoolen2019}. 
We will now show that from the general equations  (\ref{eq:ddf2=0},\ref{eq:ddg2=0},\ref{eq:ddu2=0},\ref{eq:ddv2=0},\ref{eq:ddw2=0},\ref{eq:ddtheta2=0}) one indeed recovers the results of \cite{SheikhCoolen2019}, and more (e.g. the explicit link between true and inferred association parameters). 

We will first compute the various model-dependent building blocks of the RS equations.
The relevant (functional and partial) derivatives of the logarithm of the outcome probability density are
\begin{eqnarray}
\frac{\delta}{\delta\lambda(t)}\log p(s|\xi,\lambda)&=&\frac{\delta(s\!-\!t)}{\lambda(t)}-\rme^\xi \theta(s\!-\!t),
\\
\frac{\partial}{\partial y_0}\log p(s|S\bra a\ket^{\frac{1}{2}}y_0,\lambda^\star)&=& S\bra a\ket^{\frac{1}{2}}\Big[1-\rme^{S\bra a\ket^{\frac{1}{2}}y_0}\Lambda^\star(s)\Big].
\end{eqnarray}
The function $\xi(\mu,\sigma,s,\beta_0)$
 is  here the solution of 
\begin{eqnarray}
1-\rme^\xi\Lambda(s)=(\xi-\mu)/\sigma^2.
\end{eqnarray}
Upon switching from $\xi$ to the variable $x=\mu\!-\!\xi\!+\!\sigma^2$, we can solve $x$ in explicit form:
\begin{eqnarray}
 x=W(\sigma^2\rme^{\mu+\sigma^2}\Lambda(s)).
\end{eqnarray}
Here $W(x)$ denotes the Lambert $W$-function. i.e. the inverse of $f(x)=x\exp(x)$, with derivative $W^\prime(x)=W(x)/x[1\!+\!W(x)]$.
It then follows that
\begin{eqnarray}
\xi(\mu,\sigma,s,\lambda)&=&\mu+\sigma^2-W(\sigma^2\rme^{\mu+\sigma^2}\Lambda(s)),
\\
\frac{\partial}{\partial_\mu}\xi(\mu,\sigma,s,\lambda)&=& \frac{1}{1\!+\!W(\sigma^2\rme^{\mu+\sigma^2}\Lambda(s))}.
\end{eqnarray}
\vsp

\noindent{\em 
Order parameter equations.}
We insert the above formulae into our  RS order parameter equations (\ref{eq:ddf2=0}--\ref{eq:ddtheta2=0}), and use identities such as 
$\bra \exp(S\bra a\ket^{\frac{1}{2}}y_0)\Lambda^\star(s)\ket_{s}=1$, to simplify our equations to the following set:
\begin{eqnarray}
\hspace*{-15mm}
\tilde{u}^2
&=&
\Big\bra  
\frac{a}{2\eta\!+\!\tilde{g}a}\Big\ket,
\label{eq:Cox1}
\\
\hspace*{-15mm}
v^2 &=& 
w^2\Big[
\bra a\ket
 \Big\bra \frac{a^2}{
2\eta\!+\!\tilde{g}a}\Big\ket^{\!-2}
 \Big\bra \frac{a^3}{
(2\eta\!+\!\tilde{g}a)^2}\Big\ket
-1\Big]
-
\tilde{f}\Big\bra  
\frac{a^2}{(2\eta\!+\!\tilde{g}a)^2}\Big\ket,
\label{eq:Cox2}
\\
\hspace*{-15mm}
- \zeta
\tilde{f} \tilde{u}^4
&=&
\Big\bra\!\Big\bra \Big\bra 
 \Big[
\tilde{u}^2-W(\sigma^2\rme^{wy_0+vz+\tilde{u}^2}\Lambda(s))
\Big]^2 \Big\ket_{\!s}\Big\ket\!\Big\ket, 
\label{eq:Cox3}
\\[2mm]
\hspace*{-15mm}
\zeta \tilde{g}\tilde{u}^2&=&
 \Big\bra\!\Big\bra \Big\bra 
 \frac{W(\sigma^2\rme^{wy_0+vz+\tilde{u}^2}\Lambda(s))}{1\!+\!W(\sigma^2\rme^{wy_0+vz+\tilde{u}^2}\Lambda(s))}
 \Big\ket_{s}\ket\!\Big\ket, 
 \label{eq:Cox4}
\\[1mm]
\hspace*{-15mm}
\frac{ \zeta w \tilde{u}^2 \bra a\ket^{\frac{1}{2}}}{S
 \big\bra \frac{a^2}{
2\eta+\tilde{g}a}\big\ket}
&=&
- \Big\bra\!\Big\bra\Big\bra \Big[
W(\tilde{u}^2\rme^{wy_0+vz+\tilde{u}^2}\Lambda(s))
\Big]
\Big[1-\rme^{S\bra a\ket^{\frac{1}{2}}y_0}\Lambda^\star(s)\Big]
\Big\ket_{\!s}\Big\ket\!\Big\ket,
\label{eq:Cox5}
\\[1mm]
\hspace*{-15mm}
\frac{\big\bra\!\big\bra \bra 
\delta(s\!-\!t)\ket_s\big\ket\!\big\ket}{\lambda(t)}
&=&\Big\bra\!\Big\bra \Big\bra  \rme^{wy_0+vz+\tilde{u}^2-W(\tilde{u}^2\rme^{wy_0+vz+\tilde{u}^2}\Lambda(s))}\theta(s\!-\!t)
\Big\ket_{\!s}\Big\ket\!\Big\ket.
\label{eq:Cox6}
\end{eqnarray}
The first three are immediately recognised from \cite{SheikhCoolen2019}. With the identity $\exp(-W(x))=W(x)/x$, we find also that  (\ref{eq:Cox4})
reduces to the corresponding equation in \cite{SheikhCoolen2019}. 
This leaves only the identification of equation (\ref{eq:Cox5}).
Let us start from the corresponding equation in \cite{SheikhCoolen2019}, which reads:
\begin{eqnarray}
\hspace*{-20mm}
\zeta w\Big[\tilde{g}-\bra a\ket\Big\bra \frac{a^2}{2\eta\!+\!\tilde{g}a}\Big\ket^{-1}\Big]&=&\frac{1}{\tilde{u}^2}\Big\bra\!\Big\bra y_0\Big\bra W(\tilde{u}^2\rme^{\tilde{u}^2+wy_0+vz}\Lambda(s))\Big\ket_{\!s}\Big\ket\!\Big\ket
\nonumber
\\
\hspace*{-20mm}
&=&\frac{1}{\tilde{u}^2}\Big\bra\!\Big\bra \frac{\partial} {\partial y_0}\Big\bra W(\tilde{u}^2\rme^{\tilde{u}^2+wy_0+vz}\Lambda(s))\Big\ket_{\!s}\Big\ket\!\Big\ket
\nonumber
\\[1mm]
\hspace*{-20mm}
&& \hspace*{-30mm} =\frac{1}{\tilde{u}^2}\Big\bra\!\Big\bra \Big\bra
\frac{\partial W(\tilde{u}^2\rme^{\tilde{u}^2+wy_0+vz}\Lambda(s))}{\partial y_0}
\nonumber
\\[-1mm]
\hspace*{-20mm}
&& 
+W(\tilde{u}^2\rme^{\tilde{u}^2+wy_0+vz}\Lambda(s))
\frac{\partial \log p(s|S\bra a\ket^{\frac{1}{2}}y_0,\lambda_0^\star)}{\partial y_0}
\Big\ket_{\!s}\Big\ket\!\Big\ket
\nonumber
\nonumber
\\
\hspace*{-20mm}
&& \hspace*{-30mm} = w\zeta\tilde{g}
+
\frac{S\bra a\ket^{\frac{1}{2}}}{\tilde{u}^2}\Big\bra\!\Big\bra \Big\bra W(\tilde{u}^2\rme^{\tilde{u}^2+wy_0+vz}\Lambda(s))
 \Big[1\!-\!\rme^{S\bra a\ket^{\frac{1}{2}}y_0}\Lambda^\star(s)\Big]
\Big\ket_{\!s}\Big\ket\!\Big\ket.~
\end{eqnarray}
From this we directly recover  (\ref{eq:Cox5}), as required. This confirms that from our general theory for GLMs we can indeed recover also for the example choice of regularized Cox regression the complete results of \cite{SheikhCoolen2019}. Moreover, we now have the additional identities (\ref{eq:average_beta},\ref{eq:covariance_beta}), which were not available in that earlier study, and reveal the nontrivial impact of regularization in the more realistic scenario of  correlated covariates (which cannot be extracted from overlap order parameters alone).

\section{Discussion}

In this paper we have extended to arbitrary generalized linear regression models (GLM) the replica analysis of overfitting in MAP and ML inference that was developed initially in \cite{coolen2017replica,SheikhCoolen2019} for Cox regression \cite{Cox} with time-to-event data.  Parameter inference methods such as MAP and ML were derived and work well for the regime $p\ll N$, where $p$ is the dimensionality and $N$ is the number of samples. But they can produce large inference errors when $p=\order(N)$. This seriously hampers statistical inference in high dimensions, and thereby limits progress in many data-driven scientific disciplines. 

In all GLMs,  the ML/MAP overfitting-induced parameter inference errors  consist of a  combination of a reproducible bias and excess noise, both of which disappear when $p/N\to 0$ but become more prominent as the ratio $p/N$ increases.  In the regime $p,N\to \infty$ with fixed $\zeta=p/N$, the replica method enables us to predict analytically both this inference bias and the distribution of the excess noise, expressed in terms of the true  parameters of the model that generated the data, and the distribution of covariates from which the samples were drawn. In contrast to some recent alternative approaches, such as \cite{sur2019modern,salehi2019impact,barbier2019optimal}, by using the replica method we are not restricted to uncorrelated covariates or to models with output noise only, and we can calculate in explicit form the relation between MAP/ML estimators and the true (but unknown) model parameters responsible for the data. Covariate correlations are in fact found to play an important role in this relation. Our results pave the way for correcting ML and MAP inferences in GLMs systematically for overfitting bias, and thereby extend the applicability of such models into the hitherto forbidden regime $p\!=\!\order(N)$.

We found that 
in our analysis, the choice of outcome data types and regression models can be left until after the derivation of replica symmetric order parameter equations (there is no evidence for replica symmetry breaking, which is reasonable since we have assumed absence of model mismatch). Our derivation relies only on the generalized linear form of GLMs and on choosing $L2$ priors. Hence the replica calculation need not be repeated for every new GLM model instance; as always with the replica method, it served as a relatively painless and elegant but powerful vehicle for arriving at a closed set of order parameter equations, together with formulae expressing the relation between the ML/MAP parameter estimators and the true values of these parameters. The latter equations can serve as a natural and convenient starting point for practical applications, even for those with no interest in their derivation. 
We illustrate our results in this paper by applying the general theory  to linear, logistic, and Cox regression, and find excellent agreement with simulations and earlier results. We have limited our experiments to $\zeta\leq 1$.  In ML regression this marks the point by which a phase transition will have occurred (even earlier in logistic regression), whereas in MAP regression with $\eta>0$ once can in fact continue regression and find agreement between theory and simulations into the $\zeta>1$ regime (data not shown here). 

This work can be extended in both practical and theoretical ways, several of which are presently being explored.  Our theory was built upon the idealized scenario of knowledge of the underlying data-generating model. To put it into practice, the variance $S$ of the true regression parameters and the population covariance matrix $\mathbf{A}$ need to be estimated. The former is available through the inferred MAP estimators and the existing order parameters. The latter can be estimated from the empirical covariate statistics, building on methods such as  \cite{A1,A2,burda2005spectral,el2008spectrum}. For time-to-event models, the next obvious step would be to include censored data. More general extensions of the present theory include working with alternative non-Gaussian priors, inspecting more pathological models or data where some of our mathematical assumptions no longer hold, or generalizing the theory to regression models with multiple linear predictors, such as multinomial regression \cite{obuchi2018accelerating}, multiple risks and latent classes in survival analysis \cite{Cox_book,multiple_risks}, and multilayer neural networks \cite{Coolenbook,MacKay,li2018exploring}. 
\vsp

\noindent{\bf Acknowledgements}
\\[2mm]
The authors gratefully acknowledge valuable and stimulating discussions with Sir David Cox and Dr Heather Battey. 
MS is supported by the Biotechnology and Biological Sciences Research Council (award
1668568) and GSK Ltd. AM is supported by Cancer Research UK (award C45074/A26553) and the UK's Medical Research Council (award MR/R014043/1). FAL is supported  through a scholarship from Conacyt (Mexico).

\section*{References}

\appendix

\section{Derivation of the generic saddle point form} 
\label{app:SheikhCoolen}

\subsection{Preparation}

We start with expression (\ref{eq:starting_point_of_replicas}), with the $L2$ prior $p(\bb)\propto \exp(-p\eta\bb^2)$:
\begin{eqnarray}
\hspace*{-10mm}
E_\gamma(\bbeta^\star,\theta^\star)
&=& -  \lim_{n\to 0}\frac{\partial}{\partial\gamma}\frac{1}{Nn}\log \int\!\rmd\theta^1\!\ldots\rmd\theta^n  
\int\!\rmd\bbeta^1\!\ldots\rmd\bbeta^n
 \prod_{\alpha=1}^n 
\Big[\frac{p(\bbeta^\alpha)}{p(\bbeta^\star)}\Big]^\gamma
\nonumber
\\
\hspace*{-10mm}
&&\hspace*{0mm} \times
\Big\{
\int\!\rmd\bz \rmd s~p(\bz)p(s|\bz,\bbeta^\star,\theta^\star)
 \prod_{\alpha=1}^n
\Big[\frac{p(s|\bz,\bbeta^\alpha,\theta^\alpha)}{p(s|\bz,\bbeta^\star,\theta^\star)}\Big]^\gamma\Big\}^N.
\end{eqnarray}
The covariate distribution  $p(\bz)$ is assumed to have have zero mean and covariance matrix $\A$, with entries $A_{\mu \nu} = \int\!\rmd\z~p(\z) z_{\mu} z_{\nu}$.   We consider  the regime where $N,p\rightarrow\infty$ with fixed ratio $\zz = p/N$. Following \cite{coolen2017replica,SheikhCoolen2019} we next introduce
\begin{eqnarray}
p(\y | \bb^0\!, \ldots, \bb^n) = \int\!\rmd \z~ p(\z) \prod_{\alpha=0}^n \delta \Big[ y^{\alpha} -   \frac{\bb^{\alpha}\! \cdot \z}{\sqrt{p}} \Big],
\label{eq:switch_to_y}
\end{eqnarray}
where $\y = \{ y^0 , y^1, \ldots, y^n \}\! \in\! \R^{n+1}$ (which in survival analysis would be interpreted as risk scores) and $\bbeta^0\equiv\bbeta^\star$. Now
\begin{eqnarray}
\hspace*{-15mm}
\label{eq:energy2}
E_{\gamma}(\bb^\star, \theta^\star) &=& - \frac{\partial }{\partial \gamma}  \lim\limits_{n \to 0} \frac{1}{Nn}
\log \int \!\rmd\theta^1\! \ldots \rmd\theta^n  \int \!\rmd\bb^1\! \ldots \rmd\bb^n  \prod_{\alpha=1}^n \Big[\frac{ p(\bb^{\alpha}) }{p(\bb^0)}  \bigg]^{\gamma} 
\nonumber \\
\hspace*{-15mm}
&&\hspace*{-8mm}  \times \Big\{
\int\! \rmd \y ~ p(\y | \bb^0\!, \ldots, \bb^n) \int \!\rmd s~ p(s |y^0\!, \theta^\star)
\prod_{\alpha=1}^n \Big[   \frac{p(s |y^{\alpha}\!, \theta^{\alpha}) }{p(s |y^0\!, \theta^\star) }    \Big]^{\gamma}
\Big\}^N\!.~~ 
\end{eqnarray}
 To proceed we assume that $p(\y | \bb^0, \ldots, \bb^n)$ is Gaussian, via the Central Limit Theorem. 
  Since $\int\!\rmd\z~p(\z)\z=\bnull$, the distribution $p(\y | \bb^0, \ldots, \bb^n) $ is now given by
 \begin{eqnarray}
p(\y | \bb^0, \ldots, \bb^n) = \frac{\rm\rme^{-\half \y\cdot \C^{-1}[\{\bb \}] \y}}{\sqrt{(2 \pi)^{n+1}  \det \C[\{\bb \}]}}.
\label{eq:y_mvn}
\end{eqnarray}
It is determined in full by the $(n\!+\!1)\! \times\! (n\!+\!1)$ covariance matrix $\C[\{\bb \}]$, with entries
\begin{eqnarray}
C_{\alpha \rho}[\{\bb \}] &=& \int\! \rmd\z~ p(\z) \Big(\frac{\bb^{\alpha}\!\cdot\z}{\sqrt p}\Big)\Big( \frac{\bb^{\rho}\!\cdot\z}{\sqrt p} \Big) 
~= \frac{1}{p} \bb^{\alpha}\!\cdot \A \bb^{\rho}.
\label{eq:correlated}
\end{eqnarray}
For each replica pair $(\alpha,\rho)$ we use the integral representation of the Dirac delta function, and rescale the conjugate integration parameter by $p$, substituting
\begin{eqnarray}
\hspace*{-10mm}
1 = \int\!   \rmd C_{\alpha \rho}~\delta \big[ C_{\alpha \rho} \!-\! \frac{1}{p} \bb^{\alpha}\!\cdot \A \bb^{\rho} \big] =
\int \!\frac{\rmd C_{\alpha \rho} \rmd \Chat_{\alpha \rho}}{{2 \pi}/p} \rm\rme^{\rmi p \Chat_{\alpha \rho} (C_{\alpha \rho} - \frac{1}{p} \bb^{\alpha}\!\cdot \A \bb^{\rho} )},
\end{eqnarray}
in order to simplify expression  (\ref{eq:energy2}) to
\begin{eqnarray}
\hspace*{-20mm}
 E_{\gamma}(\bb^\star, \theta^\star) &=& - \frac{\partial }{\partial \gamma}  \lim\limits_{n \to 0} \frac{1}{Nn}
\log \int \!\{\rmd \theta^1\! \ldots \rmd \theta^n\}   \int\! \rmd \C \, \rmd \hat{\C}~\frac{\rm\rme^{\rmi p \, \sum_{\alpha ,\rho=0}^n \Chat_{\alpha \rho} \, C_{\alpha \rho} }}{(2 \pi / p  )^{(n+1)^2}} \nonumber \\
\hspace*{-20mm}
&& \times \Bigg[
\int \! \frac{ \rmd \y~\rm\rme^{-\half \y\cdot \C^{-1} \y}}{\sqrt{(2 \pi)^{n+1}  \det \C}} \int \! \rmd s~ p(s | y^0\!, \theta^\star)
\prod_{\alpha=1}^n \Big[   \frac{p(s | y^{\alpha}\!, \theta^{\alpha})  }{p(s | y^0\!, \theta^\star)  }    \Big]^{\gamma}
\Bigg]^N \nonumber \\
\hspace*{-20mm}
&& \times \int\! \rmd\bb^1\! \ldots d\bb^n~  \rm\rme^{-\eta \gamma \sum_{\alpha=1}^n [ (\bb^{\alpha})^2  -  (\bb^{0})^2 ] -\rmi  \sum_{\alpha, \rho=0}^n  \Chat_{\alpha \rho}  \bb^{\alpha} \cdot\A \bb^{\rho }}.
\label{eq:energy3a}
\end{eqnarray}

\subsection{Conversion into a saddle point problem}

We next transform $\hat{\C}=-\frac{1}{2}\rmi\D$,  define $\tbb \equiv \A^{\half} \bb$ and introduce 
the $np\times np$ matrix $\bXi $ and the $np$-dimensional vector $\bxi$, with entries
\begin{eqnarray}
\Xi_{\alpha\mu;\beta\nu}= 2 \eta \gamma \delta_{\alpha \beta} (\A^{-1})_{\mu \nu} +   \delta_{\mu \nu}D_{\alpha \beta},~~~~~~ \xi_{\mu}^{\alpha}= -D_{0 \alpha}  \tb_{\mu }^{0}
\label{eq:define_xi}
\end{eqnarray}
The Gaussian integral  in (\ref{eq:energy3a}) then becomes
\begin{eqnarray}
&&\hspace*{-15mm}
\int  \Big( \prod_{\alpha=1}^n \rmd \tbb^{\alpha} \rme^{-\eta \gamma \tbb^{\alpha}\cdot \A^{-1} \tbb^{\alpha}}\Big)
\rme^{ -\frac{1}{2}  \sum_{\alpha,\rho=1}^n  D_{\alpha \rho}  \tbb^{\alpha} \cdot \tbb^{\rho }  - \sum_{\rho=1}^n D_{0 \rho}  \tbb^{0} \cdot \tbb^{\rho } }  
\nonumber
\\
&=& 
 \frac{(2\pi)^{\frac{np}{2}}}{\sqrt{\det\bXi}} \rme^{\half \bxi\cdot \bXi^{-1} \bxi}.
\label{eq:gaussian}
\end{eqnarray}
Let $\{a_\mu\}$ and $\{b_\alpha\}$  denote the eigenvalues of $\A$ and $\D$.  
The two terms $\PP$ and $\QQ$ of $\bXi$, with components 
$P_{\alpha\mu,\beta\nu}=2 \eta \gamma \delta_{\alpha \beta} (\A^{-1})_{\mu \nu} $ and $Q_{\alpha\mu,\beta\nu}= \delta_{\mu \nu} D_{\alpha \beta}$,   commute.
The eigenvectors  of $\bXi$ can therefore be written as $\{\hat{\uu}^{\mu\alpha}\}$, with components 
$\hat{u}^{\mu\alpha}_{\nu\rho}=u^\alpha_\rho v^\mu_{\nu}$, and where $\sum_{\rho\leq n} D_{\lambda\rho}u_\rho^\alpha=b_\alpha u_\rho^\lambda$ and 
$\sum_{\nu\leq p}A_{\lambda\nu} v^{\mu}_{\nu}=a_\mu v^{\mu}_{\lambda}$, and where both are normalised according to $\sum_{\rho\leq n}(u^\alpha_\rho)^2=\sum_{\nu\leq p} (v^\mu_{\nu})^2=1$. 
The eigenvalues of $\bXi$ are then $\xi_{\mu\alpha} =2 \eta \gamma/a_\mu + b_\alpha$, and 
\begin{eqnarray}
\hspace*{-15mm}
\det \bXi = \prod_{\mu=1}^p \prod_{\alpha=1}^n \big( \frac{2 \eta 
\gamma}{a_\mu} \!+\! b_\alpha  \big),~~~~~~
(\bXi^{-1})_{\alpha\mu,\alpha^\prime\mu^\prime}= \sum_{\beta=1}^n \sum_{\nu=1}^p 
\frac{u^\beta_\alpha v^\nu_{\mu} u^\beta_{\alpha^\prime} v^\nu_{\mu^\prime}}{2 \eta \gamma/a_\nu + b_\beta}.
\end{eqnarray}
  Hence the integral (\ref{eq:gaussian}) can be written as
\begin{eqnarray}
\hspace*{-18mm}
\frac{(2\pi)^{\frac{np}{2}} \rme^{\half \bxi\cdot \bXi^{-1} \bxi} }{\sqrt{{\rm det}\bXi}}&=&
\rme^{\frac{1}{2}np\log(2\pi)-\frac{1}{2}np \big\langle\!\log ( 2 \eta 
\gamma/a +b )\big\rangle+
\frac{1}{2} np\big\langle (\bxi\cdot\hat{\uu})^2
(2 \eta \gamma/a + b)^{-1}\big\rangle},~~
\label{eq:gaussian_done}
\end{eqnarray}
where the averages are over the eigenvalues and orthonormal eigenvectors of $\bXi$, i.e. 
$\langle f(a,b,\hat{\uu})\rangle\!=\!(np)^{-1}\sum_{\mu=1}^p\sum_{\alpha=1}^n f(a_\mu,b_\alpha,\hat{\uu}^{\mu\alpha})$.
Since $p=\zeta N$ with $\zeta\!>\!0$, the integrals over $\C$, $\hat{\C}$ and the base hazard rates in (\ref{eq:energy3a}) can for $N\to\infty$ be evaluated by steepest descent, 
provided the limits $n\!\to\! 0$ and $N\!\to\! \infty$ commute. 
Expression (\ref{eq:gaussian_done}) then enables us to write the result as
\begin{eqnarray}
\lim\limits_{N \to \infty} E_{\gamma}(\bb^\star, \theta^\star) &=& \frac{\partial }{\partial \gamma}\lim\limits_{n \to 0} \frac{1}{n} \mbox{extr}\, \Psi(\C, \D, \theta^1 \ldots \theta^n),
\label{eq:energy3d}
\end{eqnarray}
with
\begin{eqnarray}
 \hspace*{-20mm}
 \Psi (\C, \D, \theta^1\! \ldots \theta^n) &=&  -\frac{1}{2} \zeta \, \bigg[\sum_{\alpha, \rho=0}^n D_{\alpha \rho} C_{\alpha \rho} - 
 \frac{1}{p}D_{00} (\tbb^0)^2  \bigg]  
 +\half (n\!+\!1\!-\!n\zeta) \log(2 \pi) 
 \nonumber\\
  \hspace*{-20mm}
&&\hspace*{-10mm}
 + \half \log \det \C  - n \eta \zz \gamma S^2  
 + \half n\zz  \Big\langle\!  \log  \Big( \frac{2 \eta \gamma}{a}\! +\! b  \Big)   \Big\rangle
  - \half n\zz  \Big\langle \frac{(\bxi\cdot\hat{\uu})^2}{2 \eta \gamma/a\! +\! b} \Big\rangle
 \nonumber  \\
 \hspace*{-20mm}
&& \hspace*{-10mm}
- \log  \int \!\rmd \y ~\rme^{-\half \y\cdot \C^{-1} \y} \! \int\!\rmd s~ p(s |y^0\!, \theta^\star)
\prod_{\alpha=1}^n \Big[   \frac{p(s | y^{\alpha}\!, \theta^{\alpha})  }{p(s | y^0\!, \theta^\star)  }    \Big]^{\gamma}.
  \label{eq:energy3c}
\end{eqnarray}
where  $S^2=\lim_{p\to\infty}p^{-1}(\bb^0)^2$.
Differentiating $\Psi(\ldots)$ with respect to $D_{00}$ immediately gives $C_{00} = p^{-1}\bb^0\cdot \A \bb^0 \equiv \tS^2 $.

\subsection{Replica symmetric saddle points}

Replica symmetric (RS) saddle points are fully invariant under all permutations of the replica labels $\{1,\ldots,n\}$. For the present model the RS ansatz takes the form 
\begin{eqnarray}
\hspace*{-10mm}
\theta^{\alpha} = \theta,~~~~~~
\begin{array}{l} 
C_{0 \alpha} = c_0
\\[1mm]
D_{0\alpha}=d_0
\end{array},~~~~~~
\begin{array}{ll}
C_{\alpha \rho} & = C \delta_{\alpha \rho} + c (1 - \delta_{\alpha \rho} ),
\\[1mm]
D_{\alpha \rho} & = D \delta_{\alpha \rho} + d (1 - \delta_{\alpha \rho} ). 
\end{array}
\end{eqnarray}
Both $\C$ and $\D$ are positive definite, so $C>c$ and $D>d$.  We may now write
\begin{eqnarray}
&&
\hspace*{-10mm}
 \C = \pmatrix{%
C_{00} & c_0 &\ldots & \ldots & c_0 \cr
c_0 & C & c & \ldots & c \cr
\vdots & c & C & \ldots & c \cr
\vdots & \vdots & \vdots & \ddots & \vdots \cr
c_0 & c & c & \ldots & C \cr
},~~~~~~
\C^{-1} = \pmatrix{%
B_{00} & b_0 &\ldots & \ldots & b_0 \cr
b_0 & B & b & \ldots & b \cr
\vdots & b & B & \ldots & b \cr \nonumber
\vdots & \vdots & \vdots & \ddots & \vdots \cr
b_0 & b & b & \ldots & B \cr
}.
\label{eq:Cmatrix}
\end{eqnarray}
$\C$ has two nondegenerate eigenvalues $\lambda_{\pm}$ with 
 $\lambda_+ \lambda_- =  [C+(n\!-\!1)c ] C_{00} - nc_0^2$,  and a further $n\!-\!1$ fold degenerate eigenvalue $\lambda_0=C-c$. Hence 
 \begin{eqnarray}
\label{eq:logdetC}
\log \det \C &=&      \log C_{00} + n \log (C\!-\!c) + \frac{n\big(c\! -\! c_0^2 / C_{00}  \big)}{C-c} + \mathcal{O}(n^2).  \end{eqnarray}
The entries of $\C^{-1}$ are found to be 
\begin{eqnarray}
&& 
\hspace*{-15mm}  B_{00} =  \frac{C + (n-1)c}{C_{00} [C + (n-1)c] - nc_0^2},~~~~~~
 b_0= - \frac{c_0}{C_{00} [C + (n-1)c] - nc_0^2},
  \\  
&& 
\hspace*{-15mm}  B = b + \frac{1}{C-c},~~~~~~\hspace*{15mm}
 b =  \frac{c_0^2 - c C_{00}}{(C_{00} [C + (n-1)c] - nc_0^2)(C-c)}.
 \label{eq:Cinv}
\end{eqnarray}
Hence
\begin{eqnarray}
\label{eq:quadratic} 
\hspace*{-15mm}
\y\cdot \C^{-1} \y &=&  B_{00} (y^0)^2 + (B\!-\!b) \sum\limits_{\alpha=1}^n (y^{\alpha})^2 +  b \Big( \sum\limits_{\alpha=1}^n y^{\alpha} \Big)^2  + 2 b_0 y^0 \sum\limits_{\alpha=1}^n y^{\alpha}.
\end{eqnarray}
The  matrix $\D$ has 
one eigenvalue $D\!+\!(n\!-\!1)d$ with eigenvector $\vv = (1,\ldots, 1)$, and the $n\!-\!1$ fold degenerate  eigenvalue $D-d$ with eigenspace $ (1,\ldots, 1)^{\perp}$. Hence
\begin{eqnarray}
\hspace*{-15mm}
 \Big\langle\!  \log  \Big( \frac{2 \eta \gamma}{a}\! +\! b  \Big)   \Big\rangle
 &=&\Big\langle \log  \Big( \frac{2 \eta \gamma}{a}\! +\! D\!-\!d  \Big) \Big\rangle
 + \Big\langle\frac{da}{2 \eta \gamma\! +\! (D\!-\!d)a} \Big\rangle +O(n).
 \label{eq:RSterm1}
 \end{eqnarray}
Similarly, using the RS form of $\xi_\mu^\alpha=-d_0(\A^{\frac{1}{2}}\bb^0)_\mu$, we may write 
\begin{eqnarray}
\hspace*{-5mm}
\Big\langle \frac{(\bxi\cdot\hat{\uu})^2}{2 \eta \gamma/a\! +\! b} \Big\rangle 
&=&
d_0^2 ~
\Big\langle 
\frac{a^2 (\bb^0\!\cdot\! \vv)^2}{2 \eta \gamma\! +\! (D\!-\!d)a}\Big\rangle+O(n).
 \label{eq:RSterm2}
\end{eqnarray}
Inserting the above RS expressions into (\ref{eq:energy3c}), and using $C_{00}=\tilde{S}^2$, then gives us
\begin{eqnarray}
\hspace*{-20mm}
\frac{1}{n} \Psi (\ldots)   &=&
     -\frac{1}{2} \zeta (
 2d_0c_0+DC-dc)
 +\half (1\!-\!\zeta) \log(2 \pi)  -  \eta \zz \gamma S^2  
 +O(n)
 \nonumber\\
 \hspace*{-20mm}
&&
 + \half 
 \Big[
 \log (C\!-\!c) + \frac{c\! -\! c_0^2 /\tilde{S}^2}{C\!-\!c} 
 \Big]
   - \half \zz 
  d_0^2 ~
\Big\langle 
\frac{a^2 (\bb^0\!\cdot\! \vv)^2}{2 \eta \gamma\! +\! (D\!-\!d)a}\Big\rangle
 \nonumber
 \\
 \hspace*{-20mm}
 &&
 + \half \zz 
 \Big\langle \log  \Big( \frac{2 \eta \gamma}{a}\! +\! D\!-\!d  \Big) \Big\rangle
 +  \half \zz \Big\langle\frac{da}{2 \eta \gamma\! +\! (D\!-\!d)a} \Big\rangle 
+\frac{1}{2n}\log(\tilde{S}^2  B_{00})
  \nonumber  \\
  \hspace*{-20mm}
&& 
 -\frac{1}{n}
 \log \int\!{\rm D}z {\rm D}y_0\int\!\rmd s~ p(s |y_0/\sqrt{B_{00}}, \theta^\star)
  \nonumber
  \\
  \hspace*{-20mm}
 && ~~\times
 \Big[
  \int \!\rmd y~
  \rme^{-\half  (B-b)  y^2 + y(\rmi  z\sqrt{b}   -b_0 y_0/\sqrt{B_{00}})} 
   \frac{p^\gamma(s | y, \theta)  }{p^\gamma(s | y_0/\sqrt{B_{00}}, \theta^\star)  }  \Big]^n\!\!.~~~
\end{eqnarray}
We note that 
\begin{eqnarray}
&&\hspace*{-5mm} 
B_{00}^{-1}=\tilde{S}^2-nc_0^2/(C\!-\!c)+O(n^2),~~~~~~
B-b= 1/(C\!-\!c),
\\
&&\hspace*{-5mm} 
b_0=-c_0/\tilde{S}^2(C\!-\!c)+O(n),~~~~~~~~~~~~~~
b= \frac{c_0^2 - c\tilde{S}^2}{\tilde{S}^2 (C\!-\!c)^2}+O(n).
\end{eqnarray}
This enable us to write the limit  $\Psi_{\rm RS}(\ldots)=\lim_{n\to 0}n^{-1}\Psi(\ldots)$  in the simpler form
\begin{eqnarray}
\hspace*{-20mm}
\Psi_{\rm RS}(\ldots)&=&
    -\frac{1}{2} \zeta\Bigg\{
 2d_0c_0+DC-dc
+ \log(2 \pi) +2\eta \gamma S^2  
 \nonumber\\[-1mm]
  \hspace*{-20mm}
&&~~
 +
  d_0^2 ~
\Big\langle 
\frac{a^2 (\bb^0\!\cdot\! \vv)^2}{2 \eta \gamma\! +\! (D\!-\!d)a}\Big\rangle
-
 \Big\langle \log  \Big( \frac{2 \eta \gamma}{a}\! +\! D\!-\!d  \Big) \Big\rangle
- \Big\langle\frac{da}{2 \eta \gamma\! +\! (D\!-\!d)a} \Big\rangle 
\Bigg\}
  \nonumber  \\
 \hspace*{-20mm}
&& 
\hspace*{-14mm}
 -\! \int\!{\rm D}z {\rm D}y_0\!\int\!\rmd s~ p(s |\tilde{S}y_0, \theta^\star)
\log\!
  \int \!{\rm D} y~
   \frac{p^\gamma(s | y\sqrt{C\!-\!c}+\!z(c\!-\!c_0^2/\tilde{S}^2)^{\frac{1}{2}} \!+\! y_0c_0/\tilde{S}, \theta)  }{p^\gamma(s | \tilde{S}y_0, \theta^\star)  }.  
   \nonumber
   \\[-1mm]
 \hspace*{-20mm}&&
     \label{eq:Psi_RS_before_uvw}
\end{eqnarray}
Here the brackets denote averages over eigenvectors and eigenvalues of the covariate correlation matrix $\bA$: $\bra f(\bv,a)\ket=\lim_{p\to\infty}p^{-1}\sum_{\mu=1}^p f(\bv_\mu,a_\mu)$, with $\bA\bv_\mu=a_\mu\bv_\mu$ for all $\mu=1\ldots p$.

\subsection{Simplification of the theory}

We extremize  (\ref{eq:Psi_RS_before_uvw}) over $d_0$, which removes an order parameter, and we transform
\begin{eqnarray}
\hspace*{-15mm}
&& u= \sqrt{C\!-\!c}, ~~~~ v =\sqrt{c\! -\! (c_0/\tS)^2},~~~~ w = c_0/\tS, ~~~~f = d,~~~~ g = D\!-\!d,
\label{eq:transform}
\end{eqnarray}
with $u,v,w\in[0,\infty)$ and with the inverse transformations
\begin{equation}
c_0 = \tS w, \hsp \hsp c = v^2 \!+\! w^2, \hsp \hsp C = u^2 \!+\!v^2\! +\! w^2.
\label{eq:inverse_transformation}
\end{equation}
These steps result in 
\begin{eqnarray}
\lim\limits_{N \to \infty} E_{\gamma}(\bb^\star, \theta^\star) &=& \frac{\partial }{\partial \gamma}\mbox{extr}_{u,v,w,f,g,\lambda} \Psi_{\rm RS}(u,v,w,f,g, \theta),
\end{eqnarray}
in which
\begin{eqnarray}
\hspace*{-20mm}
\Psi_{\rm RS}(\ldots)   &=&
    -\frac{1}{2} \zeta
    (g\!+\!f)u^2 -\frac{1}{2} \zeta
 g(v^2\!+\!w^2)
-\zeta\eta \gamma S^2  
 \nonumber\\
  \hspace*{-20mm}
&&
\hspace*{0mm}
+\frac{1}{2}\zeta\Bigg\{
 \tilde{S}^2w^2~  \Big\langle 
\frac{a^2 (\bb^0\!\cdot\! \vv)^2}{2 \eta \gamma\! +\! ga}\Big\rangle^{\!-1}\!
+
 \Big\langle\! \log  \Big( \frac{2 \eta \gamma\!+\!ga}{a} \Big) \Big\rangle
+ f~\Big\langle\frac{a}{2 \eta \gamma\! +\! ga} \Big\rangle 
\Bigg\}
  \nonumber  \\
 \hspace*{-20mm}
&& 
 -\! \int\!{\rm D}z {\rm D}y_0\!\int\!\rmd s~ p(s |\tilde{S}y_0, \theta^0)
\log\!
  \int \!{\rm D} y~
   \frac{p^\gamma(s | uy \!+\! wy_0\!+\!vz, \theta)  }{p^\gamma(s | \tilde{S}y_0, \theta^0)  }.
\end{eqnarray}
We could also extremize over $f$, leading to a simple expression with which to remove $f$ and either $u$ or $g$. The true association parameters $\bb^0$ are seen to enter the asymptotic theory only in quadratic functions of $\bb^0$. In \ref{app:self_averaging} we show that, if the true associations $\{\beta_\mu^0\}$ are drawn randomly and independently from a zero-average distribution, and under mild conditions on the spectrum $\varrho(a)$ of the covariate correlation matrix $\A$, both terms will be self-averaging with respect to the realization of $\bb^0$. 
Consequently,  with $S^2=\lim_{p\to\infty}p^{-1}(\bb^0)^2$ we may then write
\begin{eqnarray}
\tilde{S}^2= S^2\bra a\ket,~~~~~~
\Big\bra \frac{a^2(\bb^0\!\cdot\vv)^2}{2\eta\gamma\!+\!ga}\Big\ket= 
\bra \frac{S^2 a^2}{2\eta\gamma\!+\!ga}\ket,
\label{eq:Stilde_to_S}
\end{eqnarray}
(where we used the fact that the eigenvectors $\vv$ of $\A$ were  normalized). Hence
\begin{eqnarray}
\hspace*{-20mm}
\lim\limits_{N \to \infty} E_{\gamma}(\bb^0, \theta^\star) &=&
\int\! {\rm D}y_0\!\int\!\rmd s~ p(s |S\bra a\ket^{\frac{1}{2}}y_0, \theta^\star)
\log p(s | S\bra a\ket^{\frac{1}{2}}y_0, \theta^\star)  
-\zeta\eta  S^2  
 \nonumber\\
  \hspace*{-20mm}
&&
 \hspace*{-20mm}
+~\eta\zeta\Bigg\{
w^2\bra a\ket  
 \Big\langle \frac{a^2}{2 \eta \gamma\! +\! ga}\Big\rangle^{\!\!-2}
 \Big\langle  \frac{a^2}{(2 \eta \gamma\! +\! ga)^2}\Big\rangle
+
 \Big\langle\! \frac{1}{2 \eta \gamma\!+\!ga} \Big\rangle- f\Big\langle\frac{a}{(2 \eta \gamma\! +\! ga)^2} \Big\rangle 
\Bigg\}
  \nonumber  \\
 \hspace*{-20mm}
&& 
\hspace*{-31mm}
 -\! \int\!{\rm D}z {\rm D}y_0\!\int\!\!\rmd s~ p(s |S\bra a\ket^{\frac{1}{2}}y_0, \theta^\star)
\frac{  \int \!{\rm D} y~
p^\gamma(s | uy \!+\! wy_0\!+\!vz, \theta)\log p(s | uy \!+\! wy_0\!+\!vz, \theta) }
{  \int \!{\rm D} y~
p^\gamma(s| uy \!+\! wy_0\!+\!vz, \theta) }.
\nonumber
\\[-1mm]
\hspace*{-20mm}
&&
\label{eq:Evalue_before_scaling}
\end{eqnarray}
The order parameters $(u,v,w,f,g,\theta\}$ are computed by extremization of the following function, from which we removed any constant terms:
\begin{eqnarray}
\hspace*{-20mm}
\Psi_{\rm RS}(\ldots)   &=&
    -\frac{1}{2} \zeta
    (g\!+\!f)u^2 -\frac{1}{2} \zeta
 g(v^2\!+\!w^2)
 \nonumber\\
  \hspace*{-20mm}
&&
\hspace*{0mm}
+\frac{1}{2}\zeta\Bigg\{
 w^2\bra a\ket  \Big\langle 
\frac{a^2}{2 \eta \gamma\! +\! ga}\Big\rangle^{\!-1}\!
+
 \Big\langle\! \log (2 \eta \gamma\!+\!ga) \Big\rangle
+ f~\Big\langle\frac{a}{2 \eta \gamma\! +\! ga} \Big\rangle 
\Bigg\}
  \nonumber  \\
 \hspace*{-20mm}
&& \hspace*{-1mm}
 -\! \int\!{\rm D}z {\rm D}y_0\!\int\!\rmd s~ p(s |S\bra a\ket^{\frac{1}{2}}y_0, \theta^\star)
\log\!
  \int \!{\rm D} y~
 p^\gamma(s | uy \!+\! wy_0\!+\!vz, \theta).
 \label{eq:RS_Psi_before_scaling}
\end{eqnarray}

\section{Self-averaging with respect to true associations}
\label{app:self_averaging}

The results of this Appendix were derived in \cite{SheikhCoolen2019}, but will be briefly recapitulated, for completeness and because they are also needed in deriving (\ref{eq:average_beta},\ref{eq:covariance_beta}). 
We investigate random variables of the form $\mathcal{R}=p^{-1}\bb^0\cdot {\bf P}\bb^0$, where the   true association vectors $\bb^0=\{\beta_\mu^0\}$ are drawn randomly from some distribution $p(\bb^0)$, and ${\bf P}$ is a fixed symmetric positive definite $p\times p$ matrix, which is independent of $\bb^0$. We wish to know the conditions under which $\mathcal{R}$ will be self-averaging, i.e. $\lim_{p\to\infty}\bra \mathcal{R}\ket>0$ exists, and $\lim_{p\to\infty}[\bra \mathcal{R}^2\ket-\bra \mathcal{R}\ket^2]=0$ (brackets denote averaging over $p(\bb^0)$). We assume:
\begin{enumerate}
\item The $\{\beta_\mu^0\}$ are independent and identically distributed, i.e. $p(\bb^0) = \prod_{\mu=1}^p p(\beta_{\mu}^0)$. 
\item $p(\beta_\mu^0)$ is symmetric in $\beta_\mu^0$,  with finite second and fourth order moments. 
\item $\lim_{p\to\infty} p^{-1}\sum_{\mu=1}^p P_{\mu\mu}\in \R$.
\item $\lim_{p\to\infty}p^{-2}\sum_{\mu\nu=1}^p P^2_{\mu\nu}=0$.
  \end{enumerate}
Given that $S^2=\lim_{p\to\infty}p^{-1}(\bb^0)^2\!$, we must identify $\bra  (\beta_\mu^0)^2\ket=S^2$. 
It was shown in \cite{SheikhCoolen2019}  that the above conditions are sufficient for $\mathcal{R} $ to be self-averaging. This enabled us to infer that the following identities hold (for $g>0$), as soon as average and width of the eigenvalue distribution $\varrho(a)$ of  $\A$ remain finite in the limit $p\to\infty$:
\begin{eqnarray}
\lim_{p\to \infty}\frac{1}{p}\bb^0\cdot\A\bb^0&=& S^2\int\!\rmd a~\varrho(a) a,
\\
\lim_{p\to\infty} \frac{1}{p} \sum_{\rho=1}^p \frac{a_\rho^2 (\bb^0\!\cdot \vv^\rho)^2 }{2\eta\gamma+ga_\rho}&=& 
\int\!\rmd a~\varrho(a)  \frac{S^2a^2}{2\eta\gamma\!+\!ga}.
\label{eq:selfav_tricky}
\end{eqnarray}

\section{Further evaluation of the RS order parameter equations}
\label{app:further}

One can take further steps in evaluating the RS order parameter equations (\ref{eq:RS_eqns_uncoupled},\ref{eq:RS_eqns_coupled}), without specifying any specific GLM model, exploiting the structural features of the theory only. For instance, the order parameter equations for $(\tilde{f},\tilde{g})$ are not model dependent, and give:
\begin{eqnarray}
&&\hspace*{-10mm}
\Big\bra  
\frac{a}{2\eta\!+\!\tilde{g}a}\Big\ket
= \tilde{u}^2,
\label{eq:ddf=0}
\\[0.5mm]
&&\hspace*{-10mm}
w^2\Bigg[
\bra a\ket
 \Big\bra \frac{a^2}{
2\eta\!+\!\tilde{g}a}\Big\ket^{-2}
 \Big\bra \frac{a^3}{
(2\eta\!+\!\tilde{g}a)^2}\Big\ket
-1\Bigg]
-
\tilde{f}\Big\bra  
\frac{a^2}{(2\eta\!+\!\tilde{g}a)^2}\Big\ket
=v^2.
\label{eq:ddg=0}
\end{eqnarray}
For uncorrelated and normalized data, where $\varrho(a)=\delta(a\!-\!1)$, this reduces to
\begin{eqnarray}
2\eta\!+\!\tilde{g}
= \tilde{u}^{-2},
~~~~~~
\tilde{f}
=-v^2/\tilde{u}^4.
\end{eqnarray}
Alternatively, for $\eta\to 0$ (ML regression)  equations (\ref{eq:ddf=0},\ref{eq:ddg=0}) become
\begin{eqnarray}
\tilde{g}=1/ \tilde{u}^2, ~~~~~~
\tilde{f}=-
 v^2/\tilde{u}^4.
\end{eqnarray}
In addition to the two partial derivatives of $\Xi_A$ with respect to $\tilde{f}$ and $\tilde{g}$, we also require derivatives with respect to $(\tilde{u},v,w)$. These are
\begin{eqnarray}
&&\hspace*{-15mm}
\frac{\partial \Xi_A}{\partial\tilde{u}}=
- \zeta
\tilde{f} \tilde{u},~~~~~~
\frac{\partial \Xi_A}{\partial v}=
- \zeta \tilde{g}v,~~~~~~
\frac{\partial \Xi_A}{\partial w}=
 \zeta w\Big[
 \bra a\ket
 \Big\bra \frac{a^2}{
2\eta\!+\!\tilde{g}a}\Big\ket^{\!-1}\!\!
\!- \tilde{g}
\Big].~
\end{eqnarray}
We also need  the partial derivatives of  (\ref{eq:Psi_B_large_gamma_compact}). We note that all partial derivatives of the argument of (\ref{eq:Psi_B_large_gamma_compact}) that are channelled indirectly via the variable $\xi$ vanish at the point $\xi=\xi(wy_0+vz,\tilde{u},s,\theta)$, by definition. Hence
\begin{eqnarray}
\frac{\partial \Xi_B}{\partial \tilde{u}}&=& \frac{1}{\tilde{u}^3}
\Big\bra\!\Big\bra\! \Big\bra 
 [\xi(wy_0\!+\!vz,\tilde{u},s,\theta)\!-\!wy_0\!-\!vz]^2\Big\ket_{\!s}\Big\ket\!\Big\ket,
 \label{eq:ddu}
\\[1mm]
\frac{\partial \Xi_B}{\partial v}&=& \frac{1}{\tilde{u}^2}\Big\{
\Big\bra\!\Big\bra z\Big\bra 
\xi(wy_0\!+\!vz,\tilde{u},s,\theta)\Big\ket_{\!s}\Big\ket\!\Big\ket-v\Big\},
 \label{eq:ddv}
\\[1mm]
\frac{\partial \Xi_B}{\partial w}&=& \frac{1}{\tilde{u}^2}\Big\{
\Big\bra\!\Big\bra y_0
\Big\bra  \xi(wy_0\!+\!vz,\tilde{u},s,\theta)\Big\ket_{\!s}\Big\ket\!\Big\ket
-w\Big\},
 \label{eq:ddw}
\\[1mm]
\frac{\partial \Xi_B}{\partial \theta}&=& 
\Big\bra\!\Big\bra \!\Big\bra \frac{\partial \log p(s|\xi,\theta)}{\partial\theta}\Big|_{\xi=\xi(wy_0+vz,\tilde{u},s,\theta)}
\Big\ket_{\!s}\Big\ket\!\Big\ket.
 \label{eq:ddtheta}
\end{eqnarray}
The remaining four order parameter equations, in addition to the previously derived pair (\ref{eq:ddf=0},\ref{eq:ddg=0}),  then become
\begin{eqnarray}
\hspace*{-15mm}
\Big\bra\!\Big\bra \!\Big\bra 
 [\xi(wy_0\!+\!vz,\tilde{u},s,\theta)\!-\!wy_0\!-\!vz]^2\Big\ket_{\!s}\Big\ket\!\Big\ket
 &=& - \zeta
\tilde{f} \tilde{u}^4,
 \label{eq:ddu=0}
\\[1mm]
\hspace*{-15mm}
\Big\bra\!\Big\bra z\Big\bra 
\xi(wy_0\!+\!vz,\tilde{u},s,\theta)\Big\ket_{\!s}\Big\ket\!\Big\ket
&=& v\Big(1- \zeta \tilde{g}\tilde{u}^2\Big),
 \label{eq:ddv=0}
\\[1mm]
\hspace*{-15mm}
\Big\bra\!\Big\bra y_0
\Big\bra  \xi(wy_0\!+\!vz,\tilde{u},s,\theta)\Big\ket_{\!s}\Big\ket\!\Big\ket
&=& w+ \zeta w \tilde{u}^2\Big[
 \bra a\ket
 \Big\bra \frac{a^2}{
2\eta\!+\!\tilde{g}a}\Big\ket^{\!\!-1}\!\!
- \tilde{g}
\Big],
 \label{eq:ddw=0}
\\[1mm]
\hspace*{-15mm}
\Big\bra\!\Big\bra \!\Big\bra \frac{\partial \log p(s|\xi,\theta)}{\partial\theta}\Big|_{\xi=\xi(wy_0+vz,\tilde{u},s,\theta)}
\Big\ket_{\!s}\Big\ket\!\Big\ket
&=& 0.
 \label{eq:ddtheta=0}
\end{eqnarray}
Equations (\ref{eq:ddv=0},\ref{eq:ddw=0}) can be simplified further upon integrating by parts over $z$ and $y_0$. We need to take care that $y_0$ appears also in the distribution $p(s|S\bra a\ket^{\frac{1}{2}}y_0,\theta^\star)$ used to define the measure  $\bra \ldots\ket_s$.
We first turn to the average in (\ref{eq:ddv=0}):
\begin{eqnarray}
\Big\bra\!\Big\bra z\Big\bra 
\xi(wy_0\!+\!vz,\tilde{u},s,\theta)\Big\ket_{\!s}\ket\!\Big\ket &=& \Big\bra\!\Big\bra \!\Big\bra 
\frac{\partial}{\partial z}\xi(wy_0\!+\!vz,\tilde{u},s,\theta)\Big\ket_{s}\Big\ket\!\Big\ket 
\nonumber
\\
&=& v\Big\bra\!\Big\bra \!\Big\bra 
(\partial_1\xi)(wy_0\!+\!vz,\tilde{u},s,\theta)\Big\ket_{\!s}\Big\ket\!\Big\ket.
\label{eq:partial_z}
\end{eqnarray}
Next we work on the average in (\ref{eq:ddw=0}):
\begin{eqnarray}
\hspace*{-20mm}
\Big\bra\!\Big\bra y_0
\Big\bra  \xi(wy_0\!+\!vz,\tilde{u},s,\theta)\Big\ket_{\!s}\Big\ket\!\Big\ket&=&
\Big\bra\!\Big\bra \frac{\partial}{\partial y_0}\int\!\rmd s~p(s|S\bra a\ket^{\frac{1}{2}}y_0,\theta^\star)
 \xi(wy_0\!+\!vz,\tilde{u},s,\theta)\Big\ket\!\Big\ket
 \nonumber
 \\
  \hspace*{-20mm}
 &=& 
 \Big\bra\!\Big\bra\int\!\rmd s\Big\{ 
  \xi(wy_0\!+\!vz,\tilde{u},s,\theta)
 \frac{\partial}{\partial y_0}p(s|S\bra a\ket^{\frac{1}{2}}y_0,\theta^\star)
 \nonumber
 \\
 \hspace*{-20mm}
 &&
\hspace*{6mm}+ w p(s|S\bra a\ket^{\frac{1}{2}}y_0,\theta^\star)(\partial_1 \xi)(wy_0\!+\!vz,\tilde{u},s,\theta)\Big\}
 \Big\ket\!\Big\ket
 \nonumber
 \\
 \hspace*{-20mm}
 &&\hspace*{-20mm} =
 \Big\bra\!\Big\bra\!\Big\bra  \xi(wy_0\!+\!vz,\tilde{u},s,\theta) \frac{\partial \log p(s|S\bra a\ket^{\frac{1}{2}}y_0,\theta^\star)}{\partial y_0}\Big\ket_{\!s}\Big\ket\!\Big\ket
 \nonumber
 \\
 \hspace*{-20mm}&&\hspace*{6mm} 
 +w  \Big\bra\!\Big\bra \!\Big\bra (\partial_1 \xi)(wy_0\!+\!vz,\tilde{u},s,\theta)\Big\ket_{\!s} \Big\ket\!\Big\ket.
\end{eqnarray}
With the above results, and upon discarding the trivial solution $v=0$ and using (\ref{eq:ddv=0}) to simplify (\ref{eq:ddw=0}),  we can rewrite our closed MAP order parameter equation set as:
\begin{eqnarray}
\hspace*{-20mm}
\Big\bra  
\frac{a}{2\eta\!+\!\tilde{g}a}\Big\ket
&=& \tilde{u}^2,
\label{eq:ddf2_app=0}
\\
\hspace*{-20mm}
w^2\Big[
\bra a\ket
 \Big\bra \frac{a^2}{
2\eta\!+\!\tilde{g}a}\Big\ket^{\!\!-2}
\!
 \Big\bra \frac{a^3}{
(2\eta\!+\!\tilde{g}a)^2}\Big\ket
\!-\!1\Big]
\!-\!
\tilde{f}\Big\bra  
\frac{a^2}{(2\eta\!+\!\tilde{g}a)^2}\Big\ket
&=&v^2,
\label{eq:ddg2_app=0}
\\
\hspace*{-20mm}
\Big\bra\!\Big\bra\!\Big\bra 
 [\xi(wy_0\!+\!vz,\tilde{u},s,\theta)\!-\!wy_0\!-\!vz]^2\Big\ket_{\!s}\Big\ket\!\Big\ket
 &=& - \zeta
\tilde{f} \tilde{u}^4,
 \label{eq:ddu2_app=0}
\\[2mm]
\hspace*{-20mm}
 \Big\bra\!\Big\bra \!\Big\bra 
(\partial_1\xi)(wy_0\!+\!vz,\tilde{u},s,\theta)\Big\ket_{\!s}\Big\ket\!\Big\ket 
&=& 1- \zeta \tilde{g}\tilde{u}^2,
 \label{eq:ddv2_app=0}
\\[1mm]
\hspace*{-20mm}
 \Big\bra\!\Big\bra\!\Big\bra  \xi(wy_0\!+\!vz,\tilde{u},s,\theta) \frac{\partial \log p(s|S\bra a\ket^{\frac{1}{2}}y_0,\theta^\star)}{\partial y_0}\Big\ket_{\!s}\Big\ket\!\Big\ket
&=& \zeta w \tilde{u}^2
 \bra a\ket
 \Big\bra \frac{a^2}{
2\eta\!+\!\tilde{g}a}\Big\ket^{\!-1}\!\!,
 \label{eq:ddw2_app=0}
\\[1mm]
\hspace*{-20mm}
\Big\bra\!\Big\bra\! \Big\bra \frac{\partial \log p(s|\xi,\theta)}{\partial\theta}\Big|_{\xi=\xi(wy_0+vz,\tilde{u},s,\theta)}
\Big\ket_{\!s}\Big\ket\!\Big\ket&=& 0.
 \label{eq:ddtheta2_app=0}
\end{eqnarray}

\section{Statistics of inferred association parameters}
\label{app:beta_stats}

 \subsection{Asymptotic form}
 
 In this Appendix we give the details of the evaluation of  the distribution (\ref{eq:betas_starting_point}), in the limit $p,N\to\infty$ with $\zeta=p/N$, first for uncorrelated and then for correlated covariates. 
 We write $\bbeta^0=\bbeta^\star$, assume flat priors for the non-association parameters $\theta$, and use the definition (\ref{eq:switch_to_y}).  Due to the limit $n\to 0$, we may also insert into the above expression without consequence quantities such as $p^{-\gamma n}(\bbeta^0)$ and $p^{-\gamma n}(s|y^0\!,\theta^\star)$, in order to bring it closer to the integrals found in 
  \ref{app:SheikhCoolen}. The result is
 \begin{eqnarray}
 \hspace*{-15mm} 
\Prob(\beta,\beta^\star) &=& 
  \lim_{\gamma\to\infty} \lim_{n\to 0}
 \frac{1}{p}\sum_{\mu=1}^p \delta(\beta^\star\!-\beta_\mu^0)
 \nonumber
 \\[-1mm]
 \hspace*{-15mm}
 &&\times
 \int\{\rmd\theta^1\ldots\rmd\theta^n\}\int\!\rmd\bbeta^1\ldots\rmd\bbeta^n~ \delta(\beta\!-\!\beta^1_\mu)
\prod_{\alpha=1}^n\Big[\frac{p(\bbeta^\alpha)}{p(\bbeta^0)}\Big]^\gamma
 \nonumber
 \\
  \hspace*{-15mm} 
 &&\hspace*{-5mm}\times 
 \Big\{
 \int\!\rmd\by~p(\by|\bbeta^0\!,\ldots,\bbeta^n)
\int\!\rmd s~p(s|y^0\!,\theta^\star) 
  \prod_{\alpha=1}^n 
\Big[\frac{p(s|y^\alpha,\theta^\alpha)}{p(s|y^0,\theta^\star)}\Big]^\gamma
 \Big\}^N.
 \end{eqnarray}
 We can now repeat the manipulations of  \ref{app:SheikhCoolen}, with slight modifications. It will in fact be useful to work with the more general family of factorizing priors $p(\bbeta)\propto \prod_{\mu\leq p}p(\beta_\mu)$, of which the Gaussian one  is a special case, but which also allows us to inspect e.g. $L1$ priors. Our expression for $\Prob(\beta,\beta^\star)$  then becomes
  \begin{eqnarray}
  \hspace*{-20mm}
\Prob(\beta,\beta^\star) &=& 
  \lim_{\gamma\to\infty} \lim_{n\to 0}\frac{1}{p}\sum_{\mu=1}^p \delta(\beta^\star\!-\beta_\mu^0)
 \int\{\rmd\theta^1\ldots\rmd\theta^n\}
 \int\!\rmd\bC\rmd\hat{\bC}~\frac{\rme^{\rmi p\sum_{\alpha,\rho=0}^n \hat{C}_{\alpha\rho}C_{\alpha\rho}}}{(2\pi/p)^{(n+1)^2}}
 \nonumber
 \\
  \hspace*{-20mm}
 &&\times 
 \Bigg[
 \int\!\frac{\rmd\by~\rme^{-\frac{1}{2}\by\cdot\bC^{-1}\by}}{\sqrt{(2\pi)^{n+1}{\rm det}\bC}}
\int\!\rmd s~p(s|y^0\!,\theta^\star) 
  \prod_{\alpha=1}^n 
\Big[\frac{p(s|y^\alpha,\theta^\alpha)}{p(s|y^0,\theta^\star)}\Big]^\gamma
 \Bigg]^N
 \nonumber
 \\
  \hspace*{-20mm}
 &&\hspace*{-5mm} \times  \int\!\rmd\bbeta^1\ldots\rmd\bbeta^n~ \delta(\beta\!-\!\beta^1_\mu)\rme^{-\rmi\sum_{\alpha,\rho=0}^n \hat{C}_{\alpha\rho}\bbeta^\alpha\cdot\bA\bbeta^\rho}
 \prod_{\alpha=1}^n\prod_{\nu=1}^p \Big[\frac{p(\beta_\nu^\alpha)}{p(\beta_\nu^0)}\Big]^\gamma.
 \label{eq:beta_beta_intermediate}
 \end{eqnarray}
We proceed to the limit $p,N\to\infty$ with fixed $\zeta=p/N$.  In view of our previous calculations we define the following quantity:
 \begin{eqnarray}
 \hspace*{-20mm}
 \Psi(\bC,\hat{\bC},\theta^1,\ldots,\theta^n)&=& -\rmi\zeta \sum_{\alpha,\rho=0}^n\hat{C}_{\alpha\rho}C_{\alpha\rho}
  + \frac{1}{2}(n\!+\!1)\log (2\pi)
 + \frac{1}{2}\log {\rm det}\bC
\nonumber
 \\
  \hspace*{-20mm}
 && \hspace*{-18mm}
- \log  \int\!\rmd\by~\rme^{-\frac{1}{2}\by\cdot\bC^{-1}\by}
\int\!\rmd s~p(s|y^0\!,\theta^\star) 
  \prod_{\alpha=1}^n 
\Big[\frac{p(s|y^\alpha,\theta^\alpha)}{p(s|y^0,\theta^\star)}\Big]^\gamma
\nonumber
\\
 \hspace*{-20mm}
&&  \hspace*{-18mm} -\frac{1}{N}\log\int\!\rmd\bbeta^1\ldots\rmd\bbeta^n~ \rme^{-\rmi\sum_{\alpha,\rho=0}^n \hat{C}_{\alpha\rho}\bbeta^\alpha\cdot\bA\bbeta^\rho}
 \prod_{\alpha=1}^n\prod_{\nu=1}^p \Big[\frac{p(\beta_\nu^\alpha)}{p(\beta_\nu^0)}\Big]^\gamma.
 \end{eqnarray}
This enables us to write  (\ref{eq:beta_beta_intermediate}) as
   \begin{eqnarray}
   \hspace*{-23mm}
\Prob(\beta,\beta^\star) &=& 
  \lim_{\gamma\to\infty} \lim_{n\to 0}
 \int\{\rmd\theta^1\ldots\rmd\theta^n\}
 \int\!\rmd\bC\rmd\hat{\bC}~
 \rme^{-N \Psi(\bC,\hat{\bC},\theta^1,\ldots,\theta^n)}
 \label{eq:beta_relation_nearly}
  \\
   \hspace*{-23mm}
 &&
 \hspace*{-17mm} \times \frac{1}{p}\sum_{\mu=1}^p \delta(\beta^\star\!-\beta_\mu^0)\frac{ \int\!\rmd\bbeta^1\ldots\rmd\bbeta^n~ \delta(\beta\!-\!\beta^1_\mu)\rme^{-\rmi\sum_{\alpha,\rho=0}^n \hat{C}_{\alpha\rho}\bbeta^\alpha\cdot\bA\bbeta^\rho}
\! \prod_{\alpha=1}^n\prod_{\nu=1}^p p^\gamma(\beta_\nu^\alpha)}
 { \int\!\rmd\bbeta^1\ldots\rmd\bbeta^n~ \rme^{-\rmi\sum_{\alpha,\rho=0}^n \hat{C}_{\alpha\rho}\bbeta^\alpha\cdot\bA\bbeta^\rho}
 \!\prod_{\alpha=1}^n\prod_{\nu=1}^p p^\gamma (\beta_\nu^\alpha)}.
 \nonumber
 \\[-2mm]
 \hspace*{-20mm}&&\nonumber
\end{eqnarray}
Integrating both sides over $(\beta,\beta^\star)$ shows that the first line of (\ref{eq:beta_relation_nearly}) on its own would equal one. Hence for $N\to\infty$ we will be left simply with the limit $p\to\infty$ of the second line, which is an $\order(1)$ object, evaluated at the the saddle point of $\Psi(\bC,\hat{\bC},\theta^1,\ldots,\theta^n)$. For Gaussian priors, the saddle point is the one computed in \ref{app:SheikhCoolen}. Hence, upon transforming 
as before $\hat{\bC}=-\frac{1}{2}\rmi \bD$, and choosing the saddle point values, 
   \begin{eqnarray}
   \hspace*{-12mm}
\lim_{N\to\infty}\Prob(\beta,\beta^\star) &=& \lim_{p\to\infty}  \lim_{\gamma\to\infty} \lim_{n\to 0}
 \frac{1}{p}\sum_{\mu=1}^p \delta(\beta^\star\!-\beta_\mu^0)
 \label{eq:link_arbitrary_A}
\\
   \hspace*{-12mm}
&&\hspace*{-15mm}\times \frac{ \int\!\rmd\bbeta^1\ldots\rmd\bbeta^n~ \delta(\beta\!-\!\beta^1_\mu)\rme^{-\frac{1}{2}\sum_{\alpha,\rho=0}^n D_{\alpha\rho}\bbeta^\alpha\cdot\bA\bbeta^\rho}
 \prod_{\alpha=1}^n\prod_{\nu=1}^p p^\gamma(\beta_\nu^\alpha)}
 { \int\!\rmd\bbeta^1\ldots\rmd\bbeta^n~ \rme^{-\frac{1}{2}\sum_{\alpha,\rho=0}^n D_{\alpha\rho}\bbeta^\alpha\cdot\bA\bbeta^\rho}
 \prod_{\alpha=1}^n\prod_{\nu=1}^p p^\gamma (\beta_\nu^\alpha)}.
 \nonumber
\end{eqnarray}
This expression depends on the choice of $p(s|\xi,\theta)$ only indirectly, via the values of the order parameters $\{D_{\alpha\rho}\}$. We will now work out  (\ref{eq:link_arbitrary_A}) first for uncorrelated and normalized covariates, followed by evaluation for arbitrary covariate correlations. 

\subsection{Uncorrelated and normalized covariates}

 This is the simplest case, where $A_{\mu\nu}=\delta_{\mu\nu}$. The integrations in the above formula now factorize over all components of $\bbeta$, giving
   \begin{eqnarray}
   \hspace*{-22mm}
\lim_{N\to\infty}\Prob(\beta,\beta^\star) &=& \Big(\lim_{p\to\infty} \frac{1}{p}\sum_{\mu=1}^p \delta(\beta^\star\!-\beta_\mu^0)\Big)\times
\\
   \hspace*{-22mm}
&&\hspace*{-23mm}\lim_{\gamma\to\infty}  \lim_{n\to 0}
\frac{ \int\!\rmd\beta^1\ldots\rmd\beta^n~ \delta(\beta\!-\!\beta^1)\rme^{-\frac{1}{2}\sum_{\alpha,\rho=1}^n D_{\alpha\rho}\beta^\alpha \beta^\rho
 -\beta^\star\sum_{\alpha=1}^n D_{\alpha 0}\beta^\alpha}
 \prod_{\alpha=1}^n p^\gamma(\beta^\alpha)}
 { \int\!\rmd\beta^1\ldots\rmd\beta^n~ \rme^{-\frac{1}{2} \sum_{\alpha,\rho=1}^n D_{\alpha\rho}\beta^\alpha \beta^\rho
  -\beta^\star\sum_{\alpha=1}^n D_{\alpha 0}\beta^\alpha}
 \prod_{\alpha=1}^n p^\gamma (\beta^\alpha)}.
 \nonumber
\end{eqnarray}
Hence
   \begin{eqnarray}&&
   \hspace*{-22mm}
\lim_{N\to\infty}\Prob(\beta|\beta^\star)~ =
\\[-1mm]
   \hspace*{-22mm}
   &&   \hspace*{-18mm}
\lim_{\gamma\to\infty}  \lim_{n\to 0}
\frac{ \int\!\rmd\beta^1\ldots\rmd\beta^n~ \delta(\beta\!-\!\beta^1)\rme^{-\frac{1}{2}\sum_{\alpha,\rho=1}^n D_{\alpha\rho}\beta^\alpha \beta^\rho
 -\beta^\star\sum_{\alpha=1}^n D_{\alpha 0}\beta^\alpha}
 \prod_{\alpha=1}^n p^\gamma(\beta^\alpha)}
 { \int\!\rmd\beta^1\ldots\rmd\beta^n~ \rme^{-\frac{1}{2} \sum_{\alpha,\rho=1}^n D_{\alpha\rho}\beta^\alpha \beta^\rho
  -\beta^\star\sum_{\alpha=1}^n D_{\alpha 0}\beta^\alpha}
 \prod_{\alpha=1}^n p^\gamma (\beta^\alpha)}.
 \nonumber
\end{eqnarray}
 We next use the replica symmetric form of the matrix $\bD$, i.e. $D_{\alpha\rho}=D\delta_{\alpha\rho}+d(1\!-\!\delta_{\alpha\rho})$ and $D_{\alpha 0}=d_0$ and  for $\alpha,\rho=1\ldots n$, and carry out a Gaussian linearization:
   \begin{eqnarray}
   \hspace*{-20mm} \lim_{N\to\infty}\Prob(\beta|\beta^\star) &=&
\nonumber
\\
 \hspace*{-20mm} &&\hspace*{-31mm}
 \lim_{\gamma\to\infty}  \lim_{n\to 0}
\frac{ \int\!{\rm D}z
\Big[ \int\!\rmd\beta^\prime~ \rme^{
 \rmi z\sqrt{d}\beta^\prime
-\frac{1}{2}(D-d) (\beta^\prime)^2
  -d_0\beta^\star \beta^\prime}
p^\gamma (\beta^\prime)\Big]^{n-1}
\!
\rme^{
\rmi z\sqrt{d} \beta
-\frac{1}{2}(D-d)\beta^2
 -d_0\beta^\star\beta}
 p^\gamma(\beta)}
 {\int\!{\rm D}z\Big[ \int\!\rmd\beta^\prime~ \rme^{
 \rmi z\sqrt{d}\beta^\prime
-\frac{1}{2}(D-d) (\beta^\prime)^2
  -d_0\beta^\star \beta^\prime}
p^\gamma (\beta^\prime)\Big]^n}
 \nonumber\hspace*{-1mm}
 \\
  \hspace*{-20mm} &=&
 \lim_{\gamma\to\infty} 
  \int\!{\rm D}z\left[ \frac{
\rme^{
-\frac{1}{2}(D-d)\beta^2
+\beta(\rmi z\sqrt{d}-d_0\beta^\star)}
 p^\gamma(\beta)}
 { \int\!\rmd\beta^\prime~ \rme^{
-\frac{1}{2}(D-d) (\beta^\prime)^2
+\beta^\prime( \rmi z\sqrt{d}
  -d_0\beta^\star )}
p^\gamma (\beta^\prime)}
\right].
\end{eqnarray}
In terms of the transformed order parameters $f=d$ and $g=D-d$ this becomes
  \begin{eqnarray}
  \hspace*{-10mm}
\lim_{N\to\infty}\Prob(\beta|\beta^\star) &=&
 \lim_{\gamma\to\infty} 
  \int\!{\rm D}z\left[ \frac{
\rme^{
-\frac{1}{2}g\beta^2
+\beta(\rmi z\sqrt{f}-d_0\beta^\star)}
 p^\gamma(\beta)}
 { \int\!\rmd\beta^\prime~ \rme^{
-\frac{1}{2}g(\beta^\prime)^2
+\beta^\prime( \rmi z\sqrt{f}
  -d_0\beta^\star )}
p^\gamma (\beta^\prime)}
\right].
\end{eqnarray}
The order parameter $d_0$, which we could remove from the general theory, here needs to be computed after all.
For the $L2$ (i.e. Gaussian) prior $p(\beta)\propto\exp(-\eta \beta^2)$ we can find $d_0$ via differentiation of (\ref{eq:Psi_RS_before_uvw}) and subsequently use (\ref{eq:selfav_tricky}), giving
\begin{eqnarray}
d_0&=& -c_0\Big\langle 
\frac{a^2 (\bb^0\!\cdot\! \vv)^2}{2 \eta \gamma\! +\! (D\!-\!d)a}\Big\rangle^{-1}
= -\frac{c_0}{S^2}
\Big\bra \frac{a^2}{2\eta\gamma\!+\!ga}\Big\ket^{-1}.
\label{eq:found_d0}
   \end{eqnarray}
For uncorrelated and normalized covariates we have $\varrho(a)=\delta(a\!-\!1)$, so $c_0=Sw$ and
   \begin{eqnarray}
d_0&=&  -\frac{c_0}{S^2}
(2\eta\gamma\!+\!g)
= -\frac{\gamma w}{S}
(2\eta\!+\!\tilde{g}).
   \end{eqnarray}
   We thus find  with $f=\tilde{f}\gamma^2$, 
    \begin{eqnarray}
    \hspace*{-10mm}
\lim_{N\to\infty}\Prob(\beta|\beta^\star) &=&
 \lim_{\gamma\to\infty} 
  \int\!{\rm D}z\left[ \frac{
\rme^{\gamma\big[
-\frac{1}{2}(2\eta+\tilde{g})\beta^2
+\beta(\rmi z\sqrt{\tilde{f}}+w
(2\eta+\tilde{g})\beta^\star/S)\big]}}
 { \int\!\rmd\beta^\prime~ 
 \rme^{\gamma\big[
-\frac{1}{2}(2\eta+\tilde{g})(\beta^\prime)^2
+\beta^\prime(\rmi z\sqrt{\tilde{f}}+w
(2\eta+\tilde{g})\beta^\star/S)\big]}
}
\right]
\nonumber
\hspace*{-10mm}
\\[0.5mm]
&=& \lim_{\gamma\to\infty} 
 \frac{\sqrt{\gamma(2\eta\!+\!\tilde{g})}}{\sqrt{2\pi}} \int\!{\rm D}z~
\rme^{-\frac{1}{2}\gamma(2\eta+\tilde{g})\big[
\beta
-\rmi z\sqrt{\tilde{f}}/(2\eta+\tilde{g})-w
\beta^\star/S\big]^2}.
\nonumber
\\[-1mm]
\hspace*{-10mm}&&
\end{eqnarray}
For $\varrho(a)=\delta(a\!-\!1)$ we also  know that $2\eta+\tilde{g}=\tilde{u}^{-2}$ and $\tilde{f}=-v^2/\tilde{u}^4$. Hence the above integral reduces to
   \begin{eqnarray}
\lim_{N\to\infty}\Prob(\beta|\beta^\star) &=&
  \frac{1}{v\sqrt{2\pi}}\rme^{-\frac{1}{2}(\beta-w\beta^\star/S)^2/v^2}
\end{eqnarray}
This confirms what was suggested by simulation data and exploited in \cite{coolen2017replica}: if we plot inferred versus true association parameters in a plane, we will find for $L2$ priors and uncorrelated covariates a linear cloud with slope $w/S$ and zero-average Gaussian noise of width $v$. We have now proved  this analytically, for {\em any} generalized linear model.

\subsection{Correlated covariates}

 This is the more tricky case. We return to (\ref{eq:link_arbitrary_A}) and implement first the replica 
  symmetry ansatz,  i.e. $D_{\alpha 0}=d_0$ and $D_{\alpha\rho}=D\delta_{\alpha\rho}+d(1\!-\!\delta_{\alpha\rho})$ for $\alpha,\rho=1\ldots n$, so that we can proceed with our calculation:
   \begin{eqnarray}
   \hspace*{-25mm}
\lim_{N\to\infty}\Prob(\beta,\beta^\star) &=& \lim_{p\to\infty}  \lim_{\gamma\to\infty} \lim_{n\to 0}
 \frac{1}{p}\sum_{\mu=1}^p \delta(\beta^\star\!-\beta_\mu^0)\times
 \nonumber
\\
   \hspace*{-25mm}
&&\hspace*{-25mm}\frac{\int\!\rmd\tilde{\bbeta}
\rme^{-\frac{1}{2}d\tilde{\bbeta}\cdot\bA\tilde{\bbeta} -d_0\tilde{\bbeta}\cdot\bA\bbeta^0}
\int\!
\prod_{\alpha=1}^n\!\Big[ \rmd\bbeta^\alpha \rme^{
-\frac{1}{2}(D-d) \bbeta^\alpha\cdot\bA\bbeta^\alpha
}\!
\prod_{\nu=1}^p p^\gamma(\beta_\nu^\alpha)\Big] \delta(\tilde{\bbeta}\!-\!\sum_{\alpha=1}^n\bbeta^\alpha) \delta(\beta\!-\!\beta^1_\mu)
}
 { \int\!\rmd\tilde{\bbeta}\rme^{-\frac{1}{2}d\tilde{\bbeta}\cdot\bA\tilde{\bbeta} -d_0\tilde{\bbeta}\cdot\bA\bbeta^0}
\int\!\prod_{\alpha=1}^n\!\Big[ \rmd\bbeta^\alpha \rme^{
-\frac{1}{2}(D-d) \bbeta^\alpha\cdot\bA\bbeta^\alpha
}\!
\prod_{\nu=1}^p p^\gamma(\beta_\nu^\alpha)\Big] \delta(\tilde{\bbeta}\!-\!\sum_{\alpha=1}^n\bbeta^\alpha) 
}
\hspace*{-25mm}
\nonumber
\\
   \hspace*{-25mm}
&=& \lim_{p\to\infty}  \lim_{\gamma\to\infty} \lim_{n\to 0}
 \frac{1}{p}\sum_{\mu=1}^p \delta(\beta^\star\!-\beta_\mu^0)\times
 \nonumber
\\
   \hspace*{-25mm}
&&\hspace*{-15mm}\frac{\int\!\rmd\tilde{\bbeta}\rmd\hat{\bbeta}~W(\hat{\bbeta},\tilde{\bbeta}) H^{n-1}\!(\hat{\bbeta})
\int\!\rmd\bbeta^1 \rme^{-\rmi\hat{\bbeta}\cdot\bbeta^1\!
-\frac{1}{2}(D-d) \bbeta^1\cdot\bA\bbeta^1
}\!
\delta(\beta\!-\!\beta^1_\mu)\prod_{\nu=1}^p p^\gamma(\beta_\nu^1) 
}
 { \int\!\rmd\tilde{\bbeta}\rmd\hat{\bbeta}~ W(\hat{\bbeta},\tilde{\bbeta})
 H^{n}(\hat{\bbeta})
},
\nonumber
\\[-1mm]
\hspace*{-25mm}&&\label{eq:beta_beta_corr_1}
\end{eqnarray}
with 
\begin{eqnarray}
W(\hat{\bbeta},\tilde{\bbeta})&=&\rme^{\rmi\hat{\bbeta}\cdot\tilde{\bbeta}-\frac{1}{2}d\tilde{\bbeta}\cdot\bA\tilde{\bbeta} -d_0\tilde{\bbeta}\cdot\bA\bbeta^0},
\\
H(\hat{\bbeta})&=&\int\! \rmd\bbeta^\prime \rme^{-\rmi\hat{\bbeta}\cdot\bbeta^\prime
-\frac{1}{2}(D-d) \bbeta^\prime\cdot\bA\bbeta^\prime
}\!
\prod_{\nu=1}^p p^\gamma(\beta_\nu^\prime).
\end{eqnarray}
Note that $\hat{\bbeta},\tilde{\bbeta}\in\R^p$. For  $n\to 0$ the denominator evaluates to $(2\pi)^p$. Hence  expression (\ref{eq:beta_beta_corr_1}) can be simplified to
  \begin{eqnarray}
  \hspace*{-10mm}
\lim_{N\to\infty}\Prob(\beta,\beta^\star) &=& \lim_{p\to\infty}  \lim_{\gamma\to\infty} 
 \frac{1}{p}\sum_{\mu=1}^p \delta(\beta^\star\!-\beta_\mu^0)\times
\\
  \hspace*{-10mm}
&&\hspace*{-21mm}\int\!\frac{\rmd\tilde{\bbeta}\rmd\hat{\bbeta}}{(2\pi)^p}~
W(\hat{\bbeta},\tilde{\bbeta})
\left\{\!
\frac{
\int\!\rmd\bbeta^\prime \rme^{-\rmi\hat{\bbeta}\cdot\bbeta^\prime\!
-\frac{1}{2}(D-d) \bbeta^\prime\cdot\bA\bbeta^\prime
}\!
\delta(\beta\!-\!\beta^\prime_\mu)\prod_{\nu=1}^p p^\gamma(\beta_\nu^\prime) }
{
\int\!\rmd\bbeta^\prime \rme^{-\rmi\hat{\bbeta}\cdot\bbeta^\prime\!
-\frac{1}{2}(D-d) \bbeta^\prime\cdot\bA\bbeta^\prime
}\!
\prod_{\nu=1}^p p^\gamma(\beta_\nu^\prime) }
\right\}.
 \nonumber
\end{eqnarray}
We choose the Gaussian prior $p(\beta)\propto \exp(-\eta\beta^2)$, we write $\delta(\beta-\!\beta^\prime_\mu) $ in integral form, we introduce the unit vector $\hat{{\bf e}}^\mu$ with components $\hat{e}^\mu_\nu=\delta_{\mu\nu}$, we use $f=d$ and $g=D-d$,  and we do the Gaussian integrals where possible.  This gives
 \begin{eqnarray}
 \hspace*{-25mm}
\lim_{N\to\infty}\Prob(\beta,\beta^\star) &=&\lim_{p\to\infty}  \lim_{\gamma\to\infty} 
 \frac{1}{p}\sum_{\mu=1}^p \delta(\beta^\star\!-\beta_\mu^0)\int\!\frac{\rmd k}{2\pi}\rme^{\rmi k\beta}
\\
&&\times\int\!\frac{\rmd\tilde{\bbeta}\rmd\hat{\bbeta}}{(2\pi)^p}~
W(\hat{\bbeta},\tilde{\bbeta})
\left\{
\frac{
\int\!\rmd\bbeta^\prime \rme^{-\rmi(\hat{\bbeta}+k\hat{\bf e}^\mu)\cdot\bbeta^\prime\!
-\frac{1}{2}\bbeta^\prime\cdot[(D-d) \bA+2\gamma\eta\one]\bbeta^\prime
}
}
{
\int\!\rmd\bbeta^\prime \rme^{-\rmi\hat{\bbeta}\cdot\bbeta^\prime\!
-\frac{1}{2}\bbeta^\prime\cdot[(D-d) \bA+2\gamma\eta\one]\bbeta^\prime}
}
\right\}
  \nonumber
 \\
  \hspace*{-25mm}
 &=&
 \lim_{p\to\infty}  \lim_{\gamma\to\infty} 
 \frac{1}{p}\sum_{\mu=1}^p \delta(\beta^\star\!-\beta_\mu^0)\int\!\frac{\rmd k}{2\pi}\rme^{\rmi k\beta-\frac{1}{2}k^2[(D-d) \bA+2\gamma\eta\one]^{-1}_{\mu\mu}}
 \nonumber
 \\
  \hspace*{-25mm}
 &&\times\!  \int\!\frac{\rmd\hat{\bbeta}}{(2\pi)^p}\rme^{-k\hat{\bf e}^\mu\cdot[(D-d) \bA+2\gamma\eta\one]^{-1}\hat{\bbeta}
}\!
 \int\!\rmd\tilde{\bbeta}~\rme^{-\frac{1}{2}d\tilde{\bbeta}\cdot\bA\tilde{\bbeta} -\tilde{\bbeta}\cdot (d_0\bA\bbeta^0-\rmi\hat{\bbeta})}
 \nonumber
\\
 \hspace*{-25mm}
&=&
\lim_{p\to\infty}  \lim_{\gamma\to\infty} 
 \frac{1}{p}\sum_{\mu=1}^p \delta(\beta^\star\!-\beta_\mu^0)\int\!\frac{\rmd k}{2\pi}
 \rme^{\rmi k\big[\beta
 + d_0
 \hat{\bf e}^\mu\cdot [(D-d) \bA+2\gamma\eta\one]^{-1}
\bA
 \bbeta^0\big]
}
 \nonumber
 \\
  \hspace*{-25mm}
 &&\hspace*{0mm} \times 
 \rme^{-\frac{1}{2}k^2\Big[ [(D-d) \bA+2\gamma\eta\one]^{-1}
-d 
  [(D-d) \bA+2\gamma\eta\one]^{-1}
\bA
[(D-d) \bA+2\gamma\eta\one]^{-1}\Big]_{\mu\mu}
  }
\nonumber
\\
 \hspace*{-25mm}
&=& 
\lim_{p\to\infty}  \lim_{\gamma\to\infty} 
 \frac{1}{p}\sum_{\mu=1}^p \delta(\beta^\star\!-\beta_\mu^0)\int\!\frac{\rmd k}{2\pi}
 \rme^{\rmi k\Big[\beta+ d_0
 [ (g\bA+2\gamma\eta\one)^{-1}
\bA
 \bbeta^0]_\mu\Big]
}
 \nonumber
 \\
  \hspace*{-25mm}
 &&\hspace*{5mm} \times 
 \rme^{-\frac{1}{2}k^2\Big[ (g\bA+2\gamma\eta\one)^{-1}
-f   (g \bA+2\gamma\eta\one)^{-1}\bA(g \bA+2\gamma\eta\one)^{-1}\Big]_{\mu\mu}
  }.
\end{eqnarray}
Next we use the scaling with $\gamma$ of the order parameters, $f=\tilde{f}\gamma^2$, $g=\tilde{g}\gamma$ and $d_0=\gamma\tilde{d}_0$. For the integrals to converge we must have $\tilde{f}<0$ (which follows from solving the order parameter equations). We can then take $\gamma\to\infty$ and do the integral over $k$, giving
 \begin{eqnarray}
 \hspace*{-20mm}
\lim_{N\to\infty}\Prob(\beta,\beta^\star) &=& 
\lim_{p\to\infty} 
 \frac{1}{p}\sum_{\mu=1}^p 
 \frac{\delta(\beta^\star\!-\beta_\mu^0)}{\sqrt{2\pi |\tilde{f}|[ 
  (\tilde{g} \bA+2\eta\one)^{-1}\bA(\tilde{g} \bA+2\eta\one)^{-1}]_{\mu\mu}}}
  \label{eq:beta_relation_derived_app}
\\
 \hspace*{-20mm}
&&
\times
 \rme^{-\frac{1}{2}\Big[\beta+ \tilde{d}_0
 [ (\tilde{g}\one+2\eta\bA^{-1})^{-1}
 \bbeta^0]_\mu\Big]^2/|\tilde{f}|[ 
  (\tilde{g} \bA+2\eta\one)^{-1}\bA(\tilde{g} \bA+2\eta\one)^{-1}]_{\mu\mu}
}.
\nonumber
\end{eqnarray}
This is expression (\ref{eq:beta_relation_derived}) in the main text.

\section{Pathologies of generalization error minimization}
\label{app:generalization_error}

Here we illustrate the dangers of using the generalization error as an objective function to be minimized, by using logistic regression as an example. The generalization error $E_g\in[0,1]$ is the expected fraction of samples for which the true and the inferred model disagree on the outcome values, for samples drawn randomly from the population (as opposed to from the training set). In logistic regression we have $s=\pm 1$ and write $p(s|\bz,\bbeta)=\frac{1}{2}+\frac{1}{2}s\tanh(\bbeta\cdot\bz)$ (rescaling by $\sqrt{p}$ is not relevant here). 
If the true and inferred parameters are $\bbeta^\star$ and $\bbeta$, the generalization error is
\begin{eqnarray}
E_g&=& \int\!\rmd\bz~p(\bz)\sum_{s,s^\prime=\pm 1}\frac{1}{2}(1\!-\!ss^\prime)p(s|\bz,\bbeta^\star)p(s^\prime|\bz,\bbeta)
\nonumber
\\
&=&  \frac{1}{2}-\frac{1}{2}\int\!\rmd\bz~p(\bz)\tanh(\bbeta^\star\!\cdot\bz)\tanh(\bbeta\cdot\bz).
\end{eqnarray}
 If we were to use $E_g$ to optimize the inferred parameters (assuming it could be estimated without explicit knowledge of the true parameters $\bbeta^\star$), we would seek to minimize $E_g$ over $\bbeta$. We note the lower bound 
\begin{eqnarray}
E_g & \geq & \frac{1}{2}-
 \frac{1}{2}\int\!\rmd\bz~p(\bz)\Big|\tanh(\bbeta^\star\!\cdot\bz)\Big|.
 \label{eq:Eg_bound}
\end{eqnarray}
While $E_g$ indeed computes the fraction of samples  for which the two models disagree on the outcome, it does {\em not} measure whether the two models also use the same outcome probabilities.
To see this, imagine  choosing $\bbeta=\kappa\bbeta^\star$. Here one would find 
\begin{eqnarray}
\hspace*{-10mm}
\frac{\rmd}{\rmd\kappa}E_g&=&  -\int\!\rmd\bz~p(\bz)\Big(\frac{\bbeta^\star\!\cdot\bz}{2\sqrt{p}}\Big)
\tanh(\frac{\bbeta^\star\!\cdot\bz}{\sqrt{p}})\Big[1\!-\!\tanh^2(\kappa\frac{\bbeta^\star\!\cdot\bz}{\sqrt{p}})\Big]
<0.
\end{eqnarray}
Hence  the of $E_g$ minimum is found for $\kappa\to\infty$, where the (diverging) estimator $\bbeta=\kappa\bbeta^\star$ satisfies the lower bound (\ref{eq:Eg_bound}). 
The inferred model $p(s|\bz,\bbeta)=\frac{1}{2}+\frac{1}{2}s~ {\rm sgn}(\bbeta^\star\cdot\bz)$ would indeed get the maximum achievable fraction  of binary outcomes  predicted correctly, but it would believe erroneously that it has 100\% prediction accuracy. 

\section{Distribution of  empirical covariance matrices}
\label{app:towards_Wishart}

Here we evaluate expression
(\ref{eq:Ahat_measure}) further, to  convert the integral in (\ref{eq:linear_betastats_1}) over all $p\times p$ matrices into an integral over positive definite and symmetric ones. 
We write symmetric and antisymmetric parts of matrices $\bM$ as  $\bM^s$ and $\bM^a$, and transform integrations over all $\hat{\bA}$ into integrations over symmetric and antisymmetric parts. The (anti)symmetrization transformations involved induce identities such as $\rmd\bM=2^{\frac{1}{2}p(p-1)}\rmd\bM^s\rmd\bM^a$ and $\delta(\bM)=2^{-\frac{1}{2}p(p-1)} \delta(\bM^s)\delta(\bM^a)$, where $\rmd\bM^s=\prod_{\mu\leq \nu}\rmd M^s_{\mu\nu}$ and $\rmd\bM^a=\prod_{\mu< \nu}\rmd M^a_{\mu\nu}$.  Moreover,  
\begin{eqnarray}
\int\!\rmd \bM^a ~\rme^{\rmi {\rm Tr}(\bA^a\bM^a)}&=&  \pi^{\frac{1}{2}p(p-1)}\prod_{\mu<\nu}\delta(A^a_{\mu\nu}),
\\
\int\!\rmd \bM^s ~\rme^{\rmi {\rm Tr}(\bA^s\bM^s)}&=& 2^p\pi^{\frac{1}{2}p(p+1)}\prod_{\mu\leq \nu}\delta(A^s_{\mu\nu}).
\end{eqnarray}
 We can now compute $P(\hat{\bA})$ for the case where  $p(\bz)=[(2\pi)^{-p}{\rm Det}\bA]^{\frac{1}{2}}\rme^{-\frac{1}{2}\bz\cdot\bA\bz}$, giving
\begin{eqnarray}
P(\hat{\bA})&=&
\int\!\frac{\rmd\bQ}{(2\pi)^{p^2}}~\rme^{\rmi {\rm Tr}(\bQ^s\hat{\bA}^s)+\rmi {\rm Tr}(\bQ^a\hat{\bA}^a)}
[{\rm Det}(\one\!+\!\frac{2\rmi}{N}\bA\bQ^s)]^{-N/2}
\nonumber
\\
&=&\delta(\hat{\bA}^a)\Big(\frac{2\rmi}{N}\Big)^{-Np/2}\!({\rm Det}\bA)^{-N/2}
\nonumber
\\
&&\times 
\int\!\frac{\rmd\bQ^s}{(2\pi)^{\frac{1}{2}p(p+1)}}~\rme^{\rmi {\rm Tr}(\bQ^s\hat{\bA}^s)}
[{\rm Det}(\bQ^s\!-\!\frac{1}{2}N\rmi\bA^{-1})]^{-N/2}.
\end{eqnarray}
Thus $P(\hat{\bA})\rmd\hat{\bA}=2^{-\frac{1}{2}p(p-1)}[\delta(\hat{\bA}^a) \rmd\hat{\bA}^a][P(\hat{\bA}^s)\rmd\hat{\bA}^s]$, where
\begin{eqnarray}
P(\hat{\bA}^s)&=& 2^{\frac{1}{2}p(p-1)}\Big(\frac{2}{N}\Big)^{\!-Np/2}\!({\rm Det}\bA)^{-N/2}
\nonumber
\\
&&\hspace*{-2mm} \times\!
\int\!\frac{\rmd\bQ^s}{(2\pi)^{\frac{1}{2}p(p+1)}}~\rme^{\rmi {\rm Tr}(\bQ^s\hat{\bA}^s)}
[{\rm Det}(\rmi\bQ^s\!+\!\frac{1}{2}N\bA^{-1})]^{-N/2}.~~~
\label{eq:PhatA_intermediate}
\end{eqnarray}
We can now forget about the antisymmetric parts of $\hat{\bA}$, and average only over all symmetric matrices. The nontrivial integral in (\ref{eq:PhatA_intermediate}) is found in \cite{Ingham}, giving
\begin{eqnarray}
P(\hat{\bA}^s)&=& \Big(\frac{2}{N}\Big)^{\!-Np/2}\!
\frac{\rme^{-\frac{1}{2}N{\rm Tr}(\hat{\bA}^s\! \bA^{-1})}({\rm Det}\hat{\bA}^s)^{\frac{1}{2}(N-p-1)}}
{\pi^{\frac{1}{4}p(p-1)}({\rm Det}\bA)^{N/2}\prod_{j=\frac{1}{2}(N-p+1)}^{N/2} \Gamma(j)}.
\label{eq:Wishart}
\end{eqnarray}
Hence $P(\hat{\bA}^s)$ is a Wishart distribution with $N$ degrees of freedom. With $\Omega_p$ denoting the space of positive definite symmetric $p\times p$ matrices, and dropping the superscript $s$,  we may then summarize our result for (\ref{eq:linear_betastats_1})  as:
\begin{eqnarray}
P(\hat{\bbeta})&=& 
\int_{\Omega_p}\!\rmd\hat{\bA} ~\Big(\frac{2}{N}\Big)^{\!-Np/2}\!
\frac{\rme^{-\frac{1}{2}N{\rm Tr}(\hat{\bA}^s\! \bA^{-1})}({\rm Det}\hat{\bA}^s)^{\frac{1}{2}(N-p-1)}}
{\pi^{\frac{1}{4}p(p-1)}({\rm Det}\bA)^{N/2}\prod_{j=\frac{1}{2}(N-p+1)}^{N/2} \Gamma(j)}
\nonumber
\\&&\hspace*{20mm}
\times {\cal N}(\hat{\bbeta}|\hat{\bG}\bbeta^\star\!,\zeta (\Sigma^\star)^2\hat{\bG}\hat{\bA}^{-1}\hat{\bG}).
\end{eqnarray}

\end{document}